\documentclass{article}
\usepackage[T1]{fontenc}
\usepackage[latin1]{inputenc}
\usepackage{graphics}
\usepackage{graphicx}
\usepackage{longtable}
\usepackage{color}
\usepackage{pslatex}
\usepackage{hyperref}
\usepackage{sverb}
\usepackage{subfigure}

\usepackage{a4p}
\usepackage{rotating}

\usepackage{graphicx}
\usepackage{atlasphysics}
\usepackage{subfigure}
\usepackage{longtable}
\usepackage{multirow}
\usepackage{textcomp}
\usepackage{authblk,colortbl}
\usepackage{units}
\usepackage{amsmath,amssymb,amsfonts} 

\usepackage{authblk}

\usepackage{rotating}

\begin{document}

\begin{titlepage}
 
\begin{flushright} 
{ \bf IFJPAN-IV-2016-11 
} 
\end{flushright}
 
\vskip 30 mm
\begin{center}
{\bf\huge  Separating Electroweak  and Strong interactions}\\
\vskip 3 mm
{\bf\huge  in Drell-Yan processes at LHC:}\\
\vskip 3 mm
{\bf\huge leptons angular distributions and reference frames. }
\end{center}
\vskip 13 mm

\begin{center}
   {\bf E. Richter-Was$^{a}$ and Z. Was$^{b}$  }\\
   \vskip 3 mm
       {\em $^b$ Institute of Physics, Jagellonian University, Lojasiewicza 11, 30-348 Cracow, Poland} \\
       {\em $^c$ Institute of Nuclear Physics, PAN, Krak\'ow, ul. Radzikowskiego 152, Poland}\\ 
\end{center}
\vspace{1.1 cm}
\begin{center}
{\bf   ABSTRACT  }
\end{center}

Among  the physics goals of  LHC experiments,  precision tests of 
the Standard Model  in the Strong and Electroweak sectors play 
an important role. Because 
of nature of the proton-proton processes, observables based on the measurement of 
the direction and energy of leptons provide the most precise signatures. 
In the present paper, we concentrate on the  angular distribution of  
Drell-Yan process leptons, in the lepton-pair rest-frame.
The vector nature of the intermediate state imposes that distributions are to a good precision described 
by spherical polynomials of at most  second order.

We show that with the proper choice of the coordinate frames, only one 
coefficient in this polynomial decomposition remains sizable, 
even in the presence of one or two high $p_T$ jets. 
The necessary stochastic choice of the  frames relies
on  probabilities independent from any 
coupling constants.

This remains true  when  
one or two partons accompany the lepton pairs. In this way electroweak 
effects can be better separated from strong interaction ones 
for the benefit of the interpretation of the measurements.

Our study exploits properties of   single gluon emission matrix elements 
which are clearly visible if a conveniently chosen  form of their
representation is used. We rely also on  
 distributions obtained from matrix element based Monte Carlo generated 
samples of  events with two  
leptons and up to  two additional partons in test samples. 
Incoming  colliding protons' partons are distributed
accordingly to PDFs and  are strictly   collinear to the corresponding beams.  
 
\vskip 1 cm

\vspace{0.2 cm}

\vspace{0.1 cm}
\vfill
{\small
\begin{flushleft}
{   IFJPAN-IV-2016-11
\\ May 2016
}
\end{flushleft}
}
\end{titlepage}

\clearpage
\section{Introduction}

The main purpose of the LHC experiments \cite{Aad:2008zzm,Chatrchyan:2008aa} is to search for the effects 
of New Physics. This programme continues after the breakthrough
discovery of the Higgs boson \cite{Aad:2012tfa,Chatrchyan:2012xdj} and measurement of its main properties~\cite{ATLAS-CONF-2015-044}. 
In parallel to searches of  New Physics, see eg.  
\cite{Aad:2015baa,Aad:2015iea,Khachatryan:2016kdk}, a program of precision measurements in the domain of Electroweak (EW) and 
Strong (QCD) interactions is on-going. 
This is the keystone of the programme of establishing the Standard Model as a fundamental theory. 
It is focused around two main directions: searches (setting upper limits)
for anomalous couplings and precision measurements of the Standard Model parameters.
Precision measurements of the production and decay of intermediate $Z$ and $W$ 
bosons represent the primary group of measurements of the second domain, see eg. ~\cite{Aad:2015auj,Aad:2015uau,CMS:2014jea,Khachatryan:2016yte}.

In the following, let us concentrate on the  phenomenology of distributions for 
leptons originating from Drell-Yan processes, and in particular on possible 
separation of electroweak and strong phenomena. There is a multitude of 
publications devoted to the theoretical preparation for such measurements, eg.~\cite{Catani:2015vma,Grazzini:2015wpa,Dittmaier:2015rxo}. 
They are mostly discussing evaluation of higher order effects of strong 
interactions. The commonly used scheme for controlling the angular distribution of leptons is
a  Collins-Soper frame \cite{Collins:1977iv} motivated by the factorisation theorem~\cite{Collins:1989gx}.

However, it is long known that the distribution of leptons produced from a vector intermediate state such as $e^+e^- \to \mu^+ \mu^-$ 
can be described as a sum of contributions consisting of Born level distributions with factorized terms independent of the relative
direction of the leptons.
The analytical form for  these distributions is a result of an explicit first 
order calculation in $O(\alpha_{QED})$, given for example in \cite{Berends:1983mi}, formula (3.4). 
This expression is directly applicable to initial state emission of gluons in case of   $q \bar q \to \ell^+ \ell^-$
hard process in proton-proton collisions as well.
It is valid all over the phase space. Let us not go into details of 
the mass effects, that is the terms of the order of $\frac{m^2}{s}$. 
This complicates details, but does not change the general principles.
In \cite{Kleiss:1990jv} it is argued that such result may hold in more general case as well.
In particular, that they break only at the level of complete calculations for two hard emitted partons.
Such considerations were essential in establishing the {\tt PHOTOS} Monte Carlo~\cite{Barberio:1990ms}
algorithm for bremsstrahlung in decays of resonances and particles 
 from its very beginning. It is of interest 
if considerations of that sort can be useful for the phenomenology of initial state parton emissions as well.

In \cite{Mirkes:1992hu} it was shown, that the distributions of leptons in the rest frame of the lepton pairs, for the processes 
of $pp$ collisions are described by spherical harmonics of at most second order. 
In the formulation of the decomposition into a sum of trigonometrical 
polynomials, up to 8 non-zero coefficients are expected, showing dependence on the lepton pair
invariant mass, transverse momenta and rapidity.
In contrast, at the Born level, only one coefficient with functional dependence on electroweak parameters suffices. 

The question is, whether the wealth of 8 coefficients brings some more information on the hard electroweak process and on QCD dynamics or is  
a consequence of geometrical alteration due to presence of accompanying jets only; thus bringing no additional physical information.
On the other hand, they are based 
on precisely measured kinematics of leptons, usually more precise than that of jets. 

By inspection of formula (3.4) from \cite{Berends:1983mi}, we find that in case of  gluon (photon) emission the distribution of the lepton  angles
in the lepton pair system consists of two Born-like contributions weighted respectively
by squares of the energies of the two incoming partons as seen in rest-frame of the lepton pair system. Such weight depends 
only on the $2 \to 3$ process kinematics (no dependence on couplings) and can be used 
over the entire range of energy and momentum of the emitted parton. That offers  easy application.
The scattering  angle is then the angle (also in the same frame) 
 between the direction of outgoing leptons and respectively first/second incoming parton. The azimuthal angle can be chosen as the same one 
for the two Born's as of formula (3.4) from \cite{Berends:1983mi}, as the angle 
between two oriented half-planes built from: (i)  the direction of the beams and 
emitted parton and (ii) the direction of the outgoing
leptons and emitted parton. It does not matter if such an angle is defined in the lepton pair frame or reaction  frame%
\footnote{In fact,
until we do not have to consider transverse spin effects of secondary
decays, out of scope of the present publication, there is a freedom in
definition of reference direction for azimuthal angle. Matrix element
does not depend on it.}.

A similar solution was used for the {\tt PHOTOS} Monte Carlo where results from
\cite{RichterWas:1994ep,RichterWas:1993ta} were instrumental.

Let us stress that the above strategy was developed in the first place for initial state QED bremsstrahlung. It is thus of interest 
to which degree it can be extended to the case of $pp$ collisions
where instead of bremsstrahlung photons, jets are emitted: not always gluonic ones, but sometimes of quark partons as well. 
For that purpose we   study  samples generated with the help of Monte Carlo programs 
featuring explicit matrix elements. 

Our paper is organized as follows. Section \ref{sec:QCD} is devoted to the presentation  of strong and electroweak 
interaction features which are important for the explanation of our results. %
In Section \ref{sec:angular} we list  properties and definitions important for the analysis, using an expansion of the angular 
distribution of decay products of Drell-Yan process in the rest-frame of the lepton pair. We elaborate on possible 
choices of the coordinate frame orientation.  The choice of orientation can be optimized with the help of  matrix element results or 
leading logarithm calculations; the  variants are presented and discussed.

In Section~\ref{sec:numerical} we collect numerical results for the distributions in 
the case   of $pp  \to l^+l^- j$ and $pp  \to l^+l^- j j$, for invariant mass of the lepton pair around the $Z$-boson peak.
The aim is to demonstrate that with the help of probabilistic splitting
of each event into configurations defined with respect to  two distinct  frames,
the results of formula (3.4) from \cite{Berends:1983mi}
are reproduced and indeed only one non-zero coefficient in the decomposition 
of the angular distribution is needed. We use series of $pp  \to l^+l^- j$ 
events generated with {\tt MadGraph}~\cite{Alwall:2011uj}. 
We first concentrate on a sample where the outgoing parton is a gluon, 
which nonetheless carries substantial transverse momentum, $p_T$. In the consecutive results and 
 subsections we release this restriction of outgoing gluon only.
Later, we turn to $pp \to l^+l^- j j$ events.  Again, a sample of events
generated by {\tt MadGraph} is studied.  Finally, we provide also results for 
{\tt Powheg+MiNLO} \cite{Nason:2004rx,Alioli:2010xd} generated sample of $pp \to Z\ j$ events at QCD NLO.

In the following Section~\ref{sec:interpretation} we discuss results obtained for different sub-groups of events, 
where different production mechanisms dominate, and how the consequences 
of choices for reference frames point to 
the general pattern. Finally, in Section~\ref{sec:summary} we conclude the paper.

\section{QCD and EW } \label{sec:QCD}

The predictive power of QCD is based on the factorisation theorem \cite{Collins:1989gx}.
It provides a framework for separating out long-distance effects in 
hadronic collisions. In consequence it allows for systematic prescriptions and tools to calculate the short-distance
dynamics perturbatively, the same time allowing for identifying the leading nonperturbative 
long-distance effects which can be extracted from experimental measurements or by numerical calculations 
in Lattice QCD.   

Let us list some references 
to illustrate the developments in this area without ambitions of their completeness.
Theoretical investigations of Drell-Yan pair production~\cite{DrellYan70} have a long history. 
It is one of the few processes in hadron-hadron collisions where the collinear QCD factorization 
has been rigorously proven \cite{Collins82, Collins84, Collins85, Collins88}. 
Within this framework, the NLO pQCD calculations of inclusive cross sections have been performed
\cite{Altarelli78,Altarelli79} and later it was done on up to NNLO accuracy 
\cite{Matsuura89, Matsuura91}.
In particular, fully exclusive NNLO pQCD calculations became available, including the leptonic decay 
of the intermediate Z boson \cite{Melnikov06a, Melnikov06b, Catani09}.

The question of the input from Electroweak sector of the Standard Model is important, 
especially for distributions of leptons originating from intermediate $Z/\gamma^*$ state. We have addressed numerical 
consequences of this point recently in \cite{Kalinowski:2016qcd} in the 
context of $\tau$ lepton polarization in Drell-Yan processes at the LHC.
A wealth of publications was devoted during last years to this point, see e.g.~\cite{Barze':2013yca,Dittmaier:2014qza,Dittmaier:2015rxo}. 
 Let us point however to constraints from limitations  to 
separating interactions into Electroweak and QCD ones, which
are well known,  since more than 15 years, see e.g.~\cite{Kulesza:1999gm}.

In the following, we will concentrate on numerical analysis of processes at tree level for parton-parton collisions 
into lepton pair and accompanying jets.
Even though  approach is limited to leading, tree-level predictions, it provides input for general discussions.
Such configurations constitute parts of the higher order corrections, or can be seen as the lowest order terms
but for observables of tagged high $p_T$ jets. 

\section{Angular decay distribution}\label{sec:angular}

The measurement of the angular distribution of leptons from the decay of a gauge boson $V \to \ell \ell$ 
where $V = W, Z$ or $\gamma^*$, produced in  hadronic collisions via a 
Drell-Yan-type process $ h_1 + h_2 \to V + X$
provides a detailed test of the production mechanism. If the  
gauge boson is produced at high transverse momentum,
 one can define an event plane 
spanned by the beam and the gauge boson directions.
This reference plane can be used for 
study of lepton-hadron current  correlation effects.

In fact, these correlations are described by the set of nine hadronic structure functions
which can be calculated within the context of the parton model using perturbative QCD.
Such measurements are already completed for Tevatron~\cite{Aaltonen:2011nr} and LHC~\cite{Khachatryan:2015paa,ATLASDIS2016}.

Following the conventions and notations of \cite{Mirkes:1992hu,arXiv9406381}, let us briefly recall that 
the lepton-hadron correlations are described by the contraction of the lepton tensor $L_{\mu \nu}$ with the
hadron tensor at the parton level $H^{\mu \nu}$, where  $L_{\mu \nu}$ acts as an analyser of the 
structure of $H^{\mu \nu}$ which carries the effective information on the polarisation
of the gauge boson produced in the interaction.
The angular dependence can be extracted introducing helicity cross-sections corresponding to the 
non-zero combinations of the polarisation density matrix elements
\begin{equation}
  H_{m m'} = \epsilon^*_{\mu}(m) H^{\mu  \nu} \epsilon_{\nu}(m')
\end{equation}
where $m, m' = +1, 0,-1$ and 
\begin{equation} 
  \epsilon_{\mu}(\pm 1) = \frac{1}{\sqrt{2}} (0; \pm 1, -i, 0), \ \ \ \ \epsilon_{\mu} (0) = (0;0,0,1)
\end{equation}
are the polarisation vectors for the gauge boson defined with respect to the chosen lepton-pair (gauge boson) rest-frame. 
The angular dependence of the differential cross-section can be written as
\begin{equation}
 \label{Eq:master1}
  \frac{ d\sigma}{dp_T^2 dY d\Omega^*} = 
     \Sigma_{ \alpha=1}^{9} g_{ \alpha}( \theta,  \phi)
     \frac{3}{ 16 \pi} \frac{d \sigma ^{\alpha }}{ dp_T^2 dY}, %
\end{equation}
where the $ g_{ \alpha }( \theta,  \phi)$ represent harmonic polynomials of the second order,
multiplied by normalisation constants and $d \sigma ^{\alpha }$ denote helicity cross-sections, corresponding
to nine helicity matrix elements.
The angle $\theta$ and $\phi$ in $d\Omega^* = d \cos\theta d\phi$ are the polar and azimuthal decay angles 
of the leptons in the lepton-pair rest-frame.

We can conveniently rewrite  Eq.~(\ref{Eq:master1}), explicitly defining polynomials and corresponding 
coefficients%
\begin{eqnarray}
 \label{Eq:master2}
  \frac{ d\sigma}{dp_T^2 dY d \cos\theta d\phi} & = & \frac{3}{ 16 \pi} \frac{d \sigma ^{ U+ L }}{ dp_T^2 dY} \\
  && [  (1 + \cos^2(  \theta)) + 1/2 A_0 (1 - 3 \cos^2(  \theta)) +  A_1 \sin( 2 \theta)\cos( \phi) + 1/2 A_2 \sin^2( \theta) \cos( 2 \phi)  \nonumber\\
  && +  A_3 \sin( \theta) \cos( \phi)+  A_4 \cos( \theta) + A_5 \sin^2( \theta) \sin( 2 \phi) + A_6 \sin(2 \theta) \sin(  \phi) + A_7 \sin( \theta)\sin( \phi) ] \nonumber 
\end{eqnarray}
where $d \sigma ^{ U+ L }$ denotes the unpolarised differential cross-section 
(a convention used in several papers of the 80's). 
The coefficients $A_i(p_T, Y)$ are related to ratios of corresponding cross-sections of  
intermediate state  helicity configurations \cite{Mirkes:1992hu}. 

 The dynamics of the production process is  hidden in the angular 
coefficients $A_i (p_T, Y)$. This allows us to treat the problem in a model independent manner. 
In particular, as we will see, all the hadronic physics is described 
implicitly by the angular coefficients  and it decouples (the $A_4$ is an exception) 
from the well understood leptonic and intermediate boson physics. 
As we have already pointed out, the spherical angles $\theta$ and $\phi$  
define the orientation of the lepton direction in 
the lepton-pair rest-frame. Let us stress that the actual choice of 
the orientation of coordinate frames one uses represents an important topic; 
we will return to it later.

The lepton angular distribution in the lepton-pair rest-frame is determined by the gauge boson polarisation
and by the range of the invariant mass of the lepton pairs, which determines indirectly the admixture of the $Z/\gamma^*$
interference to the cross-section.
The general structure of the angular distribution is given by nine helicity cross-sections corresponding to
the nine spin density matrix elements for the gauge boson \cite{Mirkes:1992hu}.
In the LO parton subprocesses $(q \bar q \to V \to \ell \ell')$ only the transverse polarisation gauge 
boson contributes to the helicity density matrix. At higher order, in general, all three polarisations 
of the gauge boson contribute.
At $O(\alpha_S)$ in perturbative QCD the angular distribution is described by six helicity cross-sections,
which, for fixed range of invariant mass of lepton pairs, are functions of the transverse momentum and rapidity of the gauge boson. At the $O(\alpha_S^2)$,
contributions from three additional helicity cross-sections are non-zero. 
It is important that higher order corrections
do not break the above picture in numerically important terms.

As we will see, for the choices of the frames discussed here,
in the limit of zero transverse momenta all coefficients,
except  $A_4$ which is defined by electroweak couplings,
vanish. 
The $(p_T, Y)$ dependence of the $A_i$ coefficients 
vary strongly with the choice of the reference frame. 
Let us present now the variants of the reference frame definition we are going to
 use.

\subsection{\bf Collins-Soper frame}

The well known and broadly used \cite{Collins:1977iv} Collins-Soper reference frame is defined as a rest-frame of the lepton-pair ($Z$-boson), with the 
polar and azimuthal angles constructed using proton directions in that frame. 
Since the $Z$-boson has a transverse momentum, 
the directions of initial protons are not collinear in the lepton-pair ($Z$-boson) rest frame. 
The polar axis (z-axis) is defined in the lepton-pair rest-frame such that it is bisecting the 
angle between the momentum of one of the proton and inverse of the momentum of the second one. 
The sign of the z-axis is defined by the sign of the lepton-pair momentum with respect to z-axis 
in the laboratory frame. To complete the coordinate system the y-axis is defined as the normal vector 
to the plane spanned by the two incoming proton momenta and the x-axis is chosen to set a right-handed 
Cartesian coordinate system with the other two axes. 
Polar and azimuthal angles are calculated with respect to the outgoing lepton 
and are labeled  $\theta$ and $\phi$ respectively. In the case of zero transverse momentum of the lepton-pair, 
the direction of the y-axis is arbitrary. 
Note, that there is an ambiguity in the definition of the $\phi$
angle in the Collins-Soper frame.
The orientation of the $x$ axis here follows convention of eg.~\cite{Mirkes:1992hu,Karlberg:2014qua,Gavin:2010az}.
  
The formula for $\cos\theta$ can be expressed using directly momenta of the outgoing 
leptons in the laboratory frame~\cite{CarloniCalame:2007cd}.

\begin{equation}
\label{eq:costhetastar}
\cos\theta=  \frac{p_z (\ell^+ \ell^-)}{|p_z (\ell^+ \ell^-)|} \frac{2}{m(\ell^+ \ell^-)\sqrt{m^2(\ell^+ \ell^-)+p_T^2(\ell^+ \ell^-)}}(P_1^+P_2^--P_1^-P_2^+)
\end{equation}

\noindent with  

\begin{equation*}\label{eq:costhetastar:P}
 P_{\textit{i}}^\pm=\frac{1}{\sqrt{2}}(E_{\textit{i}}\pm p_{z,\textit{i}})
\end{equation*}

\noindent where $E_{\textit{i}}$  and p$_{z,\textit{i}}$ are respectively the energy and longitudinal momentum of
the lepton ($i=1$) and anti-lepton ($i=2$) and p$_z (\ell^+ \ell^-)$ denotes the longitudinal momentum of the lepton system, m$ (\ell^+ \ell^-)$ its invariant mass.
The $\phi$ angle is calculated as an angle of the lepton in the plane of the x and y axes in the Collins-Soper frame. 

Only the four-momenta of outgoing leptons and incoming proton directions 
are used. That is why the frame is very convenient for experimental 
purposes.

\subsection{\bf {\tt Mustraal} frame}
\label{Sec:coefsMUSTframe}

The {\tt Mustraal} reference frame is also defined as a rest frame of the lepton
pair.
It have been proposed and used for the first time in {\tt Mustraal} Monte Carlo program~\cite{Berends:1983mi} 
for the parametrization of the phase space. 
The  {\tt Mustraal} Monte Carlo program was constructed for muon pair 
production at LEP. 
Since then, it has been  used 
successfully, over LEP time, in other  Monte Carlo programs as well.
Resulting  optimal frame was used to minimise higher order corrections from initial state radiation to the 
$e^+ e^- \to Z/\gamma^* \to f \bar f$ 
for the algorithm implementing genuine weak effects in the LEP era Monte Carlo program {\tt KORALZ} \cite{koralz4:1994}. 
A slightly different variant
 was successfully used in the {\tt Photos} Monte Carlo program ~\cite{Davidson:2010ew} for simulating QED radiation in decays of particles and 
resonances. The parametrization was useful not only for compact representation of single photon emissions
but for multi-emission configurations as well.

In this paper, implementation of the {\tt Mustraal} frame has been extended to the case of $pp$ collision and one or two partons in the final
state accompanying  Drell-Yan production of the lepton pairs. 
Let us describe in more detail the
 implementation of this phase space parametrization. As we will use it for previously generated events,
we will calculate angles and energies from the outgoing particles four-momenta, rather than the opposite way around.

The parametrization is constructed in consecutive steps, each following one, using 
information obtained from the previous. %
\begin{itemize}
\item
We start from the following information, which turns out to be sufficient: (i)  The 4-momenta and charges of outgoing 
leptons $\tau_1$, $\tau_2$. (ii) The sum of 4-momenta of all outgoing partons.
\item
The orientation of incoming beams $b_1$, $b_2$ is fixed as follows:
 $b_1$ is chosen to be always along positive $z$-axis of the laboratory 
frame and $b_2$ is anti-parallel to $z$ axis. The information on incoming 
partons of $p_1$, $p_2$ is not taken from the event record. It is 
recalculated from kinematics of outgoing particles and knowledge 
of the center of mass energy of colliding protons. 
In this  convention the  energy
 fractions $x_1$ and $x_2$ of $p_1$, $p_2$  carried by colliding partons,
 define also  the  3-momenta which are along  $b_1, b_2$ respectively.
\item
The flavour of incoming partons (quark or antiquark) is  attributed as follows:
 incoming parton of larger $x_1$ ($x_2$) 
is assumed to be the quark. This is equivalent to choice that the quark 
follow direction of the outgoing $\ell \ell$ system,
similarly as it is defined for the Collins-Soper 
frame\footnote{Improvements are possible but this would require 
discussion of systematic errors due to PDFs parametrisation.}. This choice is necessary 
to fix sign of $\cos \theta_{1.2}$ defined later. 
\item
The 4-vectors of incoming partons and outgoing leptons are boosted into lepton-pair rest frame.
\item To fix orientation of the event we use versor $\hat x_{lab}$ of the laboratory reference frame.
It  is boosted into lepton-pair rest frame as well.
It will be used in definition  of azimuthal angle $\phi$, which has to extend 
over the range $(0, 2\pi)$.
\item
We first  
calculate  $\cos \theta_1$ (and  $\cos \theta_2$) of the angle between the outgoing lepton and incoming quark (outgoing anti-lepton and incoming anti-quark) directions. 
\begin{equation}
 \cos\ \theta_1 = \frac{\vec{\tau_1} \cdot \vec{p_1}}{|\vec{\tau_1}| | \vec{p_1}|},  \ \ \ \ \ \ \ \ \ \
 \cos\ \theta_2 = \frac{\vec{\tau_2} \cdot \vec{p_2}}{|\vec{\tau_1}| | \vec{p_2}|}
\end{equation}
\item
The azimuthal  angles $\phi_1$ and $\phi_2$ corresponding to $\theta_1$ and $\theta_2$  are defined as follows. We first define $\vec{e_{y_{1,2}}}$ versors and with their help  later $\phi_{1,2}$  as: 
\begin{equation}
\vec{e_y} = \frac{\vec{x_{lab}} \times \vec{p_2}}{|\vec{e_y|}}, \ \ \ \ \ 
\vec{e_x} = \frac{\vec{e_y} \times \vec{p_2}}{|\vec{e_x}|} \nonumber
\end{equation}
\begin{equation}
 \cos\ \phi_1 = \frac{\vec{e_x} \cdot \vec{\tau_1}}{\sqrt{(\vec{e_x} \cdot \vec{\tau_1})^2 + (\vec{e_y} \cdot \vec{\tau_1})^2}} \ \ \ \ \ \
 \sin\ \phi_1 = \frac{\vec{e_y} \cdot \vec{\tau_1}}{\sqrt{(\vec{e_x} \cdot \vec{\tau_1})^2 + (\vec{e_y} \cdot \vec{\tau_1})^2}} \ \ \ \ \ \
\end{equation}
and similarly for $\phi_2$:
\begin{equation}
\vec{e_y} = \frac{\vec{x_{lab}} \times \vec{p_1}}{|\vec{e_y}|}, \ \ \ \ \ 
\vec{e_x} = \frac{\vec{e_y} \times \vec{p_1}}{|\vec{e_x}|} \nonumber
\end{equation}
\begin{equation}
 \cos\ \phi_2 = \frac{\vec{e_x} \cdot \vec{\tau_2}}{\sqrt{(\vec{e_x} \cdot \vec{\tau_2})^2 + (\vec{e_y} \cdot \vec{\tau_2})^2}} \ \ \ \ \ \
 \sin\ \phi_2 = \frac{\vec{e_y} \cdot \vec{\tau_2}}{\sqrt{(\vec{e_x} \cdot \vec{\tau_2})^2 + (\vec{e_y} \cdot \vec{\tau_2})^2}}. \ \ \ \ \ \
\end{equation}
\item
Each event contributes with  two Born-like kinematics configurations $\theta_1\phi_1$, ($\theta_2\phi_2$),
 respectively with
$wt_1$ (and $wt_2$) weights;  $wt_1+wt_2=1$ where
\begin{equation}
\label{Eq:weights}
wt_1 = \frac{E_{p1}^2 (1 + \cos^2\theta_1)}{E_{p1}^2 (1 + \cos^2\theta_1)+E_{p2}^2 (1 + \cos^2\theta_2)}, \ \ \ \ \ \
wt_2 = \frac{E_{p2}^2 (1 + \cos^2\theta_2)}{E_{p1}^2 (1 + \cos^2\theta_1)+E_{p2}^2 (1 + \cos^2\theta_2)}.
\end{equation}
In the calculation of the weight, incoming partons energies $E_{p1}, E_{p2}$ in the rest frame of lepton pair are used, 
but not their couplings or flavours. That is also why, instead of $\sigma_B(s, \cos\theta)$ the
simplification  $(1 + \cos^2\theta)$
is used in Eq.~(\ref{Eq:weights}). Dependence on the sign of  $\cos\theta$ drops out\footnote{Stronger simplification is explored in \cite{Peng:2015spa}.
As a consequence they predict eg. different dominant dependence on electroweak couplings of $A_7$ than in \cite{Mirkes:1992hu}.
In our work we evaluate an effect of approximation with the help of numerical studies of results from Monte Carlo
programs based on explicit matrix element calculations.
Our choice of frame does not depend on couplings present in the matrix elements. However 
it depends on some general geometrical properties of the emission amplitudes. 
That is why our observations
are not in conflict with statements of reference \cite{Faccioli:2011pn}:
we take into account  dominant parts of intrinsic properties of the production mechanism 
because they are of a geometrical nature.   }.
\item
The above definitions of $\phi_{1,2}$  are not used directly.
The $\phi_1$ and  $\phi_2$ are retrieved as $\mathrm{acos}(\cos\ \phi_1)$ and  $\mathrm{acos}(\cos\ \phi_2)$
as explained above. However,
to extend to the full  range $(0, 2\pi)$, 
the directions of the incoming partons $p_1$, $p_2$ and outgoing parton (sum of outgoing partons) $p_3$ 
of the laboratory frame 
and $\tau_1$, $\tau_2$ of the $\tau$-pair rest  frame are used to calculate only sign of the following 
\begin{equation}
  \mathrm{sign}\Bigl((\vec{p_3} \times (\vec{p_3} \times \vec{\tau_1})) \cdot (\vec{p_3} \times \vec{p_1})\Bigr). \ \ \ \ \ 
\end{equation}
 For $\mathrm{sign} < 0$, azimuthal angle $\phi = 2 \pi - \mathrm{acos}(\cos\ \phi)$ is taken.
\item
Let us point out once again, that our weights $wt_{1,2}$ are  independent from
coupling constants of matrix element. 
The four-momenta of incoming partons are not
needed; necessary information is obtained from final states.
That is also why the solution is stable in cases where further 
sources of $p_T$ beyond one or two reconstructed jets would be present. 
\end{itemize}

The algorithmic logic of the above construction, the extended {\tt Mustraal} frame, is quite 
similar to the one developed for the {\tt PHOTOS} Monte Carlo \cite{Barberio:1990ms} 
or for the inclusion of genuine weak corrections 
(and final state bremsstrahlung) in {\tt KORALZ} \cite{koralz4:1994}. 
Experience from unfinished projects \cite{RichterWas:1987gk,RichterWas:1988gi} was of importance as well.

Unlike the case of the Collins-Soper frame, 
the {\tt Mustraal} frame requires not only information on 4-momenta of 
outgoing  leptons but also on outgoing jets (partons). 
The information on jets (partons),
is used to approximate the directions and energies of incoming partons
for calculation of $wt_{1,2}$ and  $\theta$, $\phi$ angles. 
This does not have to be very precise but can introduce 
additional experimental systematics, and thus requires attention.

\section{Numerical results} \label{sec:numerical}

Let us now present numerical results. We will use samples of events generated with the
{\tt MadGraph} matrix element Monte Carlo for Drell-Yan production 
of $\tau$-lepton pairs, with $m_{\tau \tau} = 80 - 100$ GeV and 13 TeV $pp$ collisions. Lowest order spin amplitudes 
are used in this program for the parton level process. 
For the EW scheme we have used default initialisation of the {\tt MadGraph} with on-shell definition of $\sin^2\theta_W=1-m_W^2/m_Z^2 = 0.2222$,
which determines value of the axial coupling for leptons and quarks to the Z-boson. 
The incoming partons are distributed accordingly to PDFs (using CTEQ6L1 
PDFs \cite{Pumplin:2002vw} linked through LHAPDF v6 interface)
and remain precisely collinear to the beams. Two samples of 2M  and 20M events were prepared\footnote{The analysed sample of events 
show a rather large instability of the numerical predictions for the $A_i$ coefficients which by far exceeds statistical
fluctuations. This may be the consequence of limitations due to pretabulations. That is why we have not pursued generation of higher statistic samples.}, 
the first one features $\tau\tau j$ final states and the other one  $\tau\tau jj$.
At this level, jet (j) denotes outgoing parton of unspecified flavour. 
In the generated sample information on incoming and outgoing partons flavours
are stored and we will use this information to define subsamples.

\begin{figure}
  \begin{center}                               
{
   \includegraphics[width=7.5cm,angle=0]{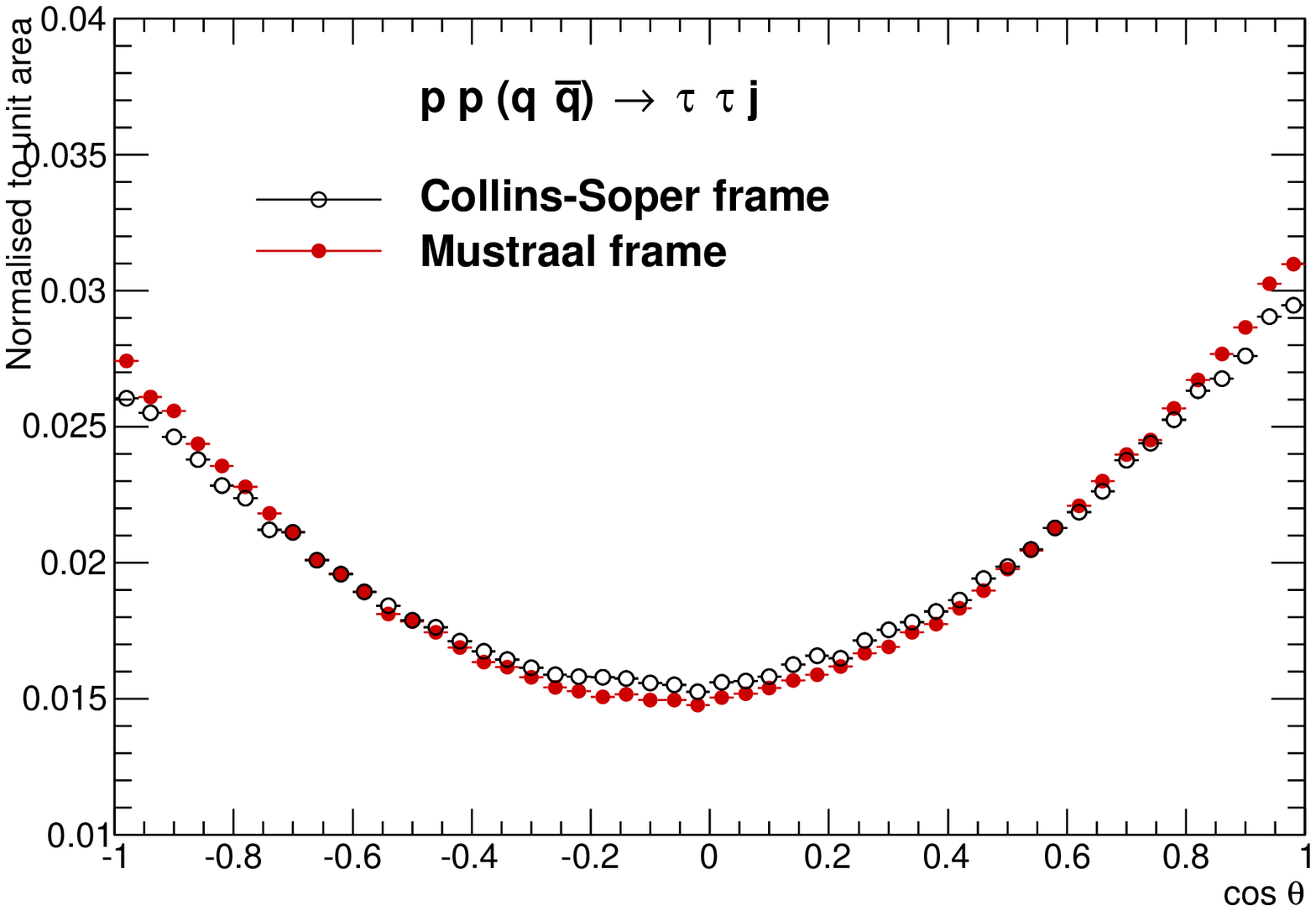}
   \includegraphics[width=7.5cm,angle=0]{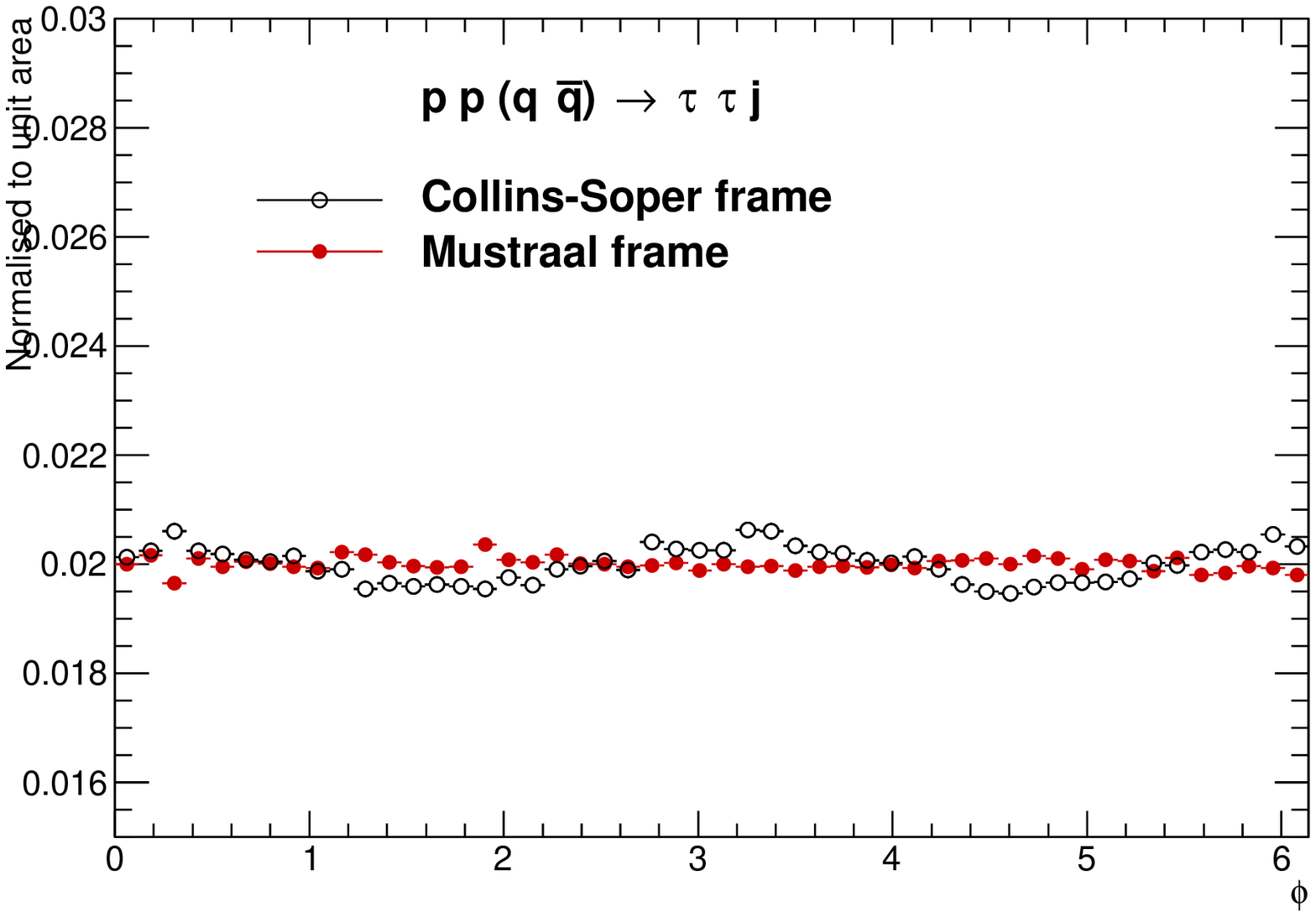}
}
{
   \includegraphics[width=7.5cm,angle=0]{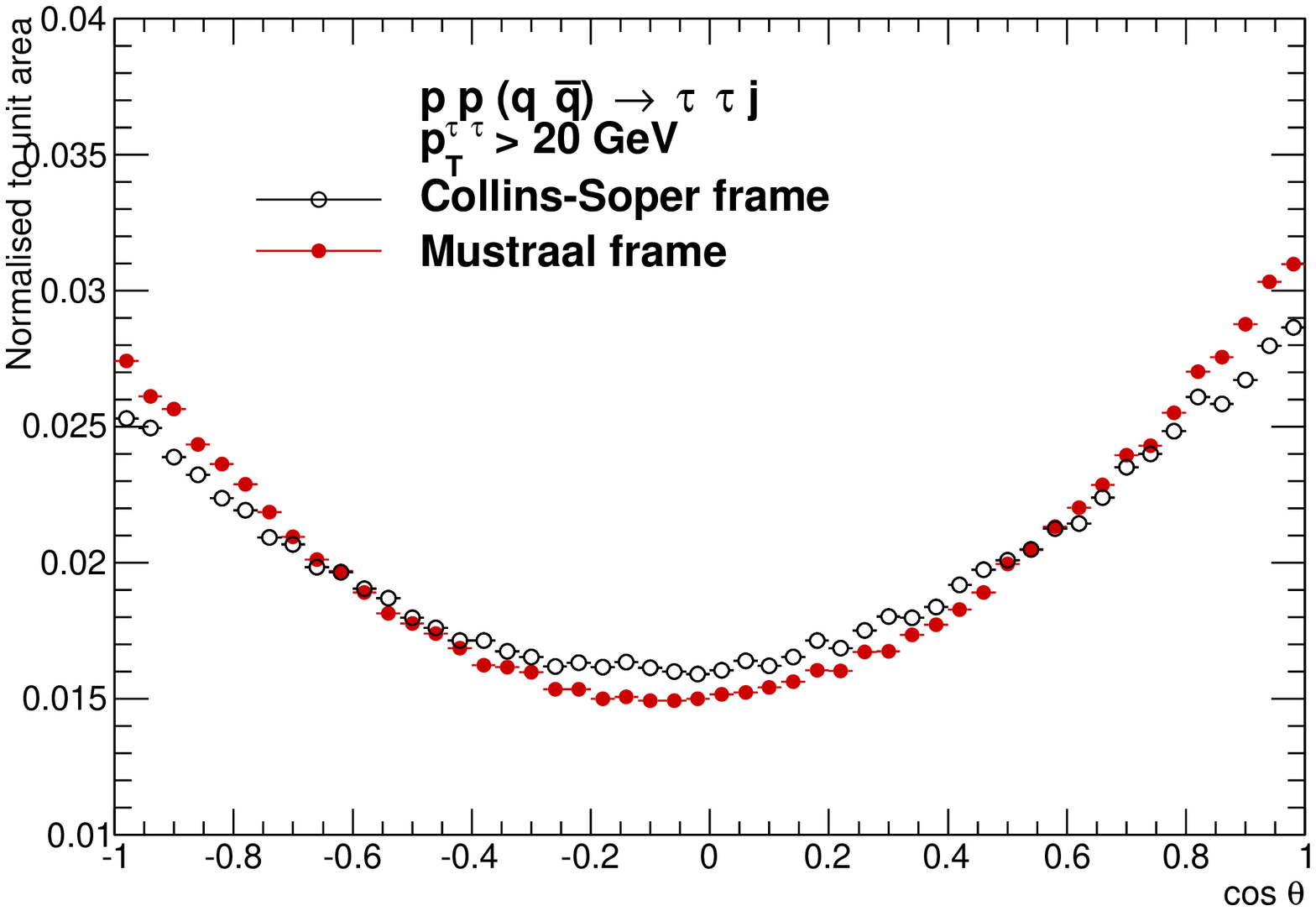}
   \includegraphics[width=7.5cm,angle=0]{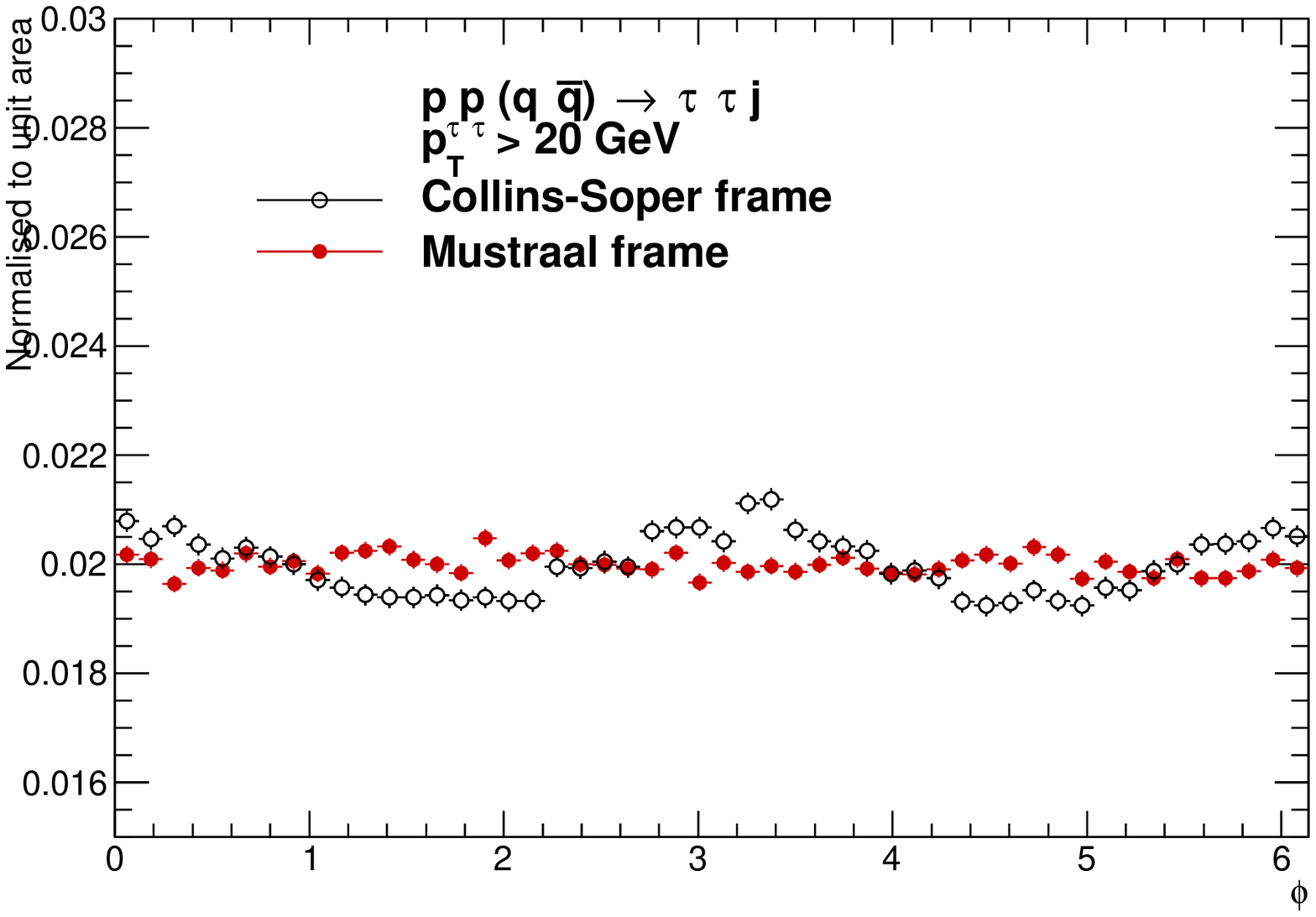}
}
{
   \includegraphics[width=7.5cm,angle=0]{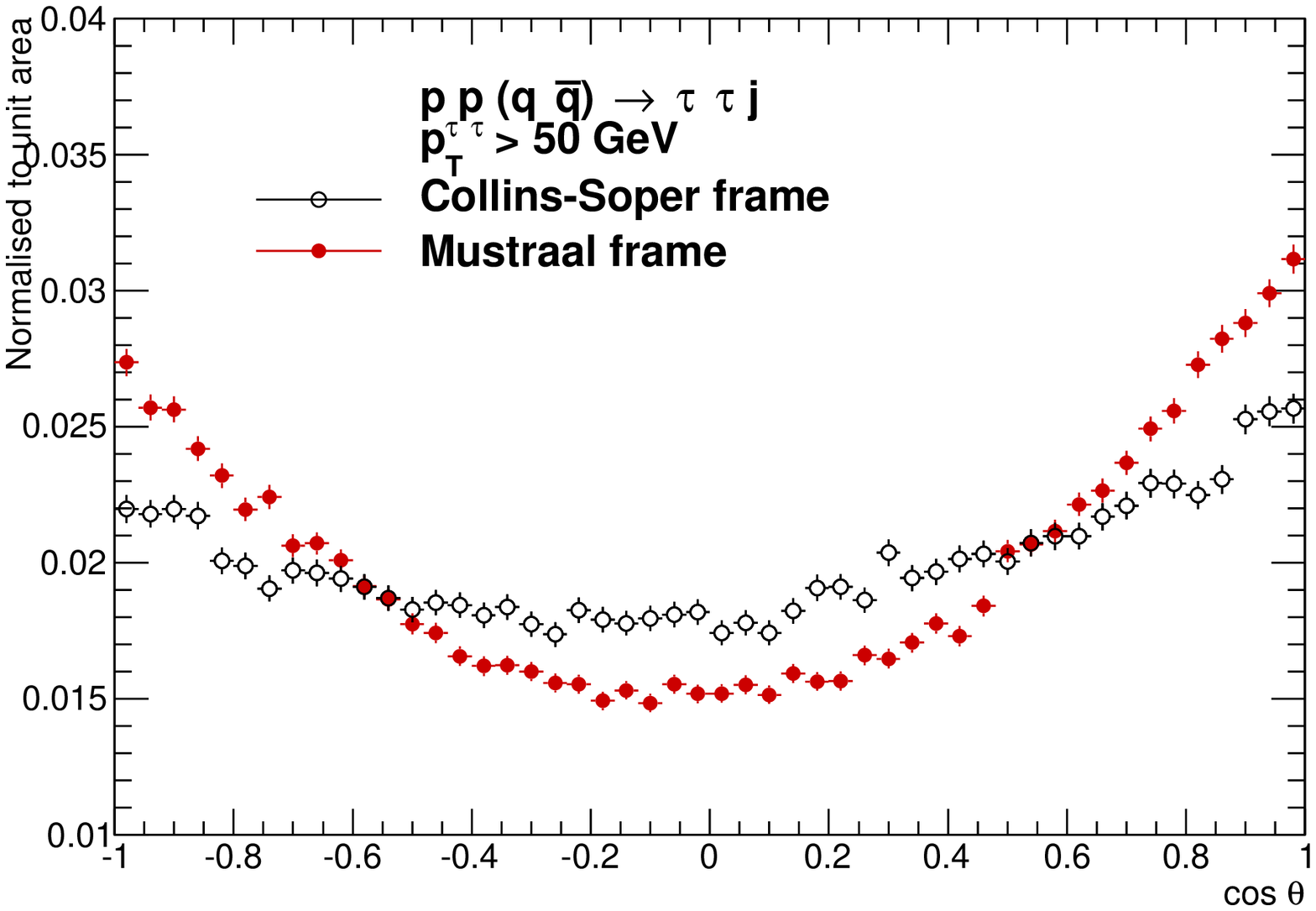}
   \includegraphics[width=7.5cm,angle=0]{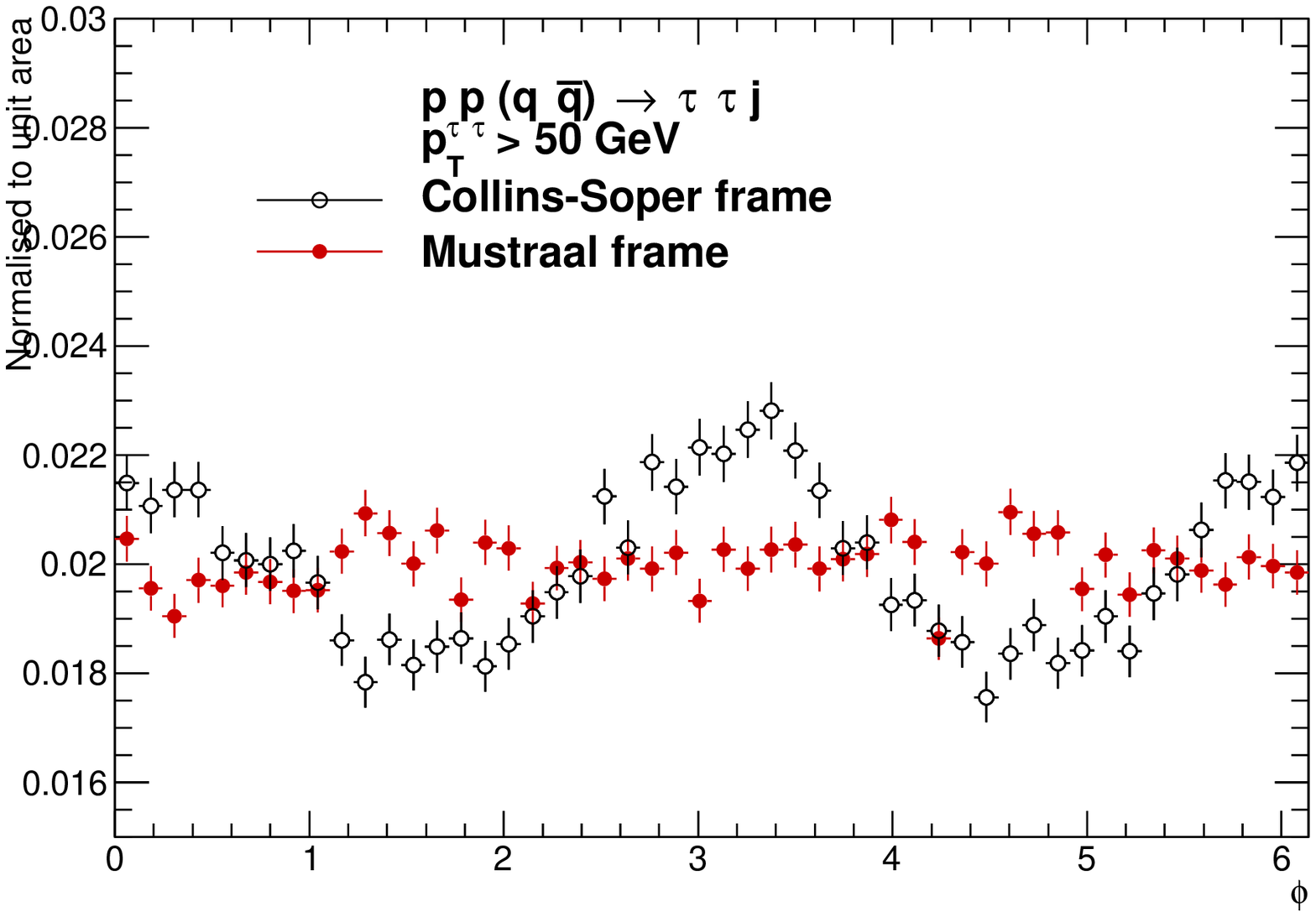}
}
\end{center}
\caption{ Distributions of $ \cos\theta$ and $\phi$ calculated in the Collins-Soper (black) and {\tt Mustraal} (red) frames. 
Case of $p p (q \bar q) \to \tau \tau j$ process generated with {\tt MadGraph}.
Results  for three thresholds of $\tau \tau$ system transverse momenta
are presented. Details of initialization are given in Section \ref{sec:numerical}.
\label{Fig:CosThetaPhi_qqbar} }
\end{figure}

\begin{figure}
  \begin{center}                               
{
   \includegraphics[width=7.5cm,angle=0]{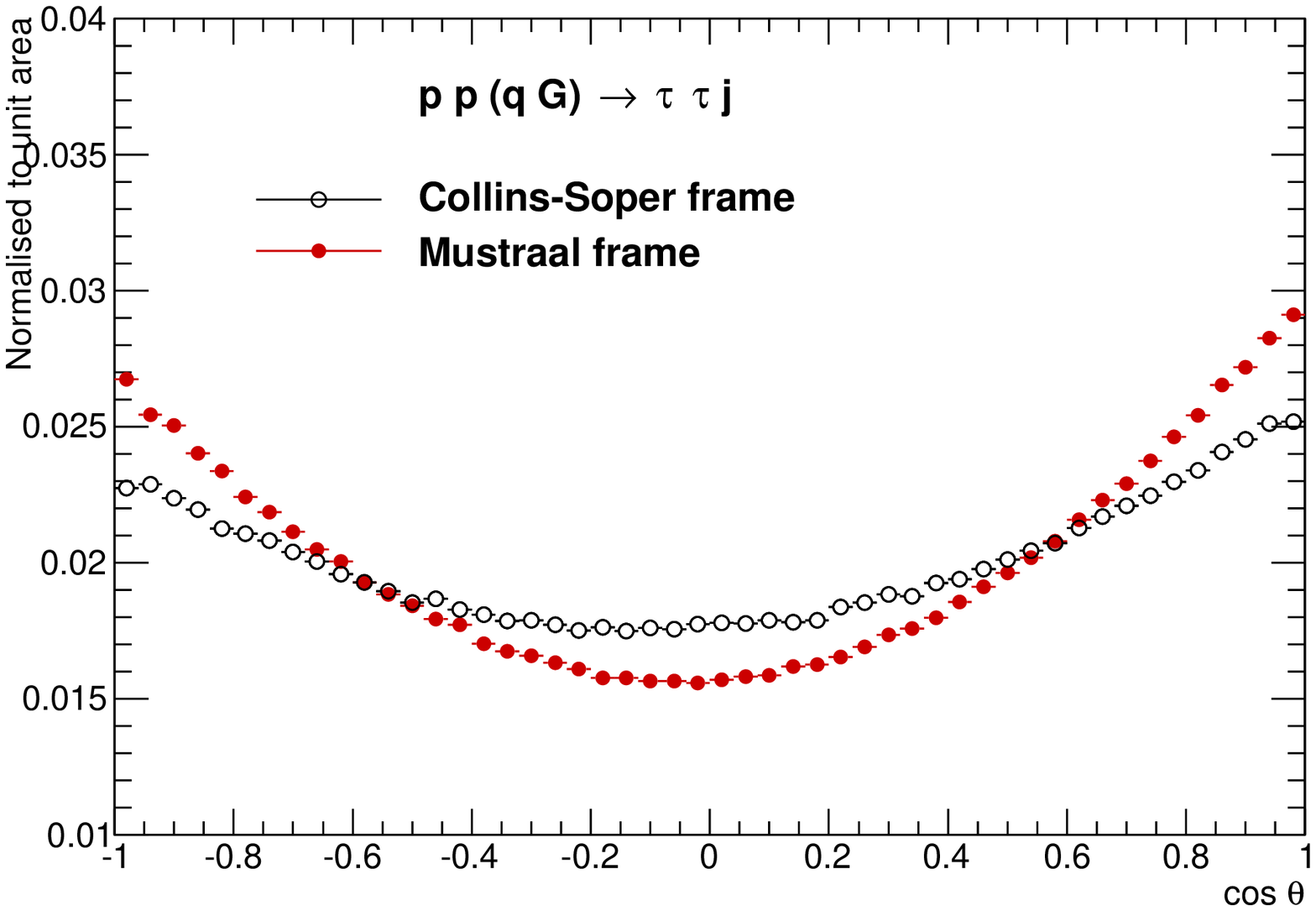}
   \includegraphics[width=7.5cm,angle=0]{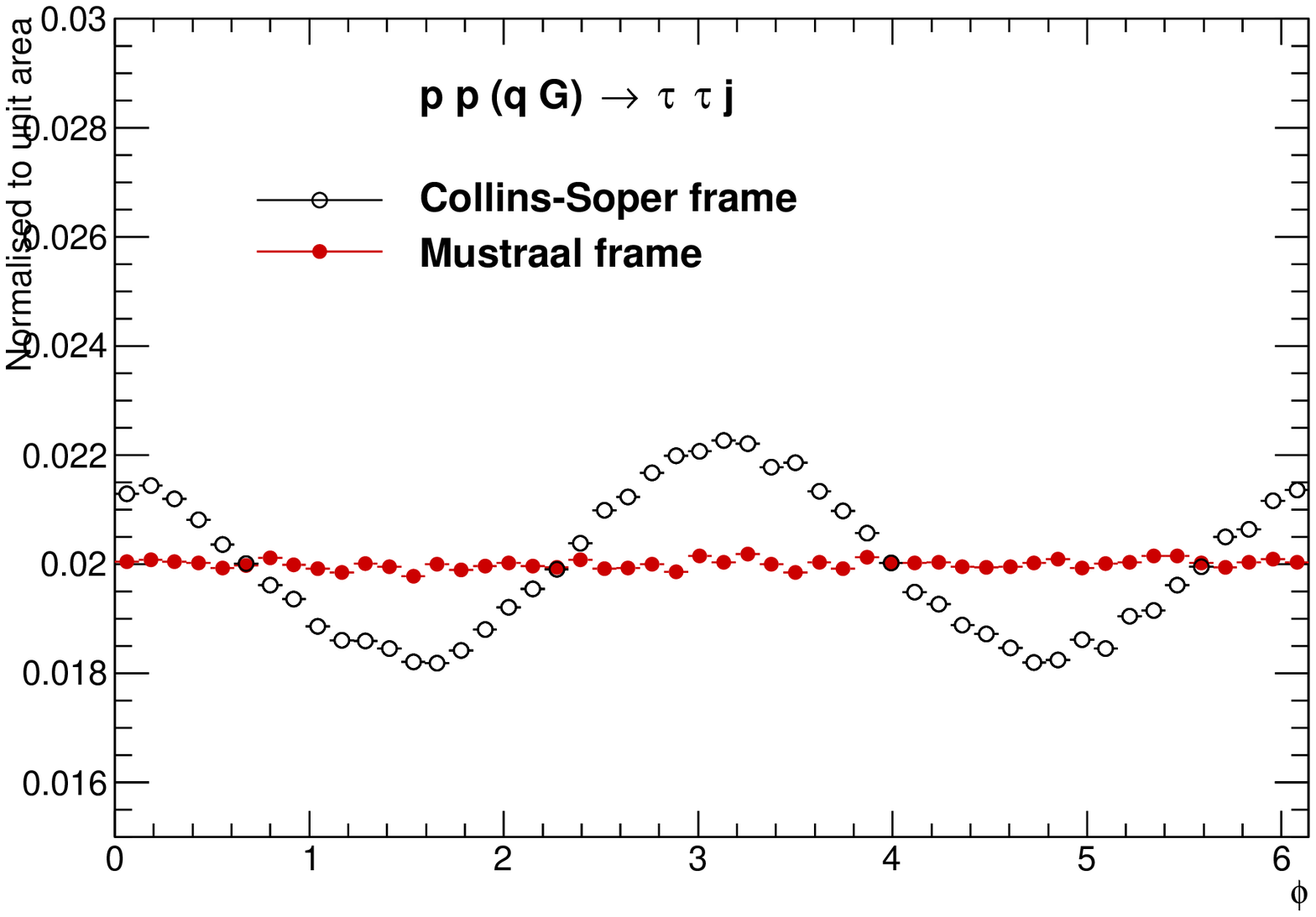}
}
{
   \includegraphics[width=7.5cm,angle=0]{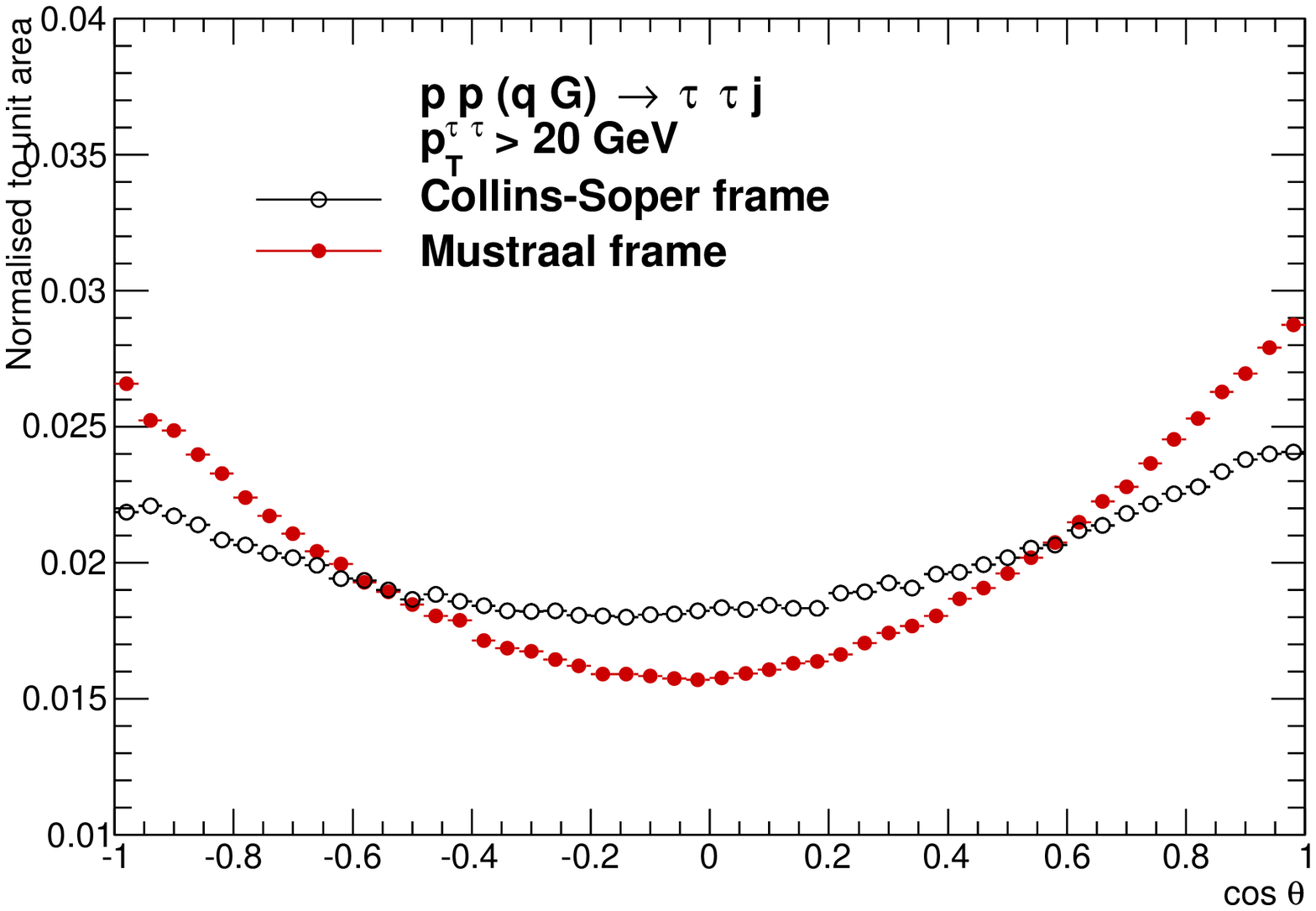}
   \includegraphics[width=7.5cm,angle=0]{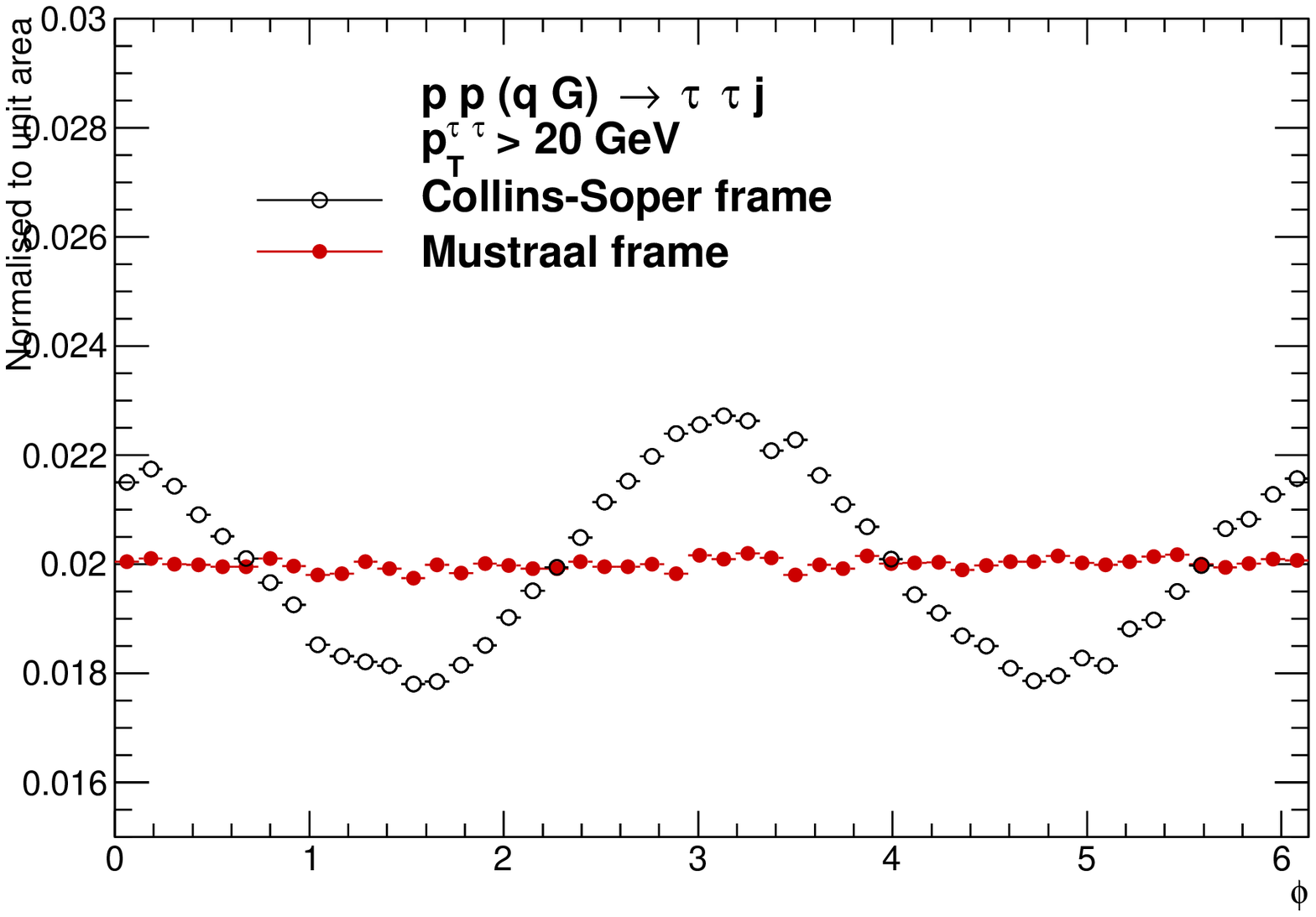}
}
{
   \includegraphics[width=7.5cm,angle=0]{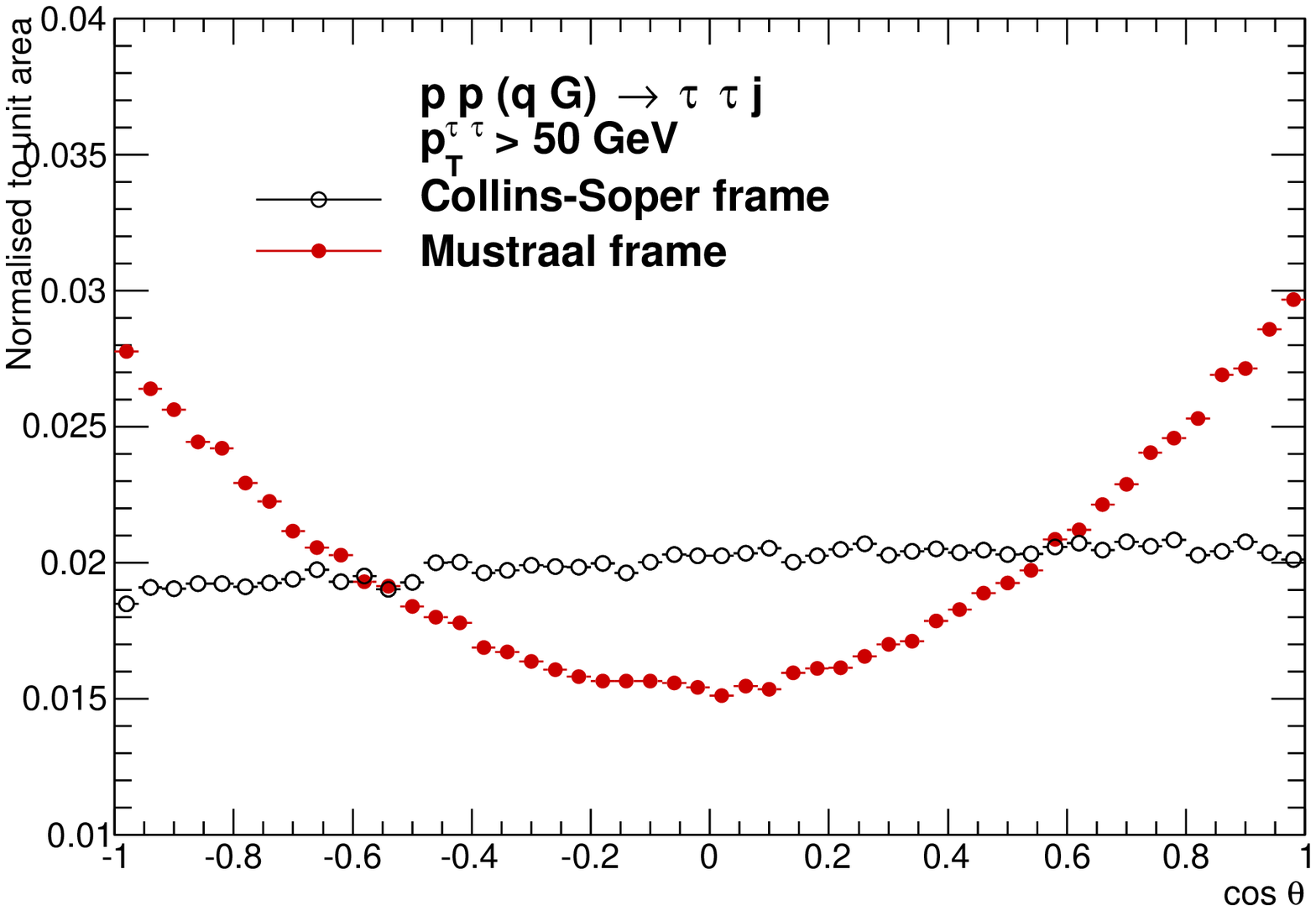}
   \includegraphics[width=7.5cm,angle=0]{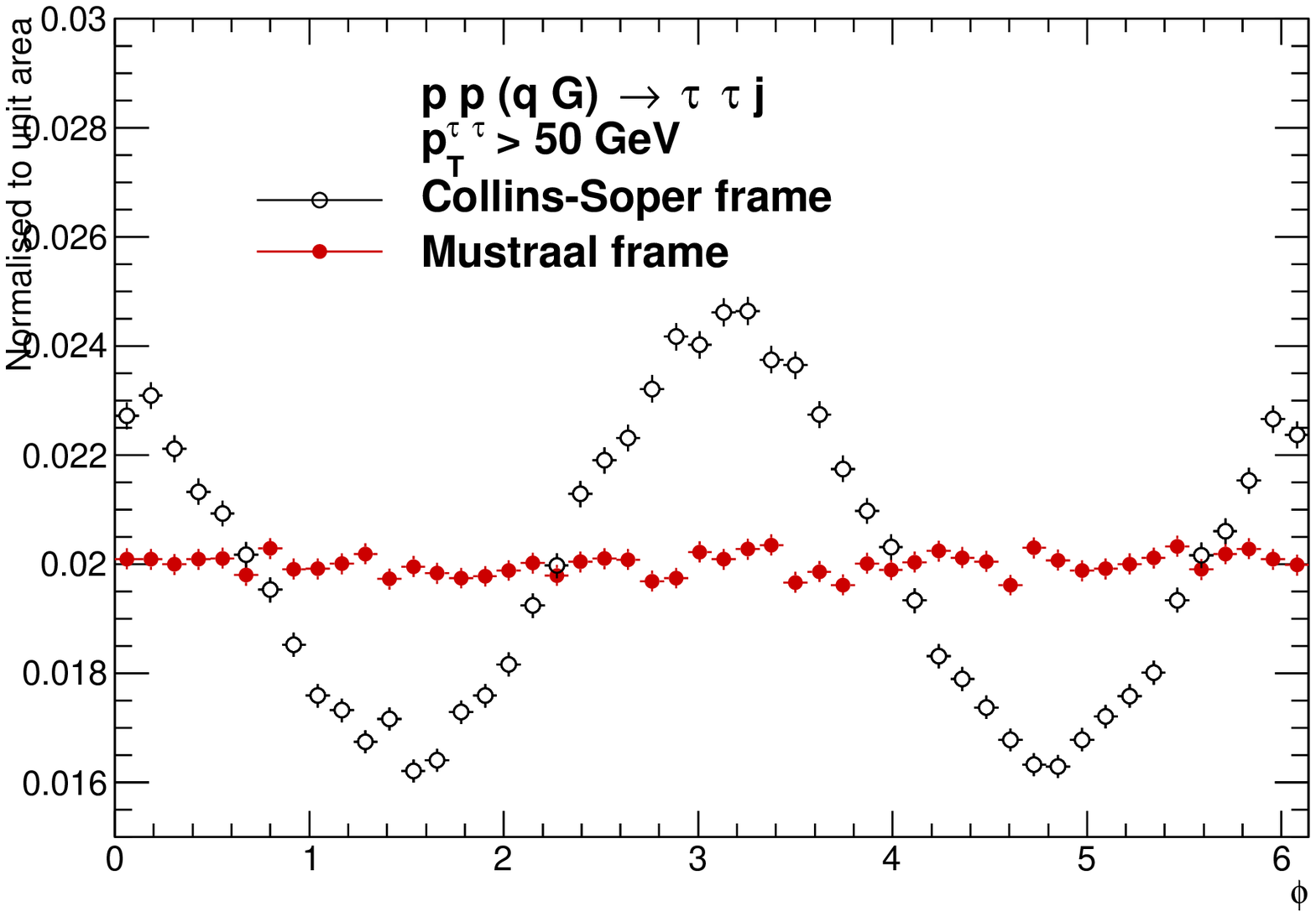}
}
\end{center}
\caption{ Distribution of $ \cos\theta$ and $\phi$ calculated in the Collins-Soper (black) and {\tt Mustraal} (red) frames. 
Case of $p p ( q G) \to \tau \tau j$ process generated with {\tt MadGraph}. 
Results for three thresholds of $\tau \tau$ system transverse momenta are shown.
Details of initialization are given in Section \ref{sec:numerical}.
\label{Fig:CosThetaPhi_qglu}} 
\end{figure}

Fig.~\ref{Fig:CosThetaPhi_qqbar} and~\ref{Fig:CosThetaPhi_qglu}  show distributions of $ \cos\theta$ and $\phi$ calculated respectively in the  
Collins-Soper and {\tt Mustraal} frames for $pp \to \tau \tau j$ process in case of $q \bar q$ annihilation or $q G$  scattering. 
Clearly the asymmetric shape of $ \cos\theta$ is preserved despite an 
increasing threshold on $p_T^{\tau \tau}$ transverse momenta and 
the $\phi$ modulation is almost completely removed with the choice of {\tt Mustraal} frame. 
One can conclude from those two figures that in the {\tt Mustraal} frame angular correlations remain robust against QCD 
initial state effects, preserving information on the V-A structure of electroweak couplings in the Z-boson vertex,
not only in case of outgoing gluonic jet but also of outgoing quark. No deformation  of the shape in $ \cos\theta$ distribution, which remains as at the Born level, is observed.
In the case of the Collins-Soper frame, deformations are present.
It was expected that choice of {\tt Mustraal} frame preserves the Born shape
in case of gluonic emission: in the definition of this frame properties of $q \bar q \to \tau \tau j$  matrix 
elements were explored. The case of an outgoing quark, where a similar pattern 
is observed, is beyond what we would expect from the matrix element representation which was used for the {\tt Mustraal}
Monte Carlo construction. We will return to this point in Section~\ref{Sec:coefsMUSTframe}.

Nonetheless, the above comparison is inspiring for more checks, namely  if such properties will hold in more general cases,
like when more partons are present in the final state, or when higher order corrections (QCD NLO)
are taken into account.
Let us continue with preparation of the phenomenological picture for such studies before 
continuing with further presentation of numerical results.

\subsection{Angular coefficients in the Monte Carlo samples}

The angular coefficients $A_i(p_T,Y)$ are not explicitly input to the theoretical calculations nor to the MC~event generators. 
They can however be extracted from the shapes of the angular distributions with the method proposed in~\cite{arXiv9406381}, 
owing to the orthogonality of the spherical polynomials. 
The weighted average of the angular distributions with respect to any 
specific polynomial isolates an average reference value or moment of its corresponding coefficient. 
 Ref.~\cite{arXiv9406381} argues that only the coefficients 
of spherical harmonics of the second order will contribute, unless complete
effects of $\alpha_s^2$ are taken into account.
The  moment of a polynomial $P(\cos\theta,\phi)$ over a specific range of~$p_T$, $Y$ is defined as follows:

\begin{equation}
\langle P(\cos\theta,\phi)\rangle = \frac{\int  P(\cos\theta,\phi) d\sigma(\cos\theta,\phi) d\cos\theta d\phi}{\int d\sigma(\cos\theta,\phi) d \cos\theta d\phi} .
\end{equation}

As a consequence of Eq.~(\ref{Eq:master2})) we obtain:

\begin{equation}
\begin{split}
\langle\frac{1}{2}(1-3 \cos^2\theta)\rangle & = \frac{3}{20} (A_0 - \frac{2}{3} ); \ \ \ 
\langle\sin2\theta \cos\phi\rangle  = \frac{1}{5} A_1; \ \ \
\langle\sin^2\theta \cos2\phi\rangle  = \frac{1}{10} A_2; \\
\langle\sin\theta \cos\phi\rangle & = \frac{1}{4} A_3; \ \ \
\langle\cos\theta\rangle  = \frac{1}{4} A_4; \ \ \
\langle\sin^2\theta \sin 2 \phi\rangle = \frac{1}{5} A_5; \\
\langle\sin 2\theta \sin \phi\rangle & = \frac{1}{5} A_6; \ \ \
\langle\sin \theta \sin \phi\rangle  = \frac{1}{4} A_7 .
\end{split}
\label{Eq:moments}
\end{equation}

Thanks to the discussion in~\cite{arXiv9406381}, we expect that terms beyond  Eq.~(\ref{Eq:master2}), i.e. of higher order polynomials, 
 should be of order  0.01 or smaller. Let us point out that two jets of
large $p_T$ constitute contributions to corrections to the Drell-Yan process of such an order, 
so the effect at some corners of the phase space may be noticeable.

From now on, we will monitor the $A_i$ coefficients rather than the 
$\theta$, $\phi$ distributions themselves. We will not discuss higher than second 
order coefficients. We have checked, that  they are indeed small and  
remain within expected range. The size of those terms 
can be deciphered from the presented plots as well, that is why we do not devote 
to these point much attention. Our cross checks of those aspects are 
left  unpublished, as at present they are of limited interest 
 and are also easy to obtain. 

\subsection{Numerical results for configurations beyond single gluon emission.}  

\begin{figure}
  \begin{center}                               
{
   \includegraphics[width=7.5cm,angle=0]{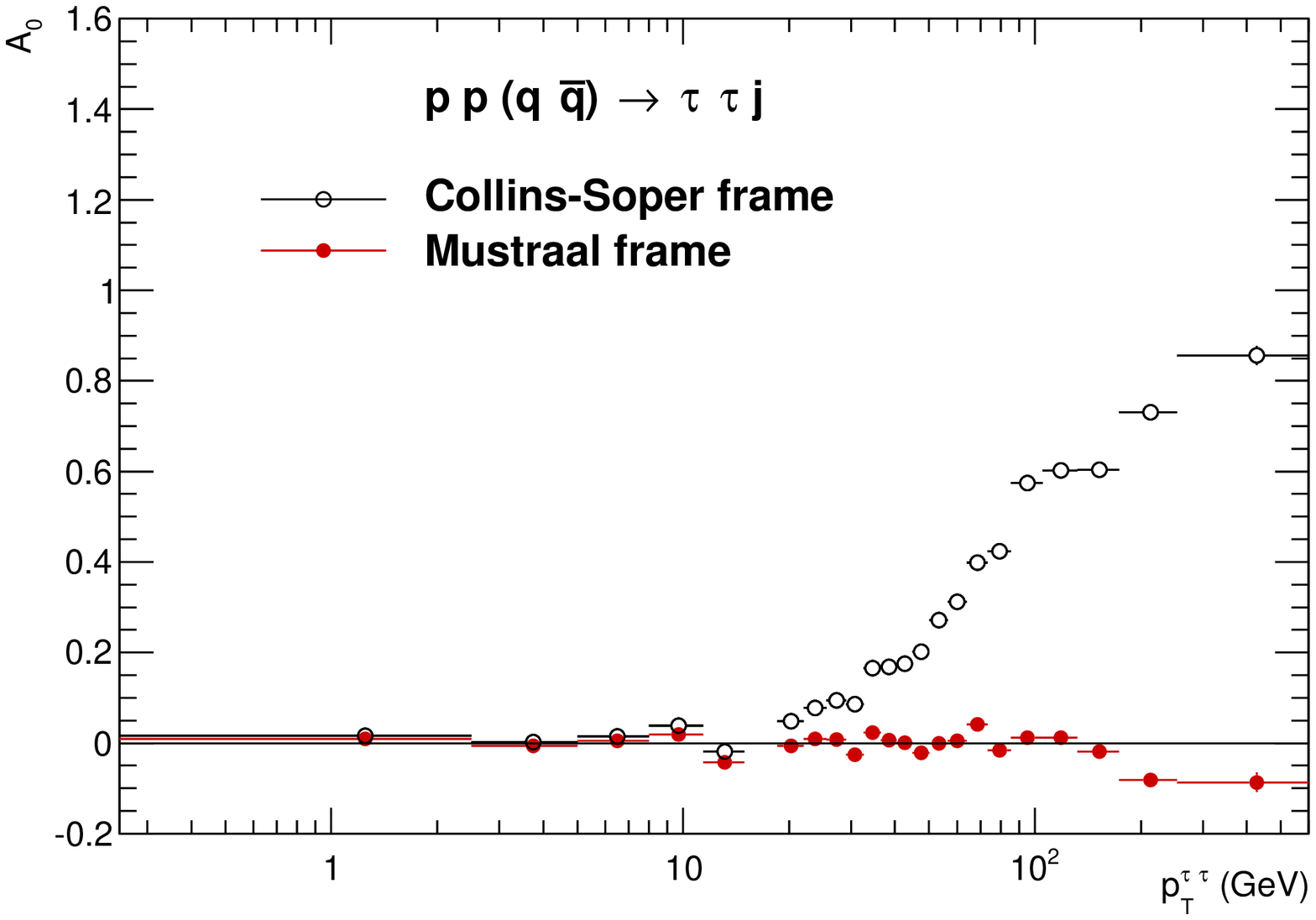}
   \includegraphics[width=7.5cm,angle=0]{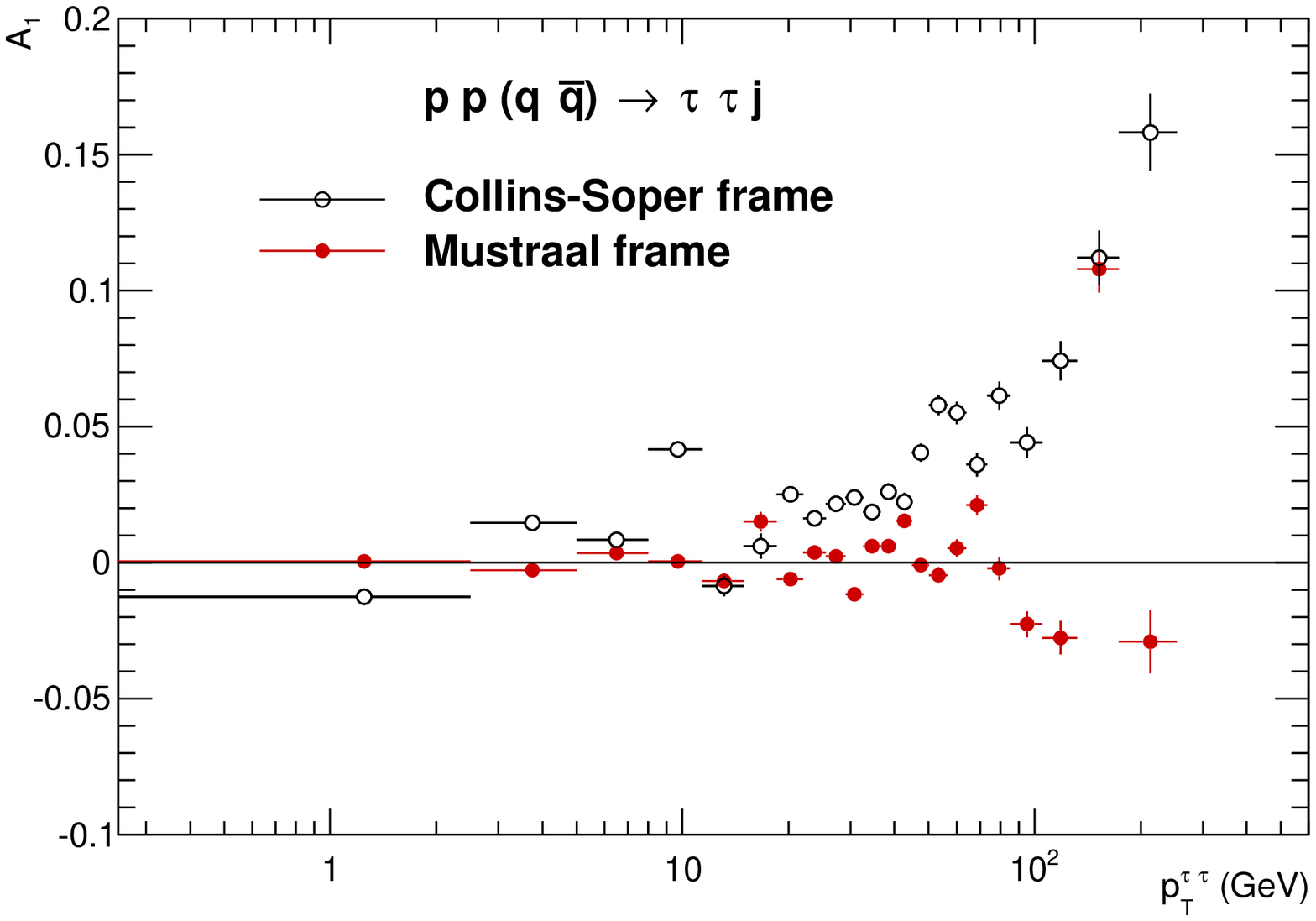}
   \includegraphics[width=7.5cm,angle=0]{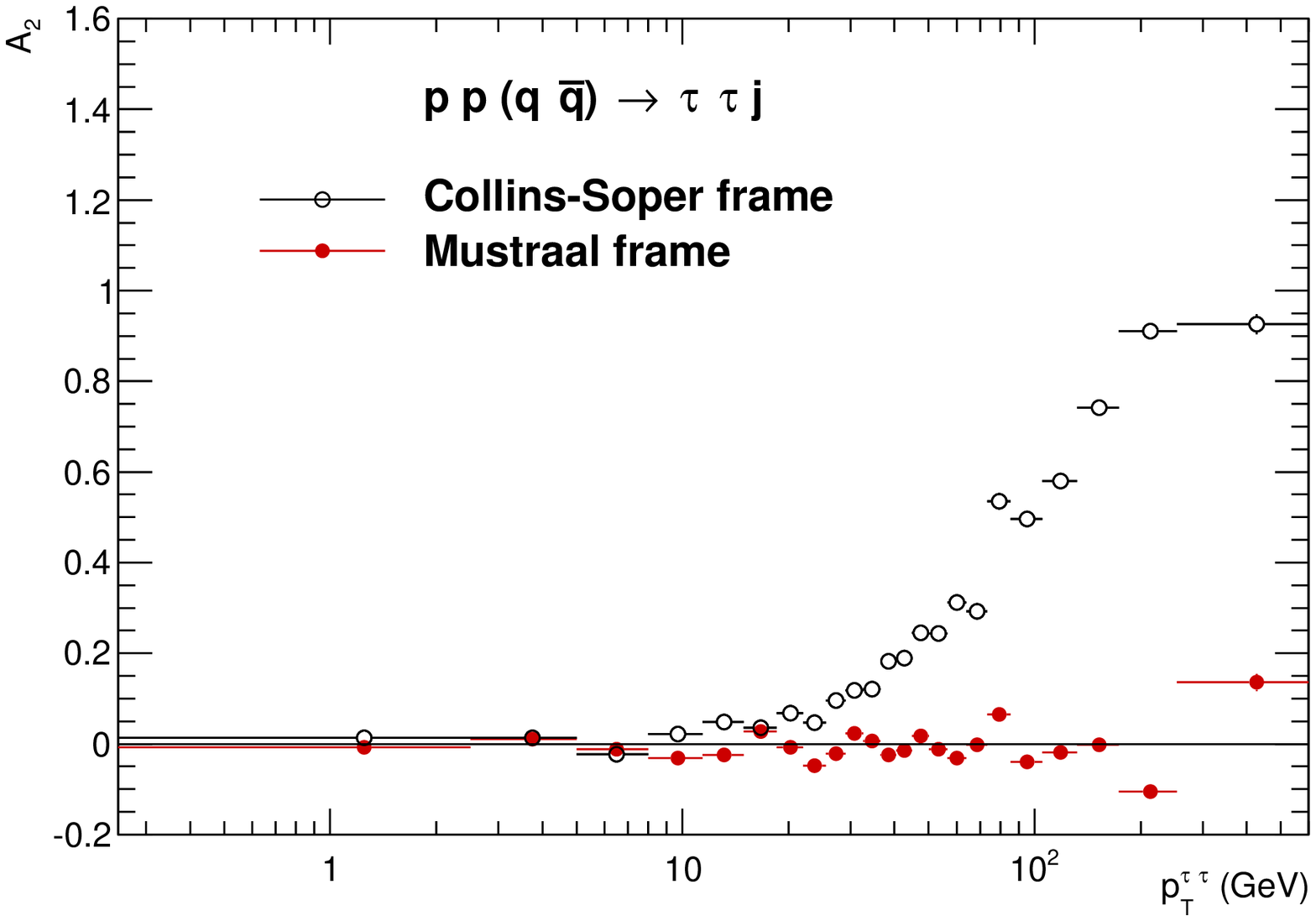}
   \includegraphics[width=7.5cm,angle=0]{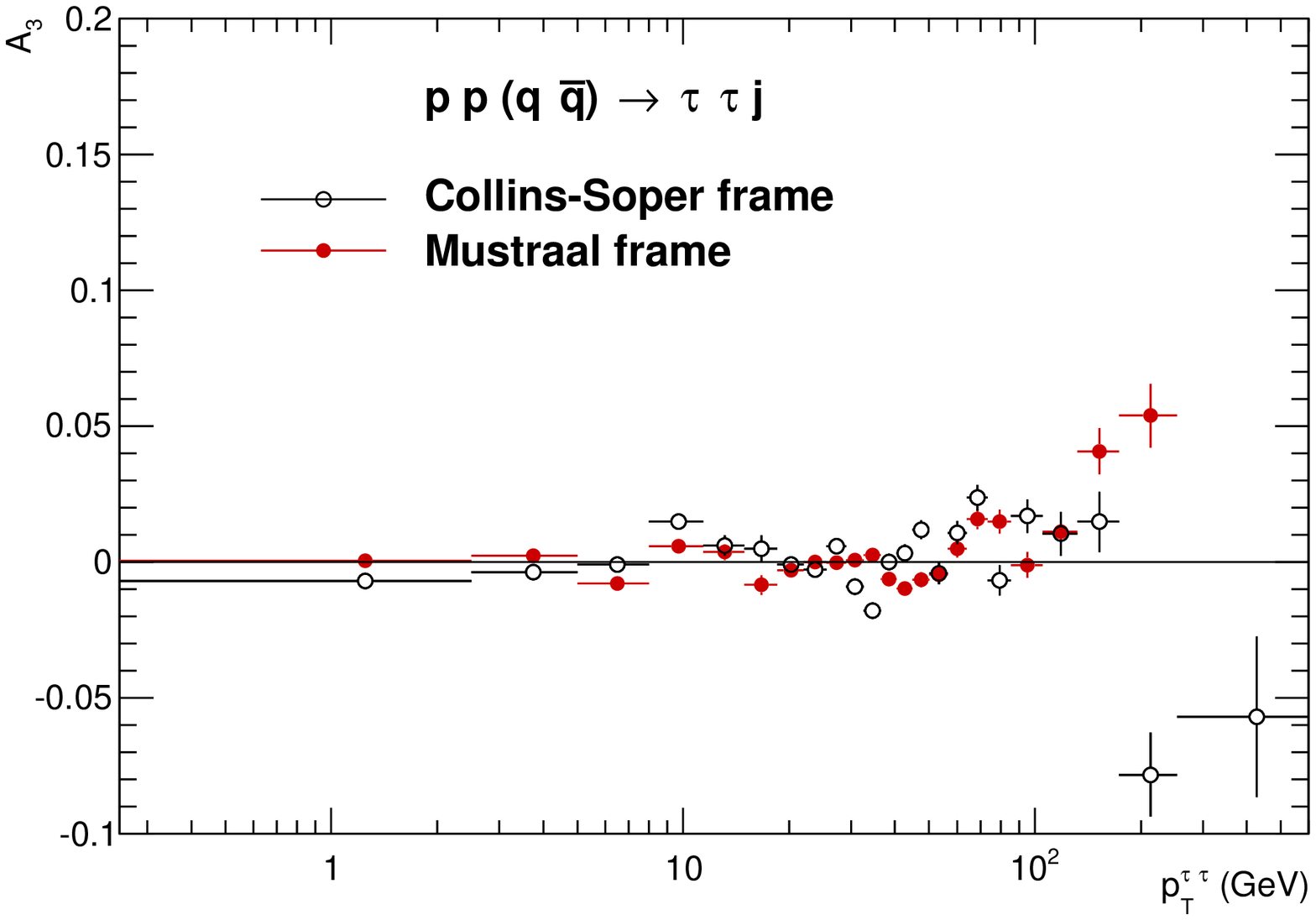}
   \includegraphics[width=7.5cm,angle=0]{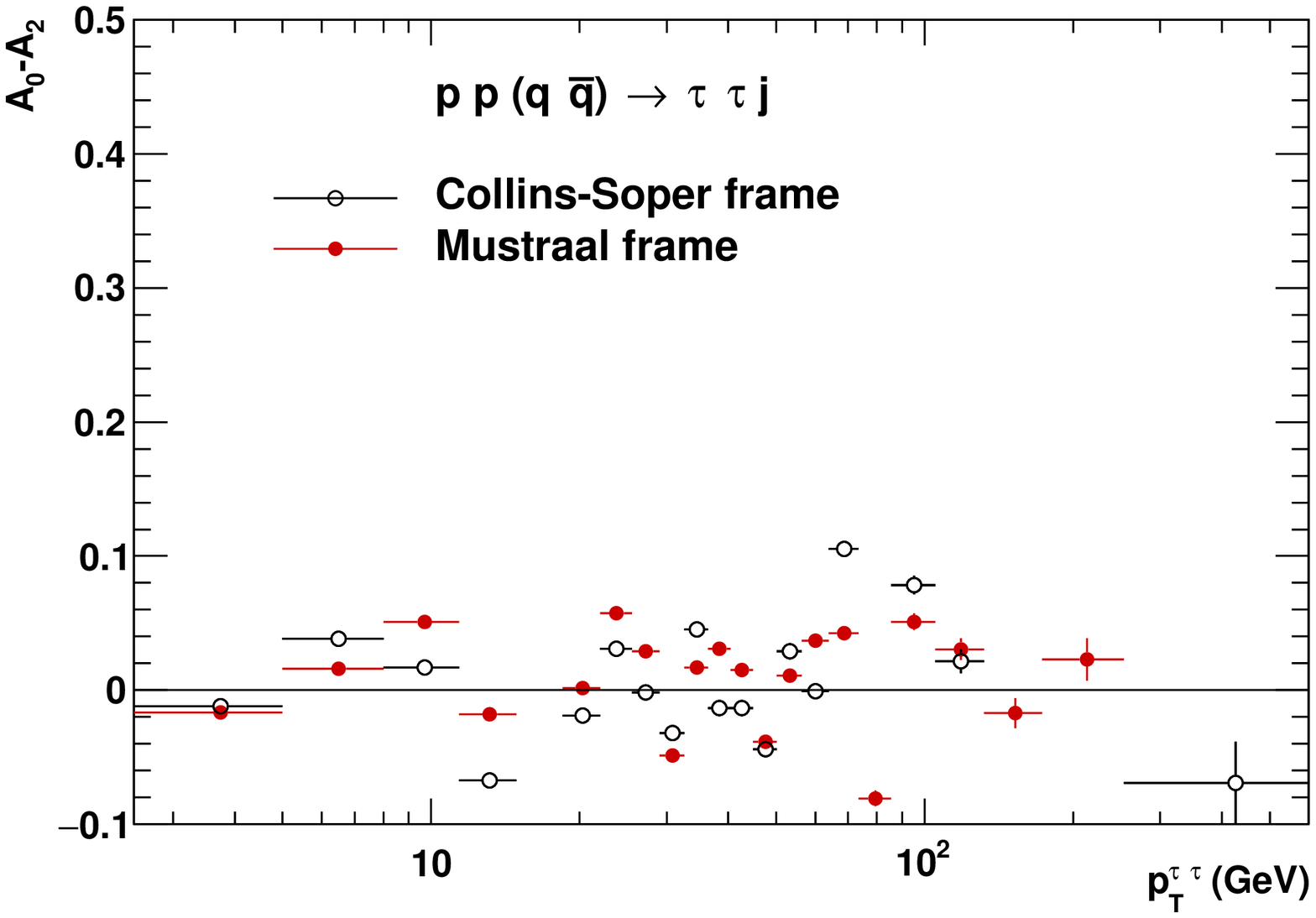}
   \includegraphics[width=7.5cm,angle=0]{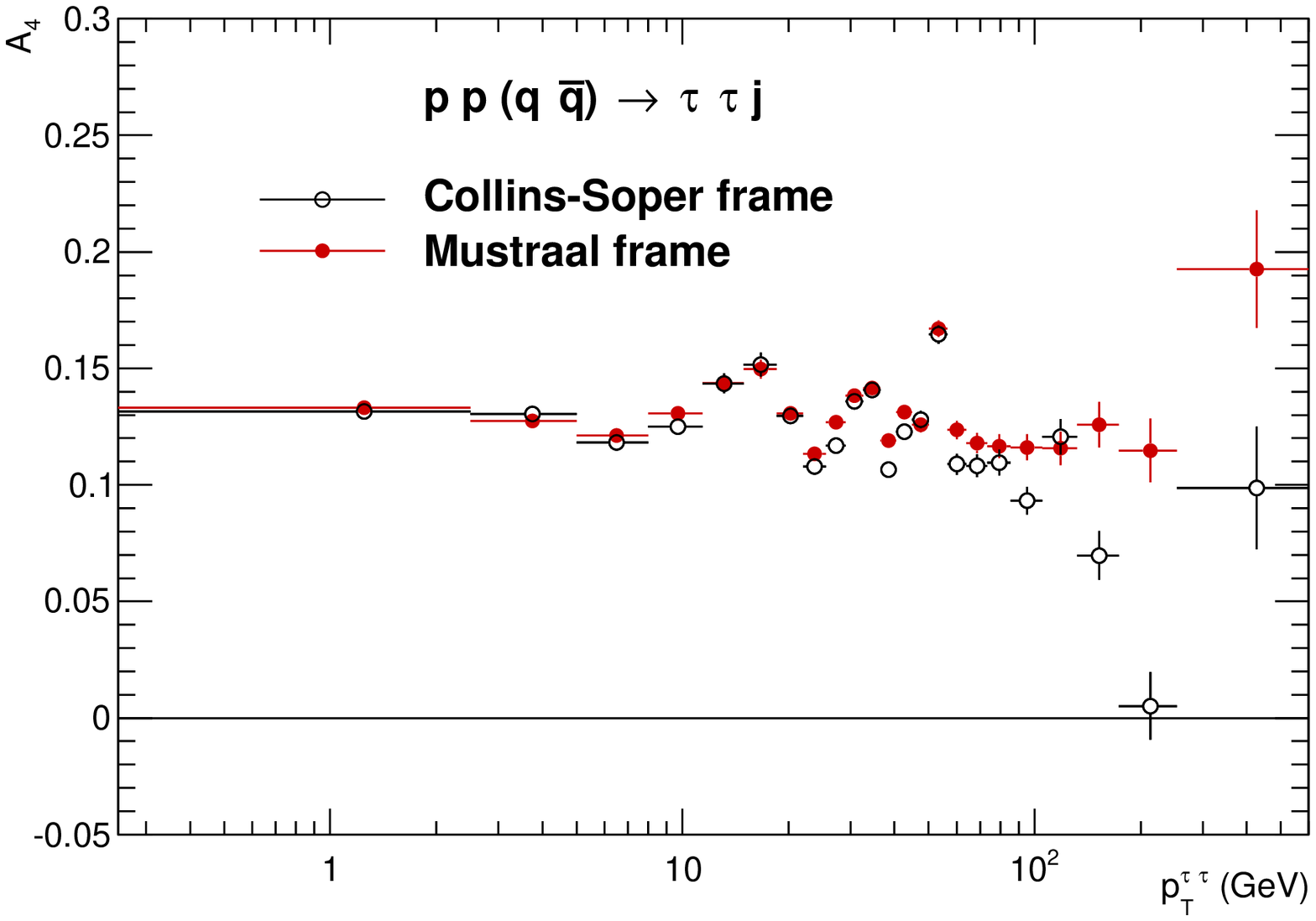}
}
\end{center}
\caption{ 
The $A_i$ coefficients of Eq.~(\ref{Eq:master2}))  calculated in Collins-Soper (black) and in {\tt Mustraal} (red) frames 
for $p p ( q \bar q) \to \tau \tau j$ process generated with {\tt MadGraph}.
Details of initialization are given in Section \ref{sec:numerical}.
\label{fig:Ai1gluon} }
\end{figure}

\begin{figure}
  \begin{center}                               
{
   \includegraphics[width=7.5cm,angle=0]{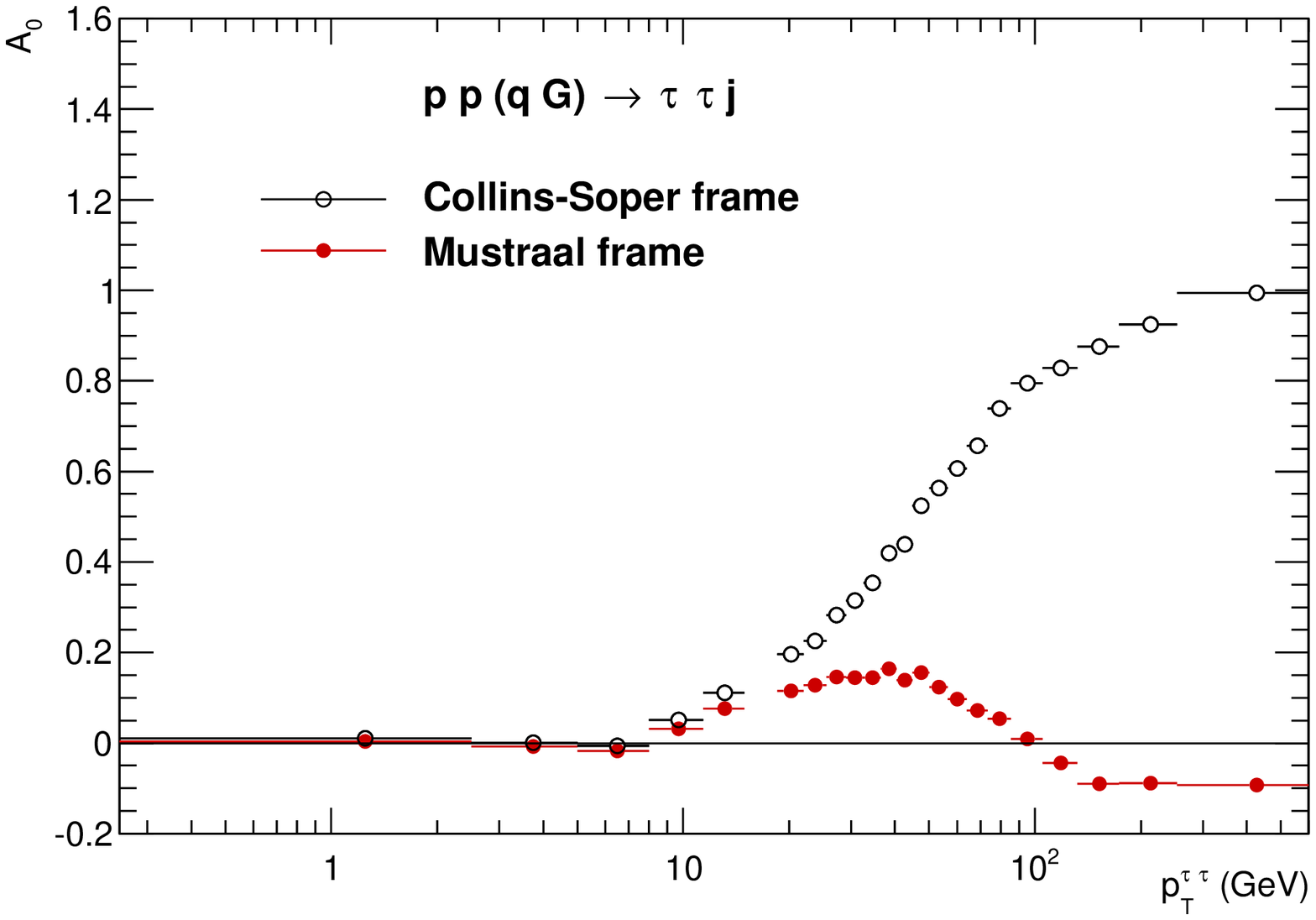}
   \includegraphics[width=7.5cm,angle=0]{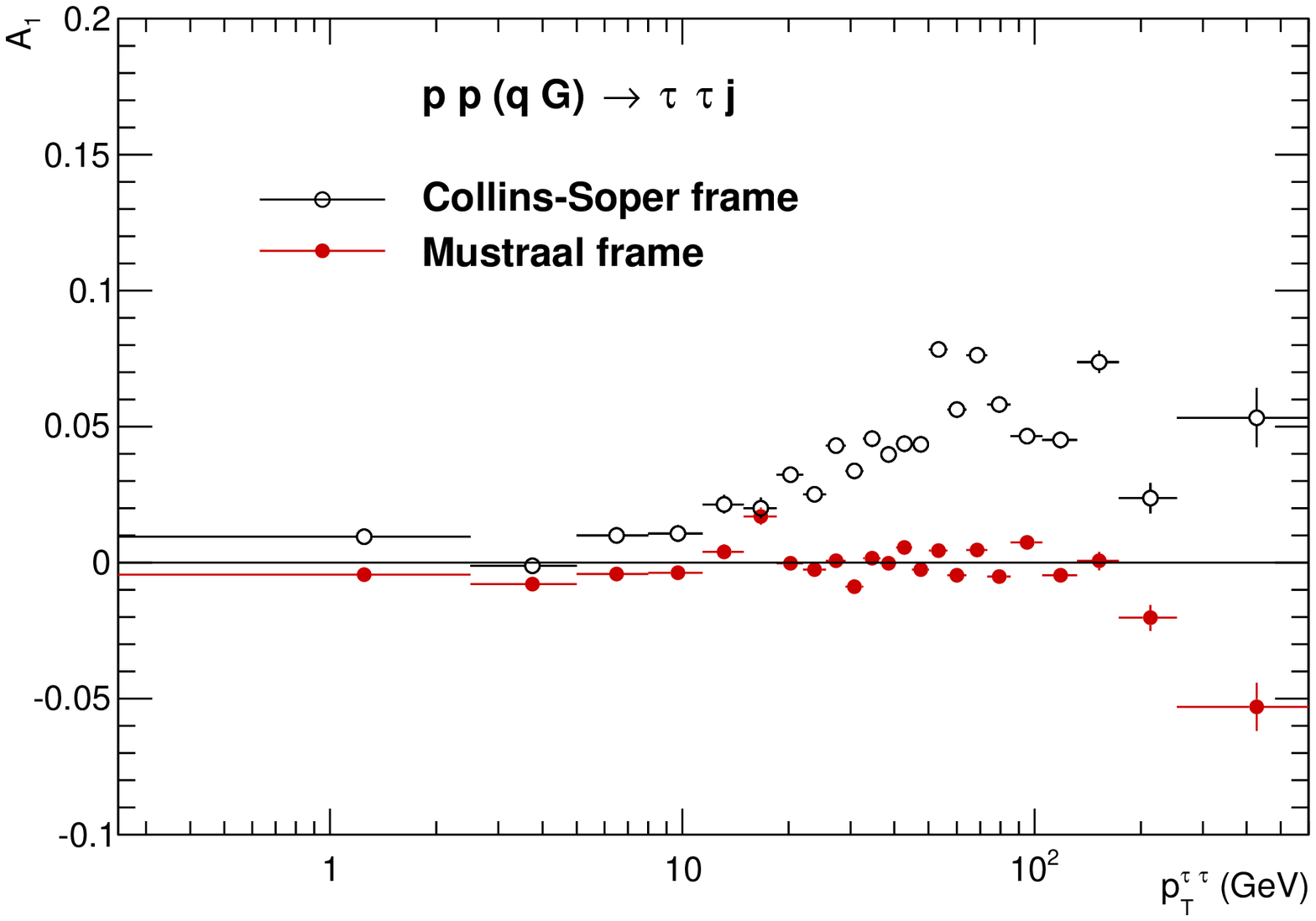}
   \includegraphics[width=7.5cm,angle=0]{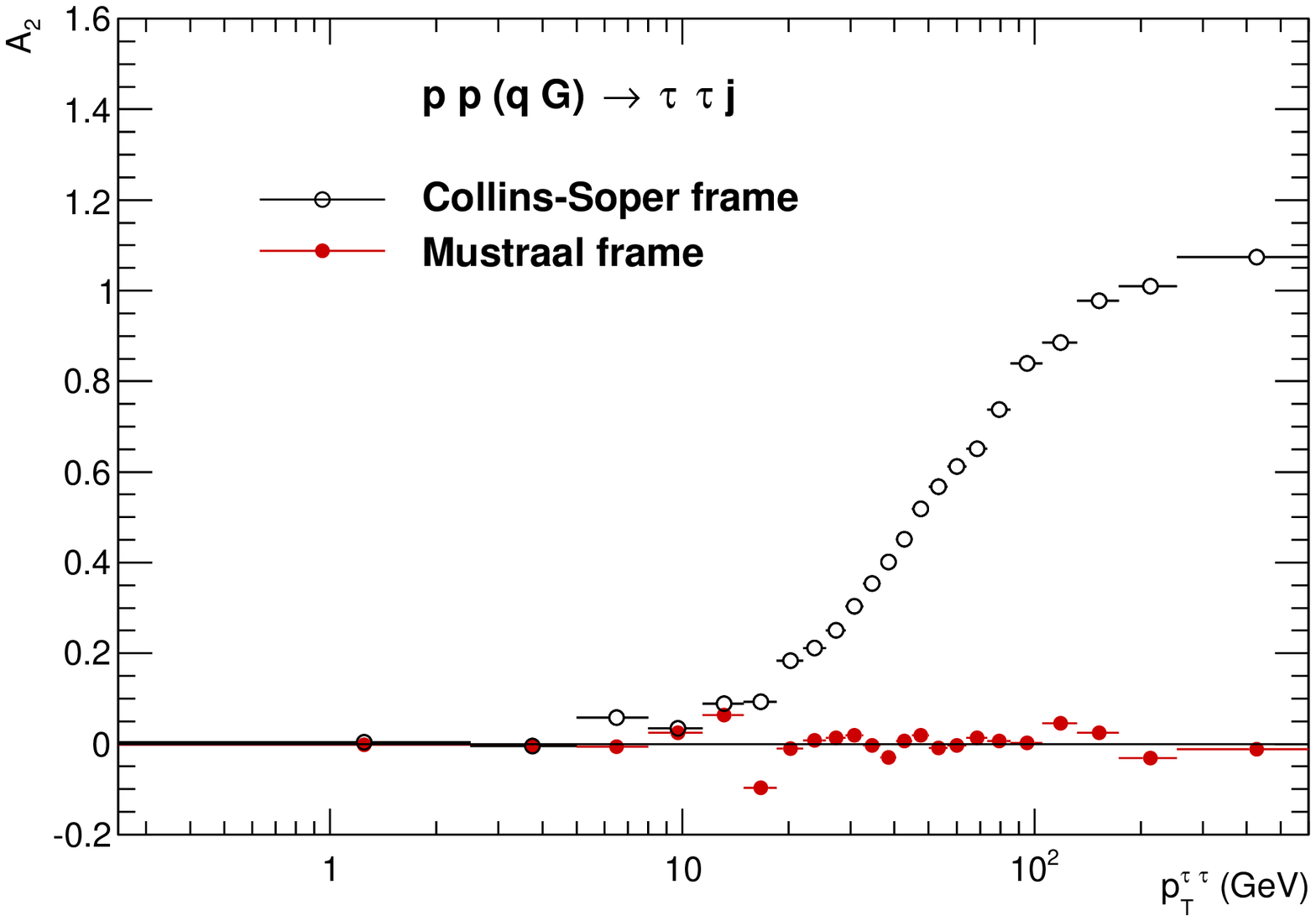}
   \includegraphics[width=7.5cm,angle=0]{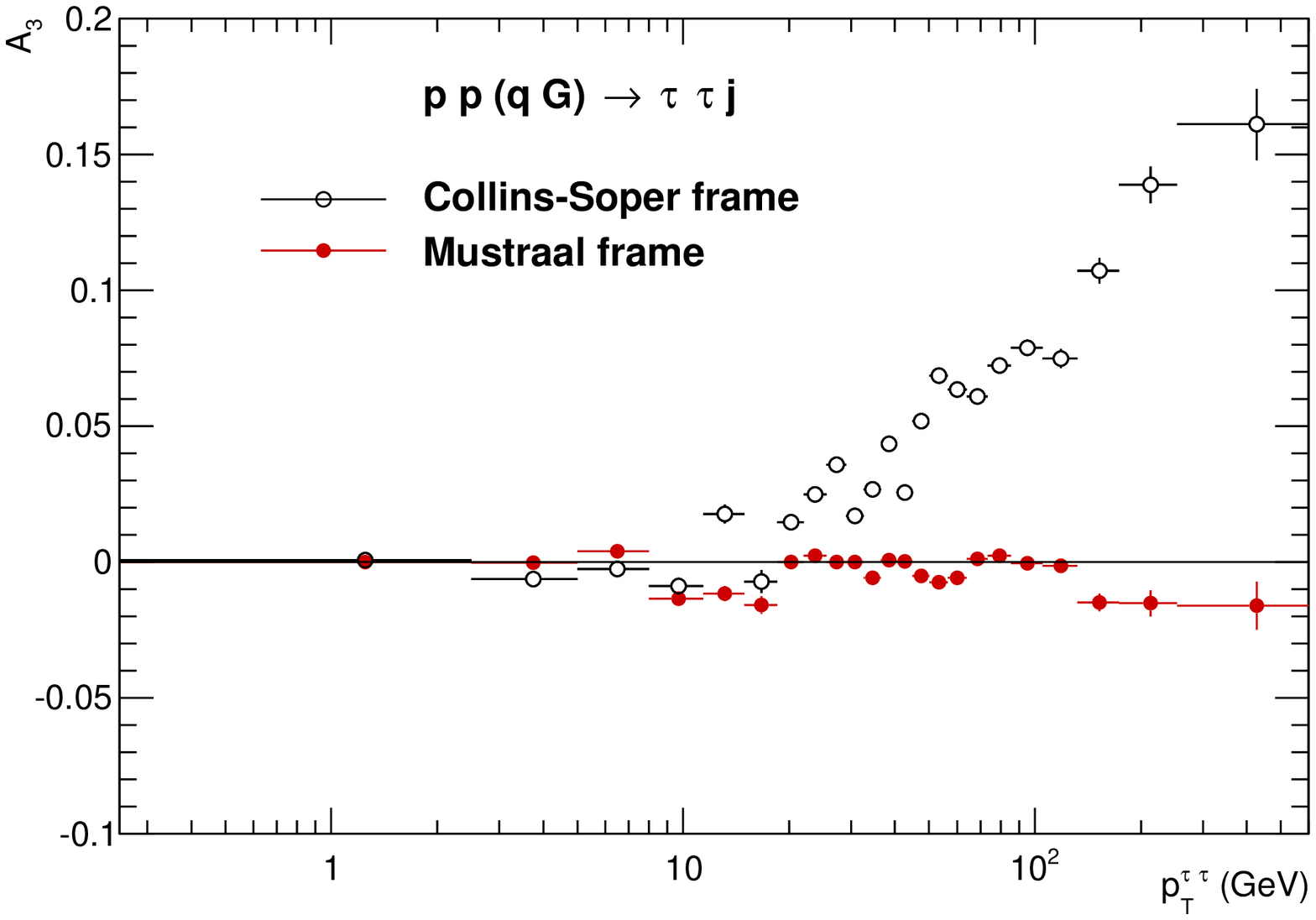}
   \includegraphics[width=7.5cm,angle=0]{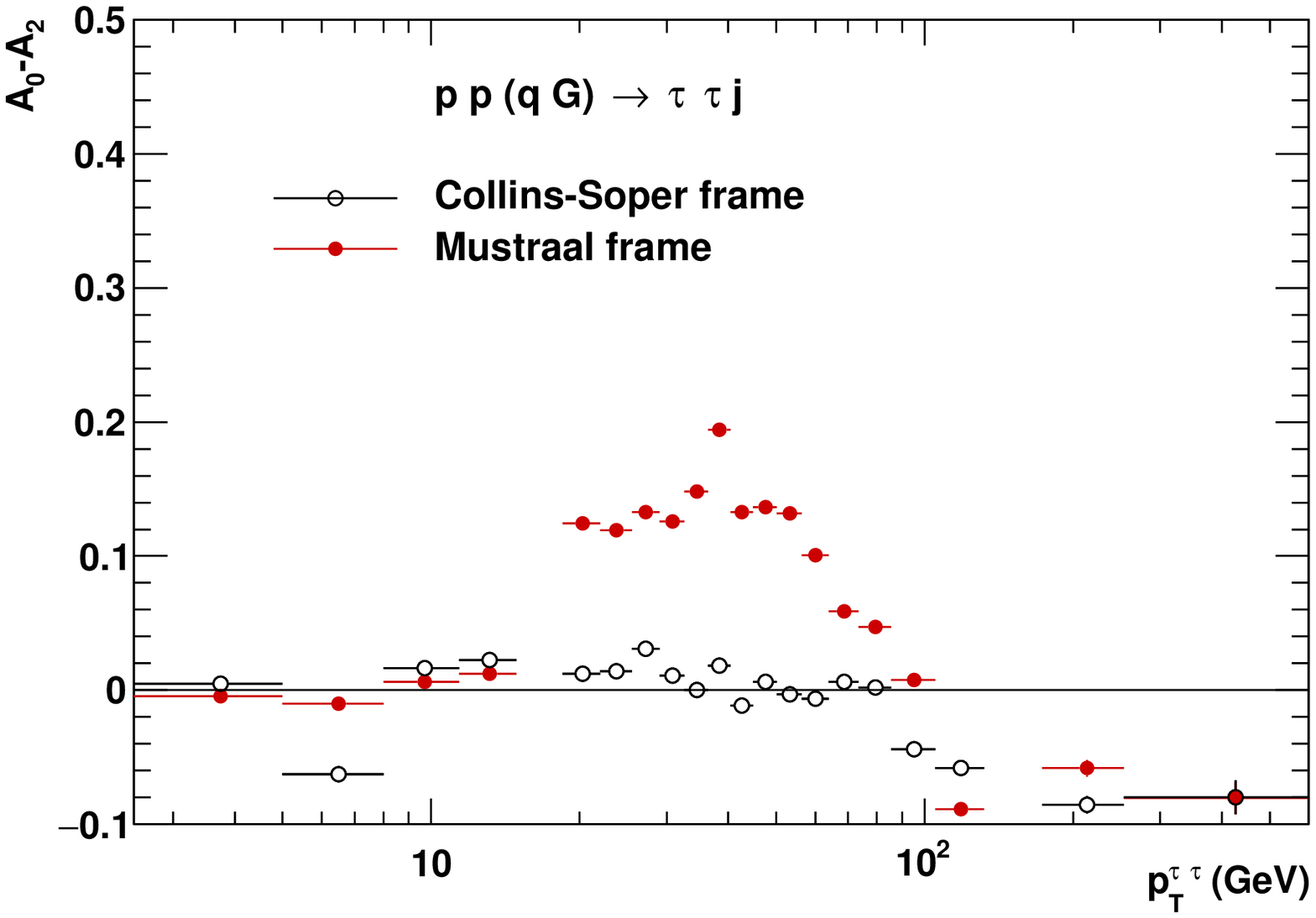}
   \includegraphics[width=7.5cm,angle=0]{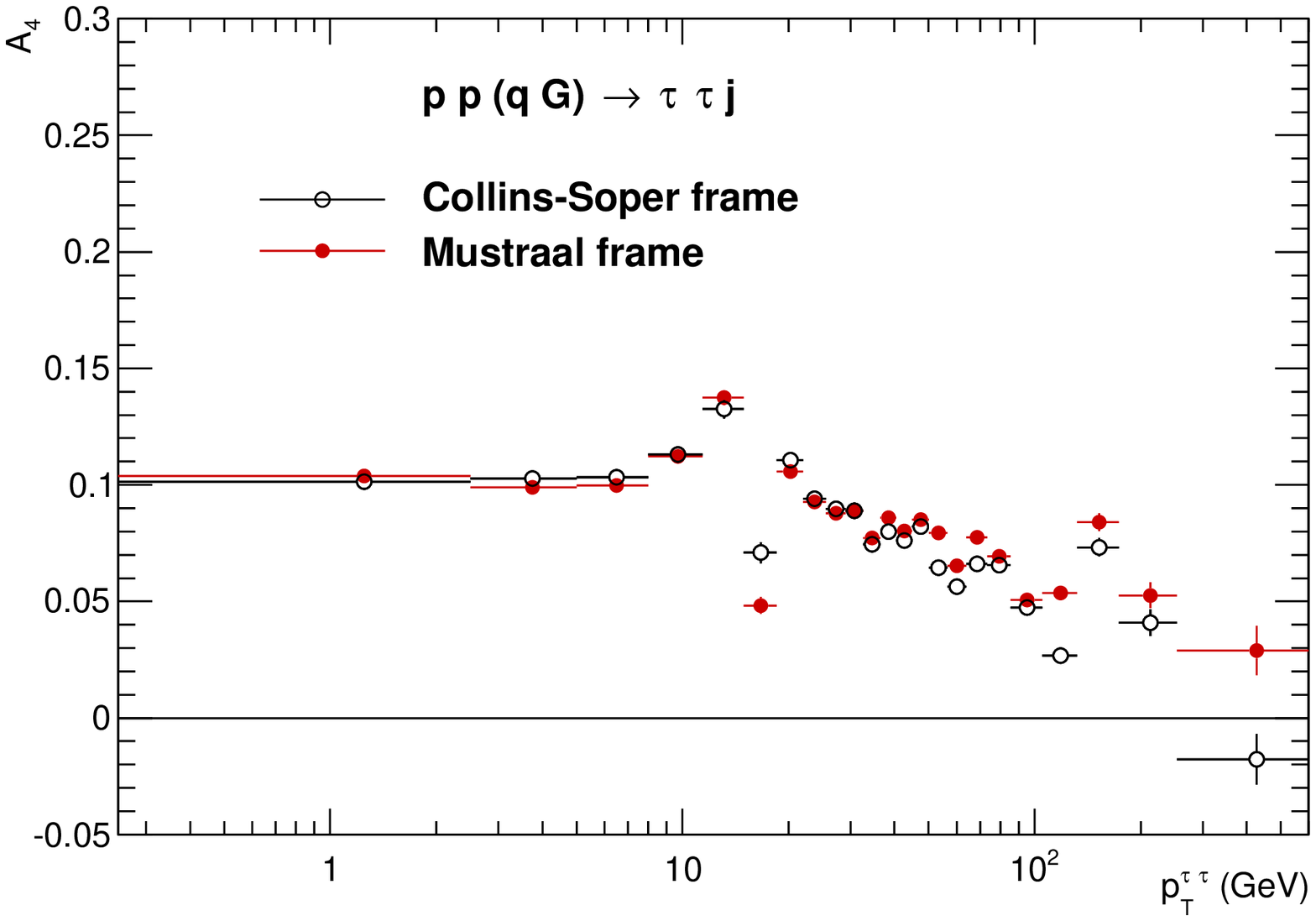}
}
\end{center}
\caption{ 
 The $A_i$ coefficients of Eq.~(\ref{Eq:master2})) calculated in Collins-Soper (black) and in {\tt Mustraal} (red) frames 
for $p p (q ( \bar q) G) \to \tau \tau j$ process generated with {\tt MadGraph}.
Details of initialization are given in Section \ref{sec:numerical}.
\label{fig:Ai1quark} }
\end{figure}

\begin{figure}
  \begin{center}                               
{
   \includegraphics[width=7.5cm,angle=0]{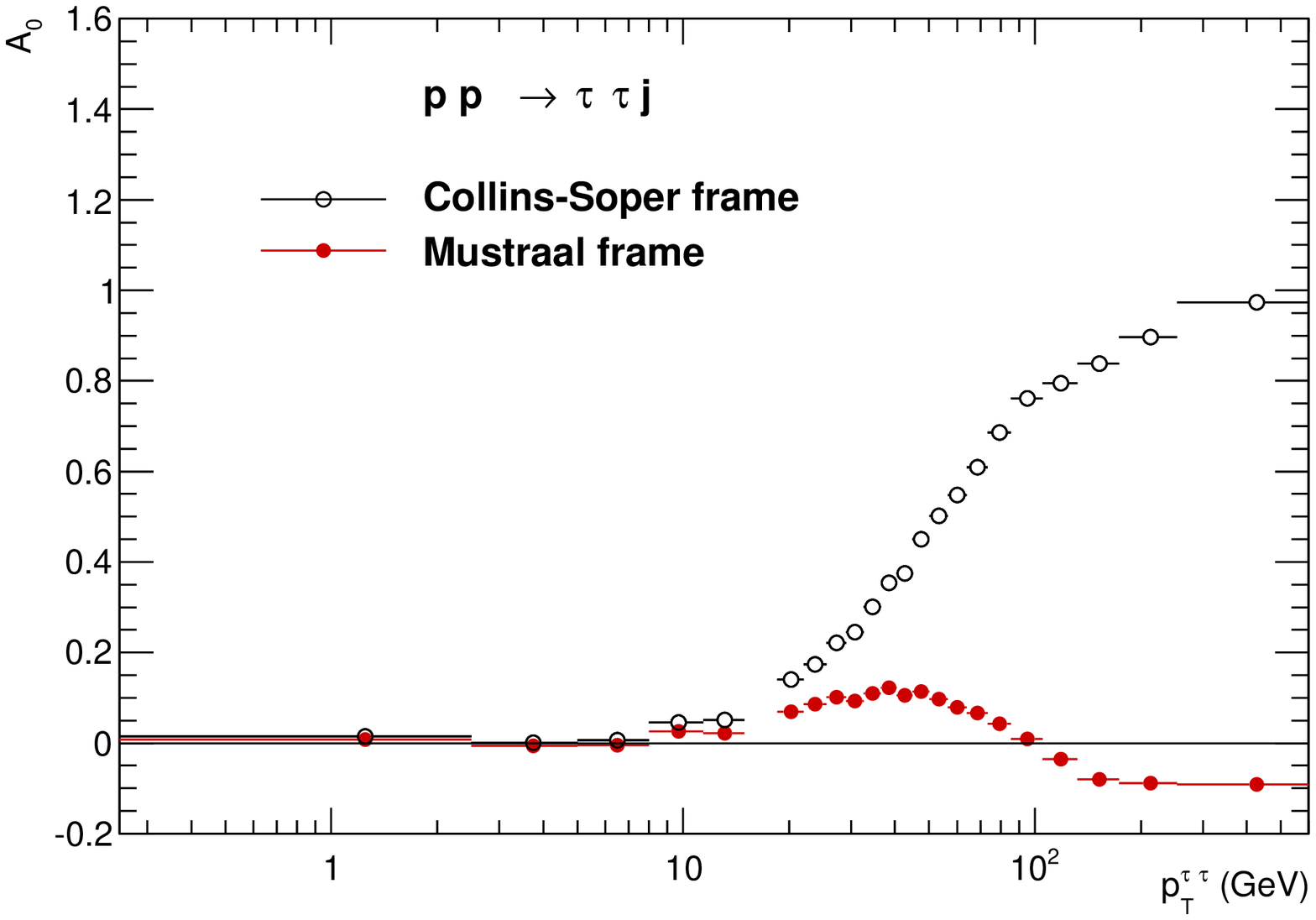}
   \includegraphics[width=7.5cm,angle=0]{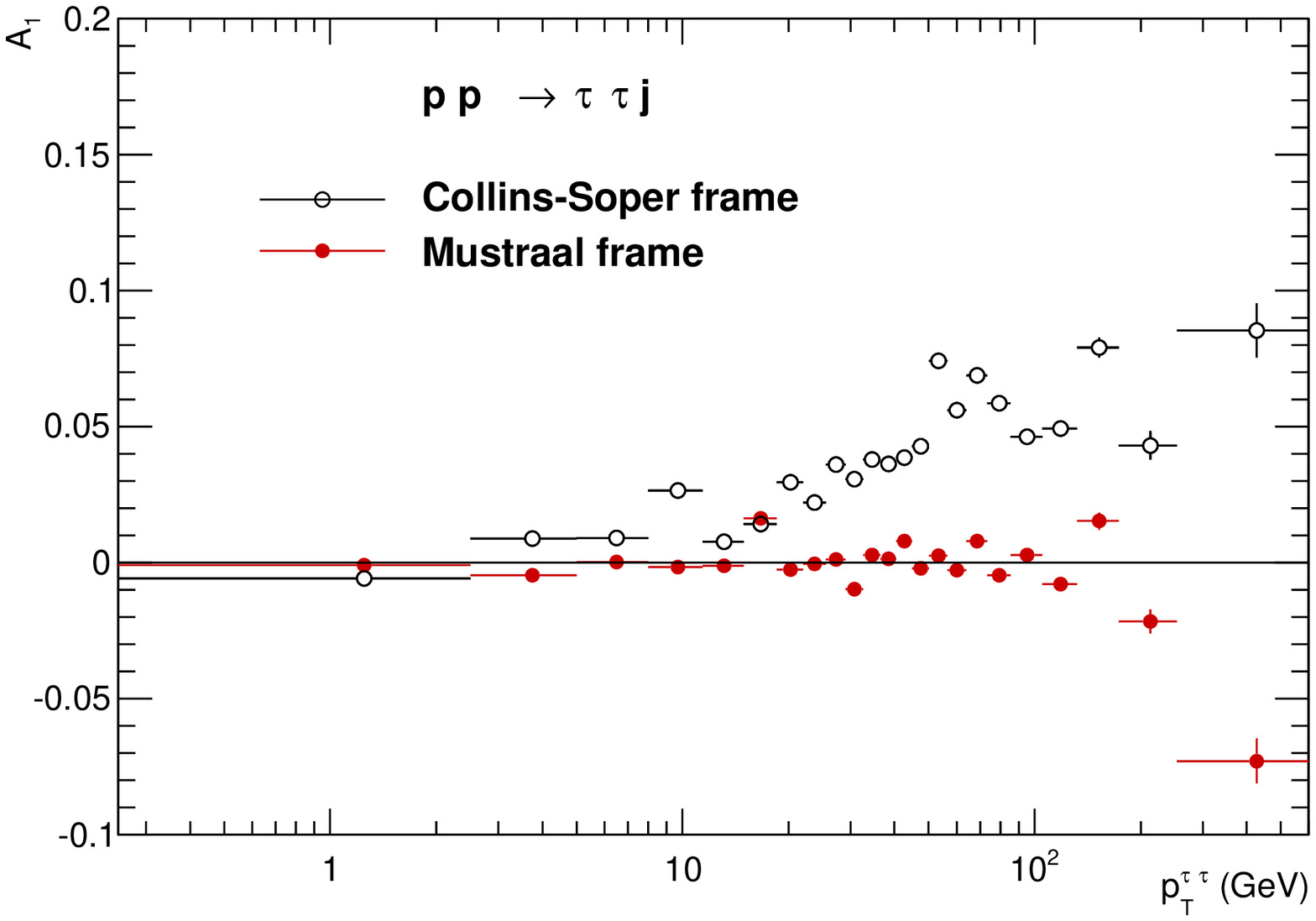}
   \includegraphics[width=7.5cm,angle=0]{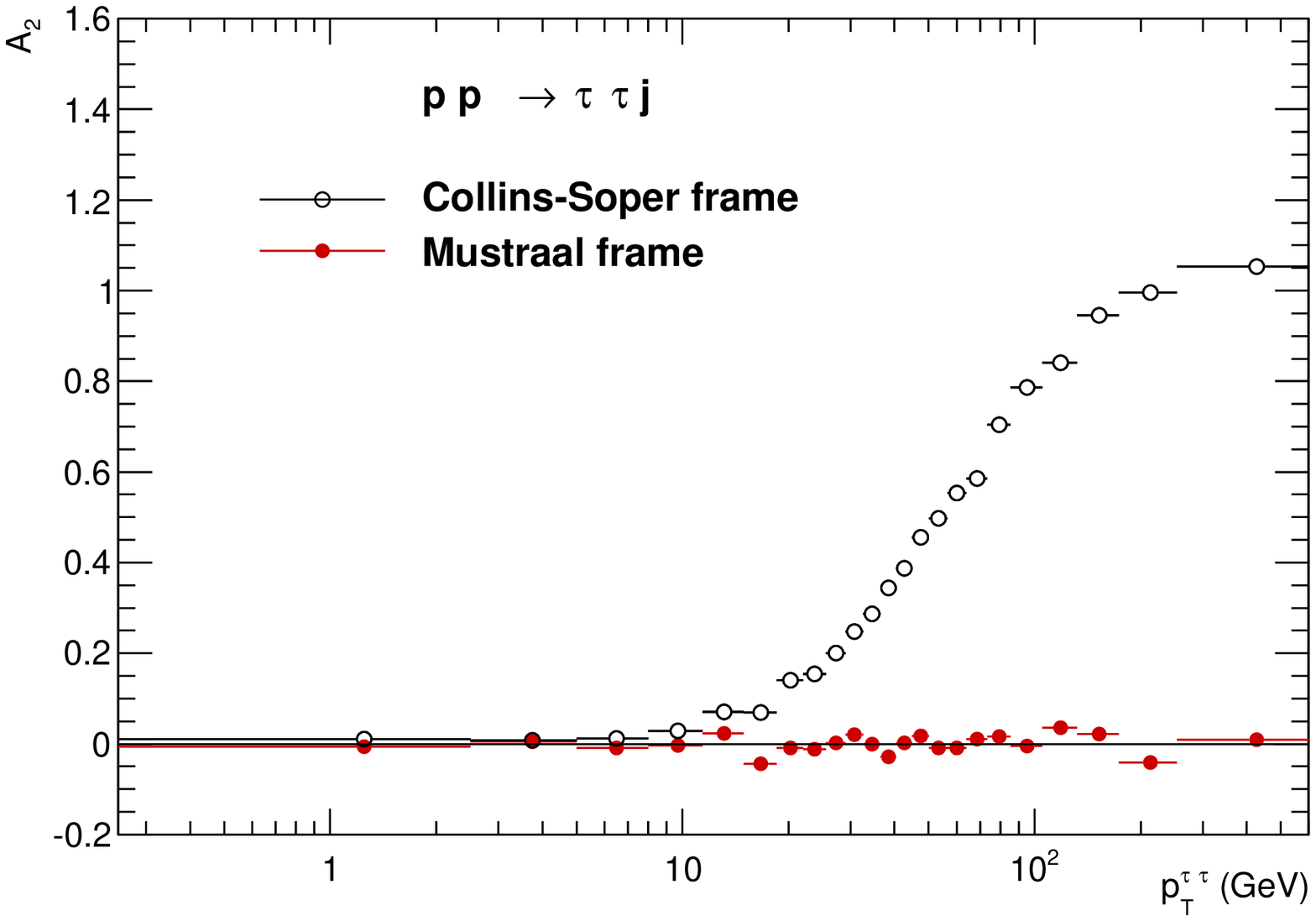}
   \includegraphics[width=7.5cm,angle=0]{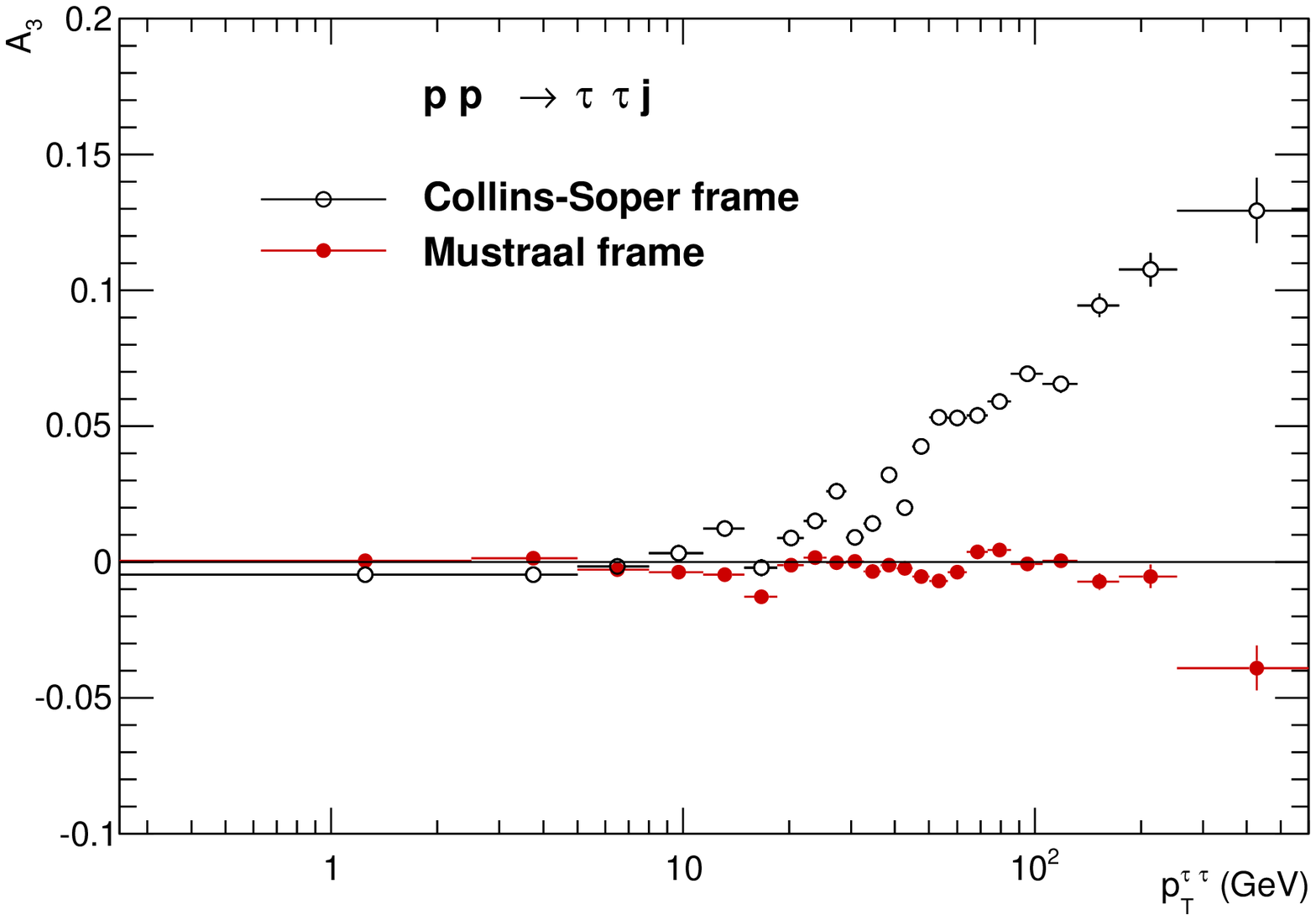}
   \includegraphics[width=7.5cm,angle=0]{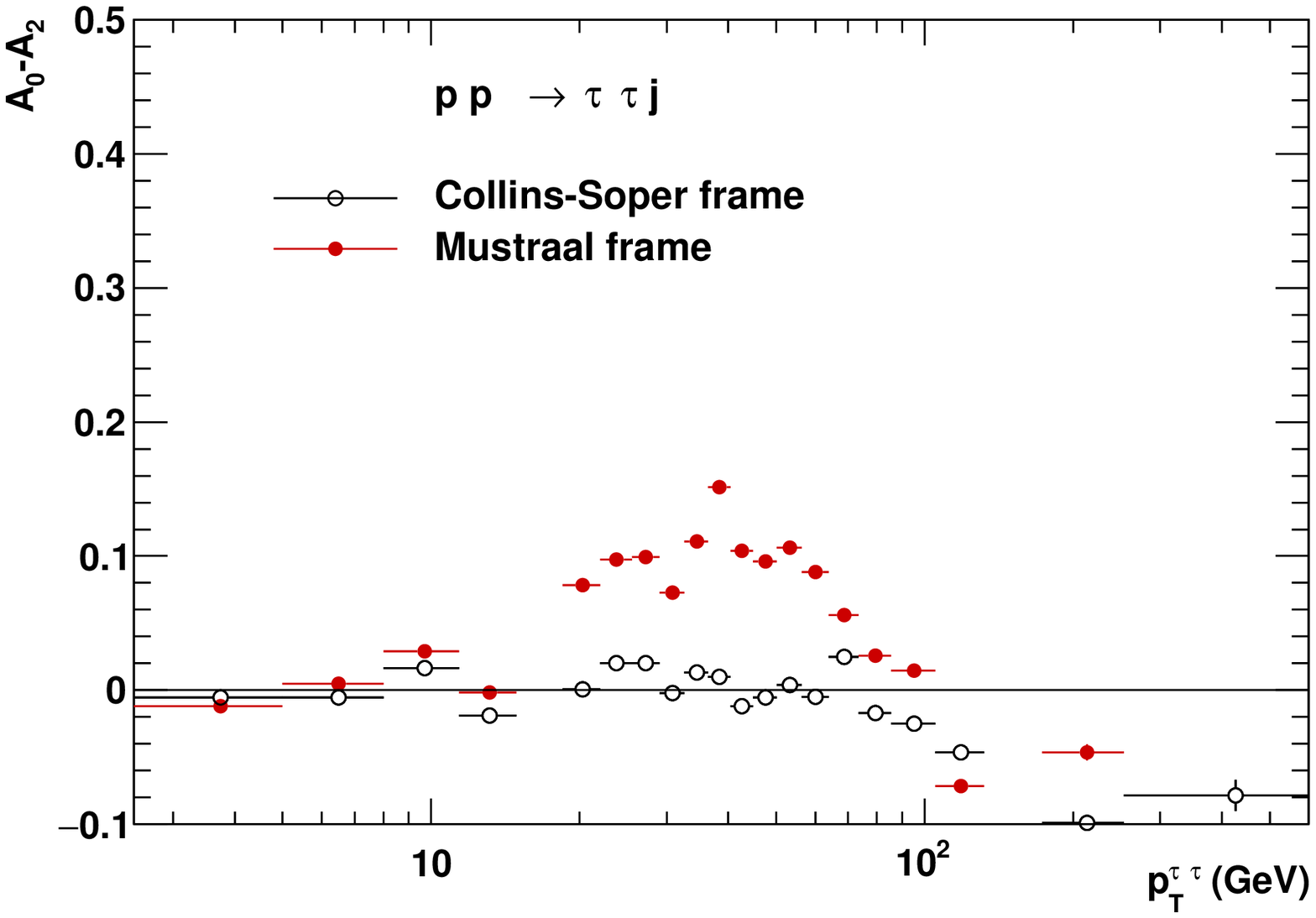}
   \includegraphics[width=7.5cm,angle=0]{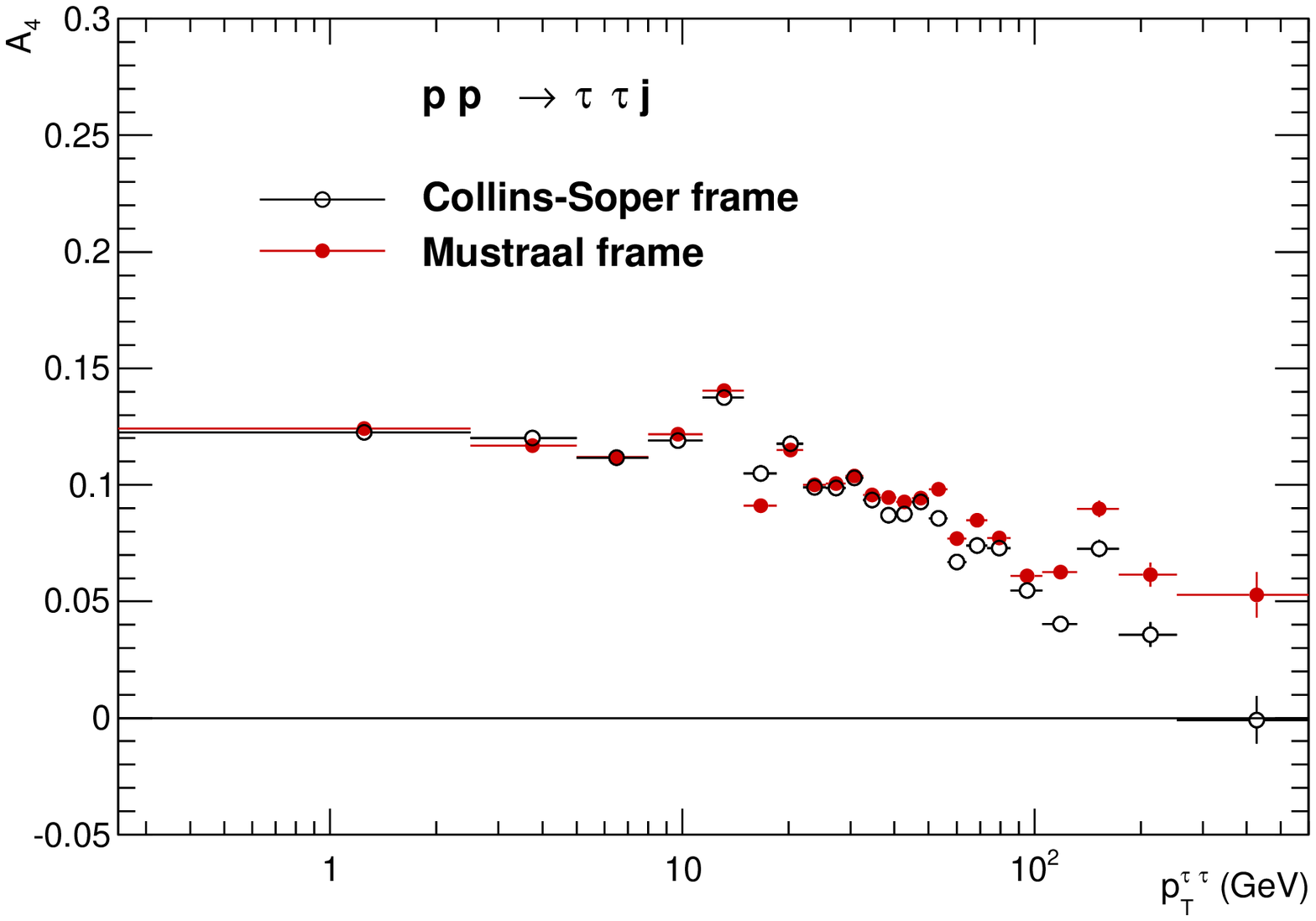}
}
\end{center}
\caption{ 
The $A_i$ coefficients of Eq.~(\ref{Eq:master2}))  calculated in Collins-Soper (black) and in {\tt Mustraal} (red) frames 
for $p p \to \tau \tau  j$ processes generated with {\tt MadGraph}.
Details of initialization are given in Section \ref{sec:numerical}.
\label{fig:Ai1jet} }
\end{figure}

In the three figures~\ref{fig:Ai1gluon},~\ref{fig:Ai1quark} and~\ref{fig:Ai1jet} we collect 
results for angular coefficients $A_i$ in the processes
of one jet  in the final state: first when it is from single gluon emission,
later when 
it can be initiated by quark or anti-quark, finally when both subprocesses 
are combined. In the above and in the following plots, 
we show sets of five angular coefficients $A_0 -A_4$ only; the remaining ones
$A_5 -A_7$ are close to zero over the full $p_T^{\tau \tau}$ range  for 
 both definition of frames, Collins-Soper and {\tt Mustraal}. 
We  present also $\Delta A_{02} = A_0 - A_2$, it is known to be sensitive to the QCD sector
and should be precisely zero at the LO QCD and the Collins-Soper frame,
because of  the Lam-Tung relation \cite{Lam:1978pu}.

\begin{figure}
  \begin{center}                               
{
   \includegraphics[width=7.5cm,angle=0]{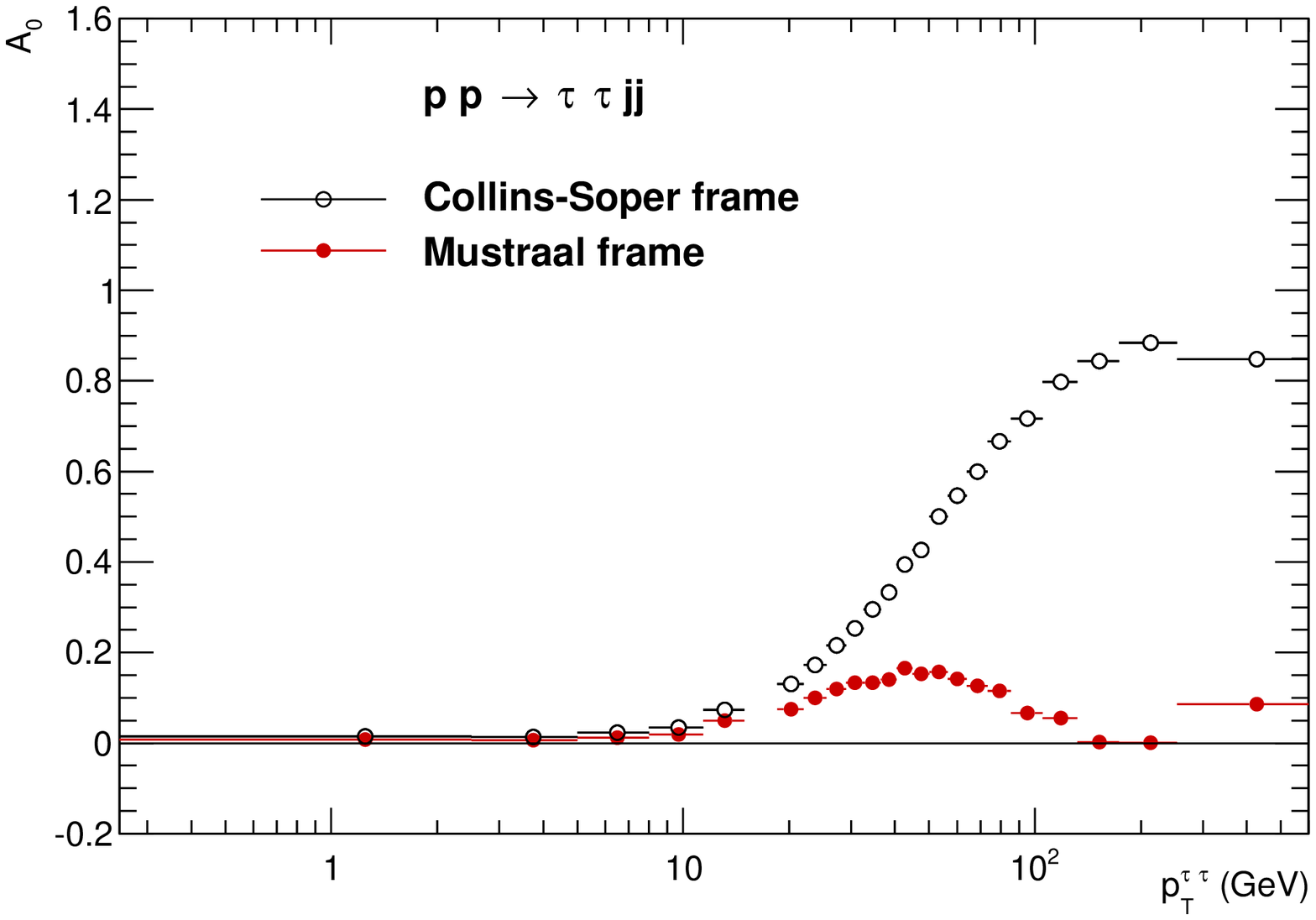}
   \includegraphics[width=7.5cm,angle=0]{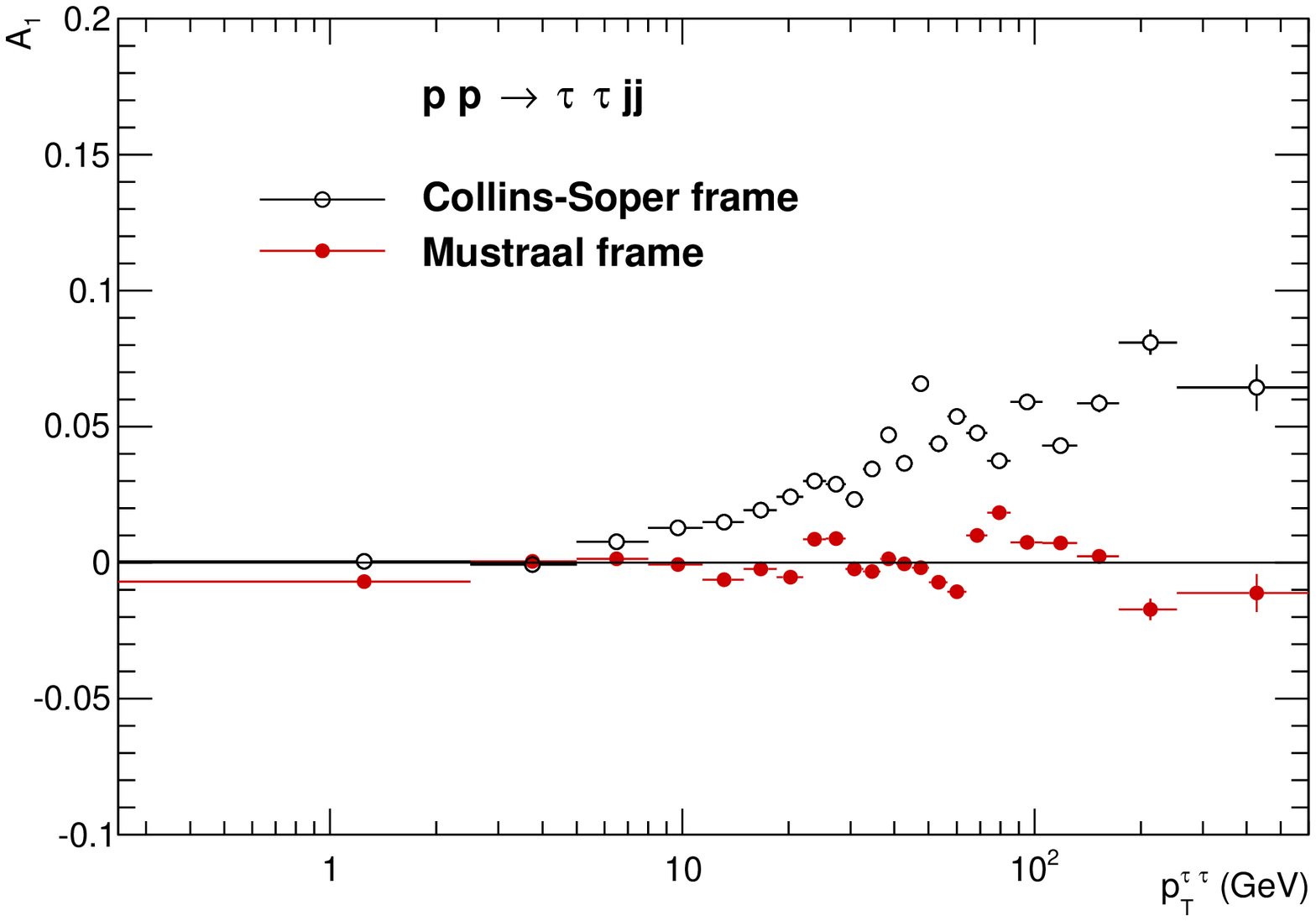}
   \includegraphics[width=7.5cm,angle=0]{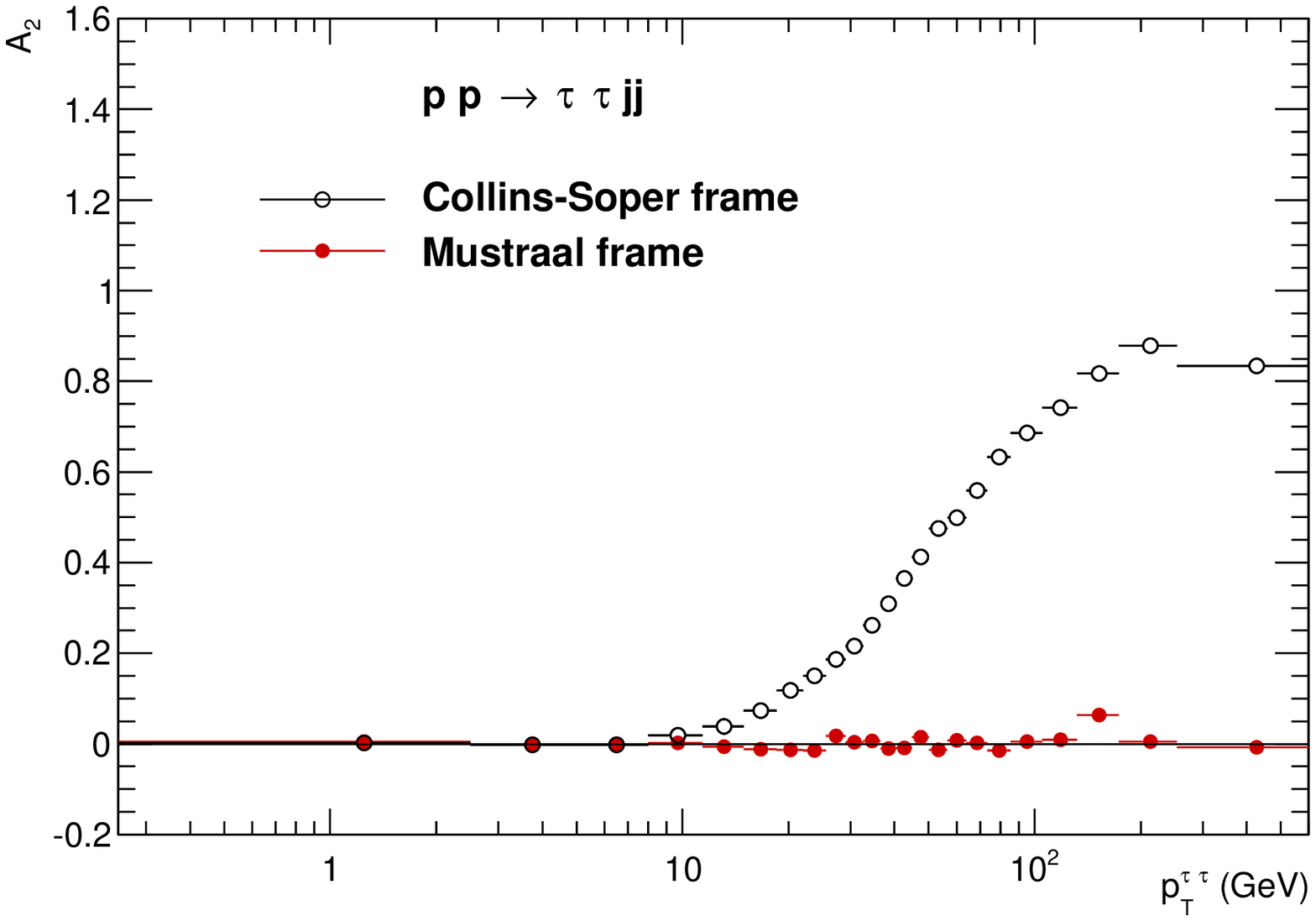}
   \includegraphics[width=7.5cm,angle=0]{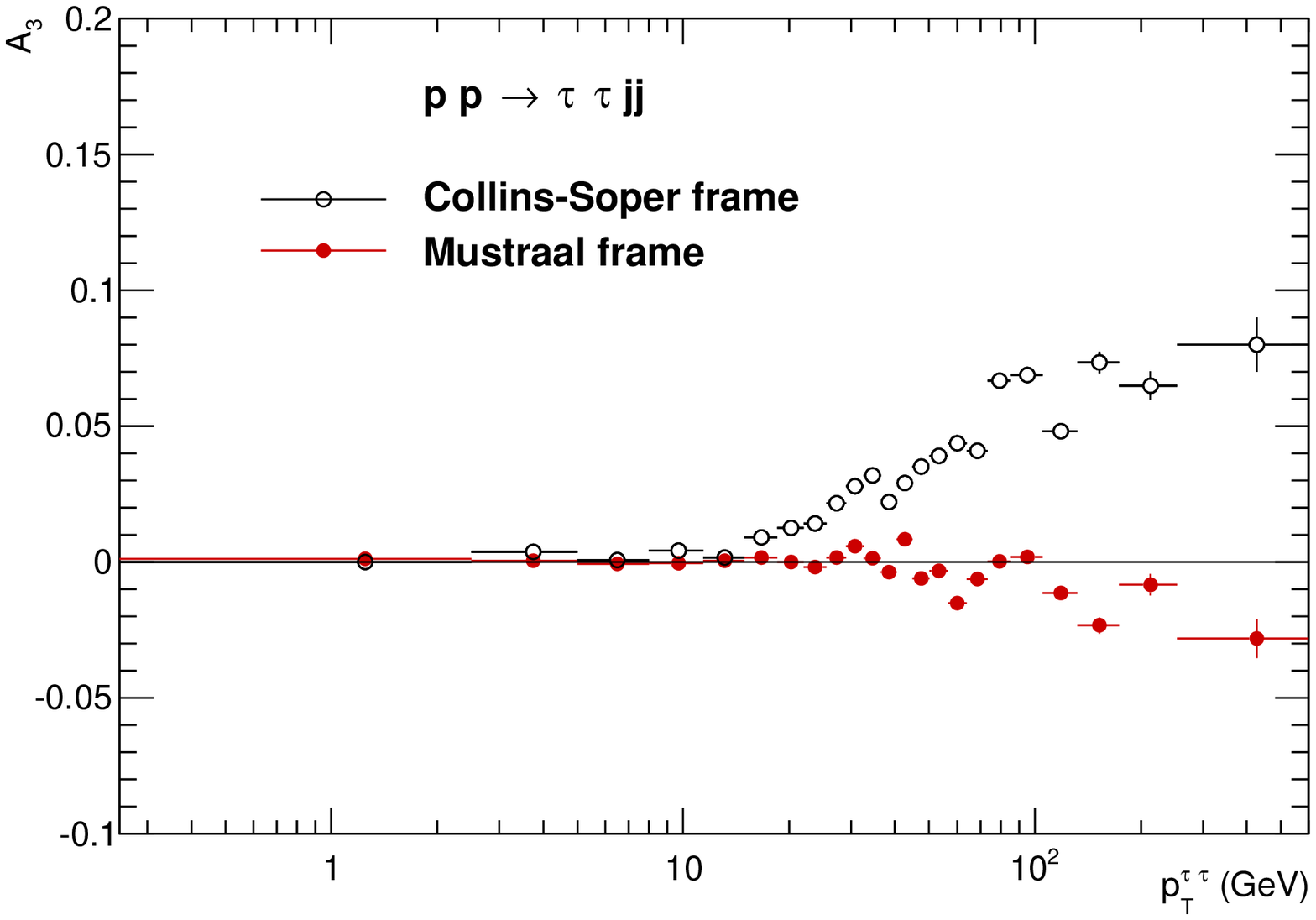}
   \includegraphics[width=7.5cm,angle=0]{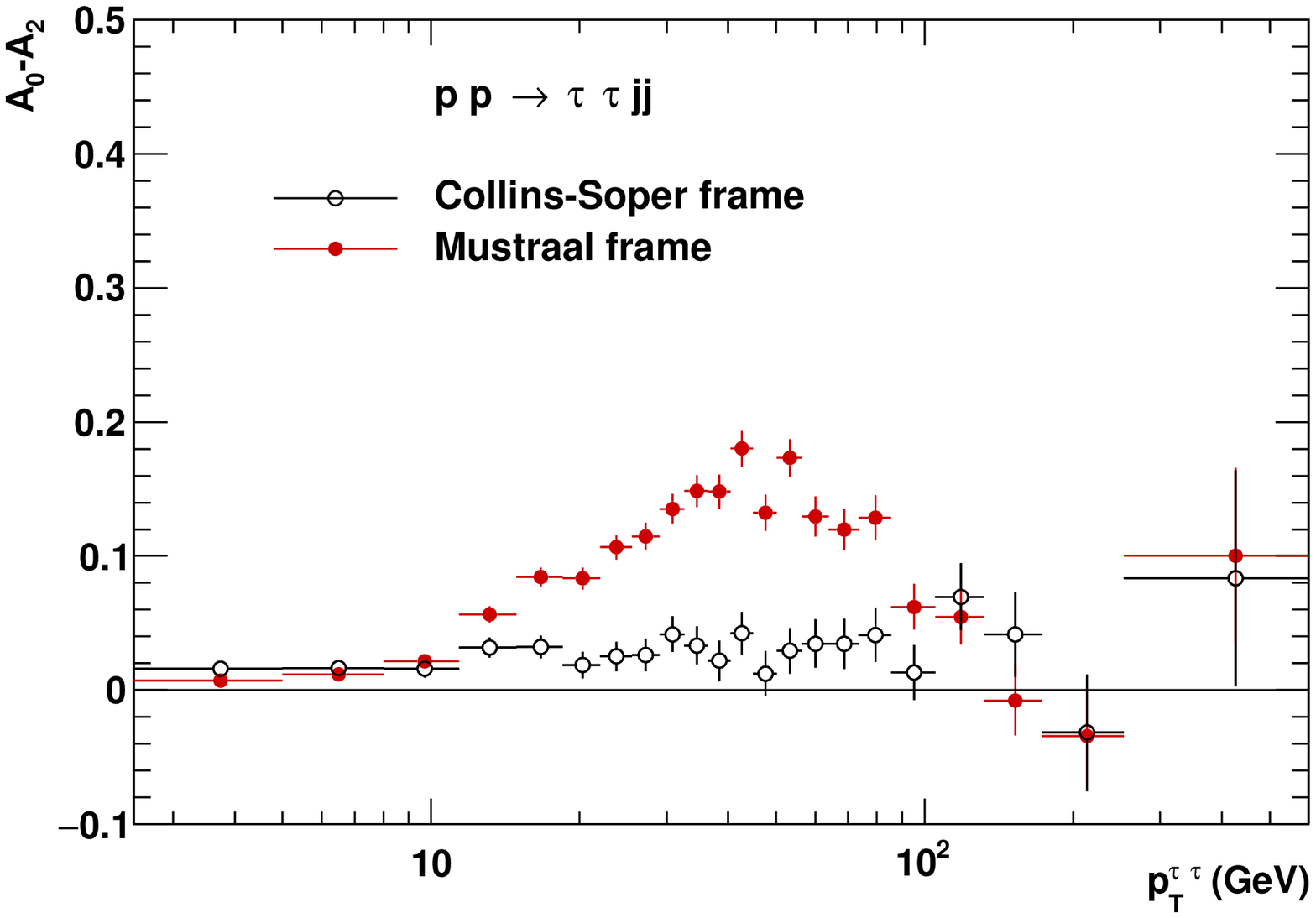}
   \includegraphics[width=7.5cm,angle=0]{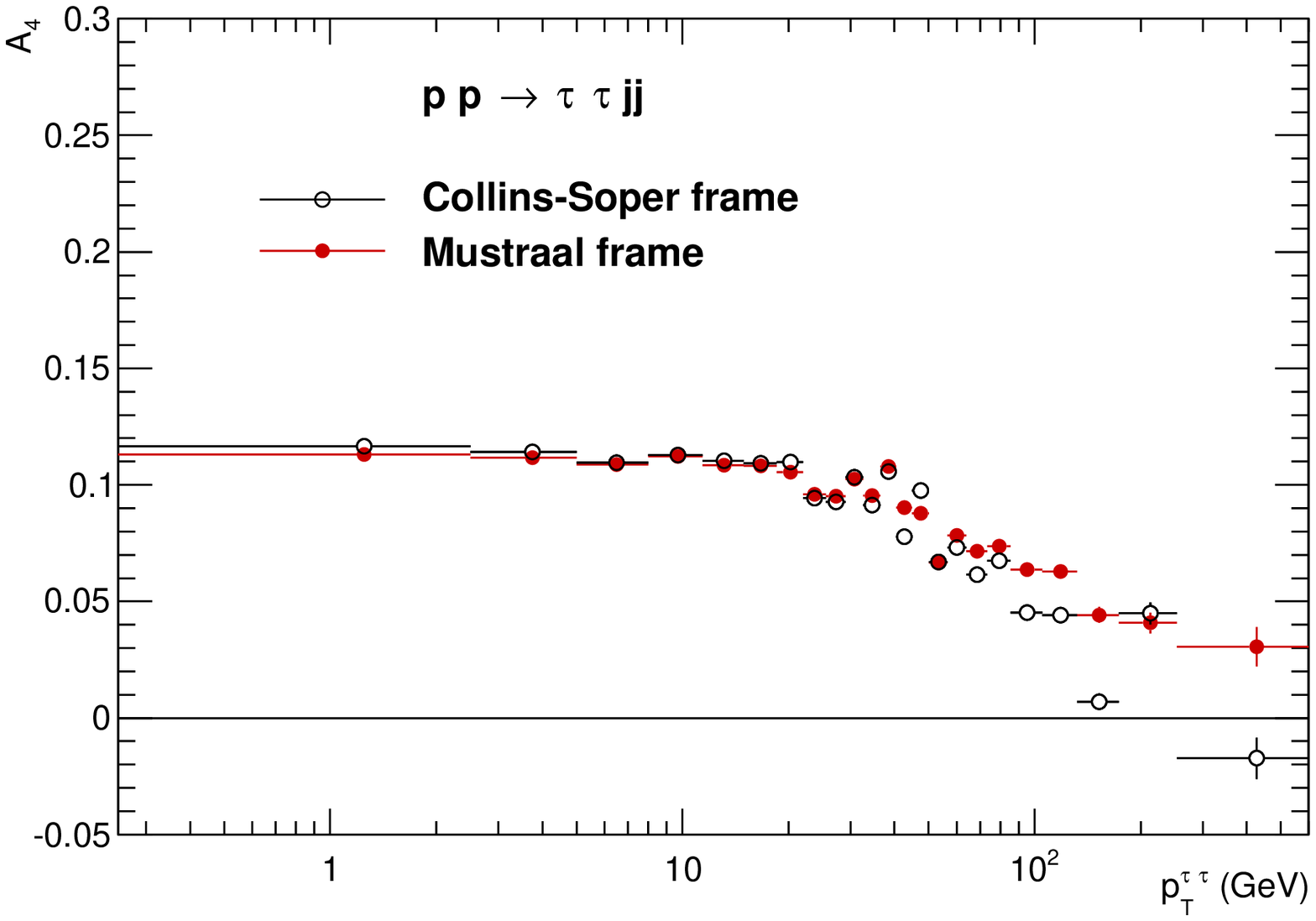}
}
\end{center}
\caption{ 
The $A_i$ coefficients of Eq.~(\ref{Eq:master2}))  calculated in Collins-Soper (black) and in {\tt Mustraal} (red) frames 
for $p p  \to \tau \tau j j$ process generated with {\tt MadGraph}.
Details of initialization are given in Section \ref{sec:numerical}.
\label{fig:Ai2jets} }
\end{figure}

\begin{figure}
  \begin{center}                               
{
   \includegraphics[width=7.5cm,angle=0]{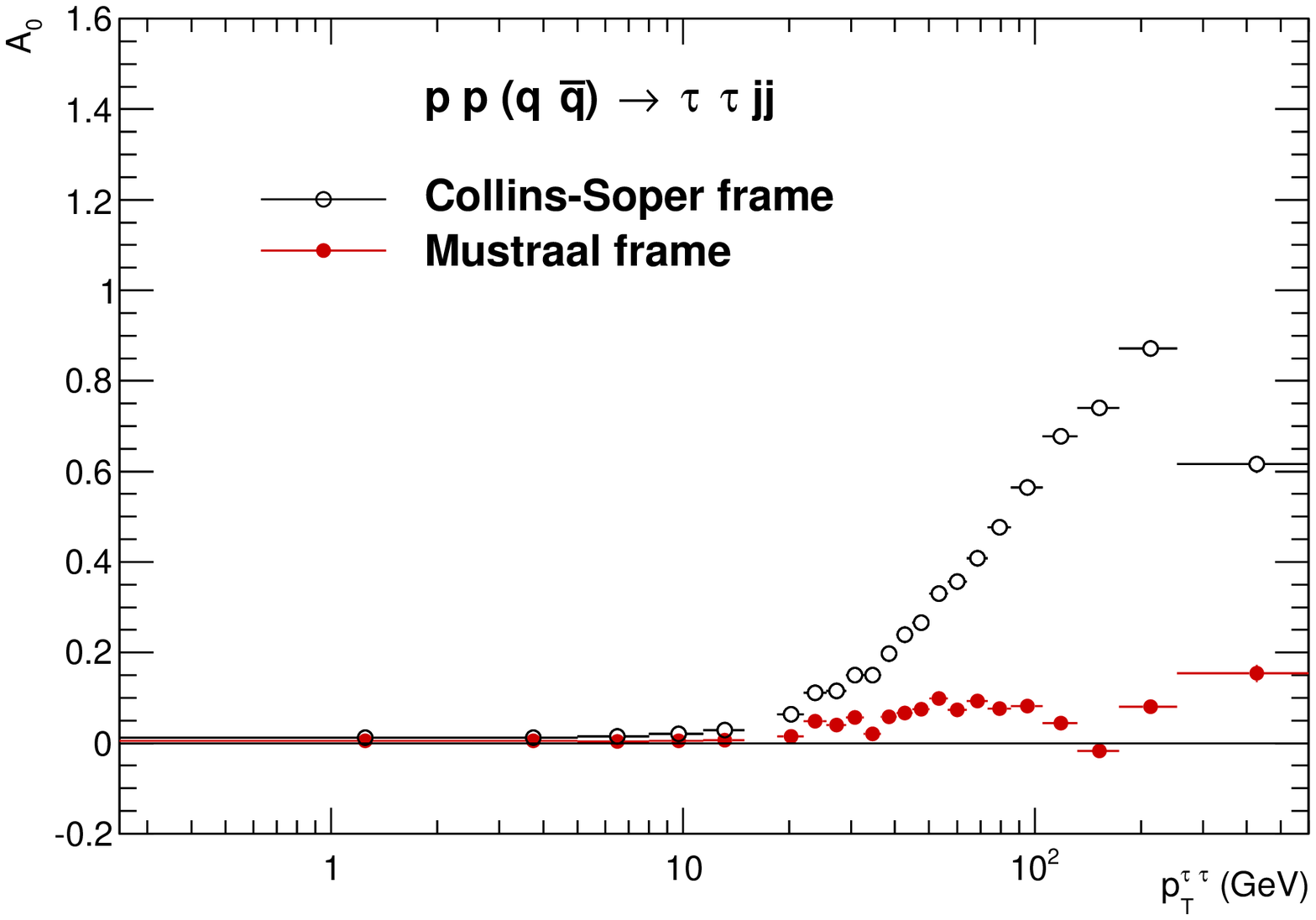}
   \includegraphics[width=7.5cm,angle=0]{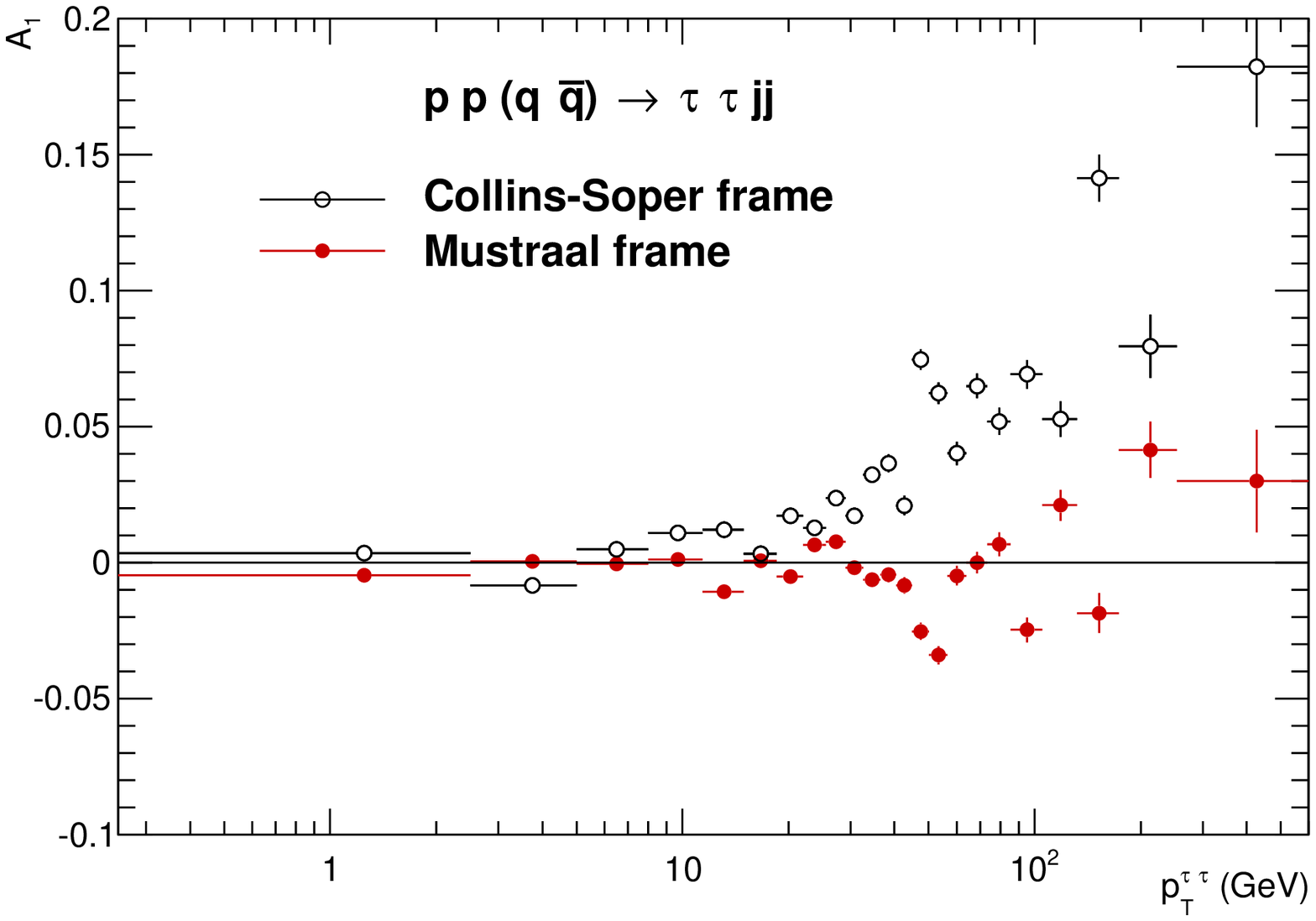}
   \includegraphics[width=7.5cm,angle=0]{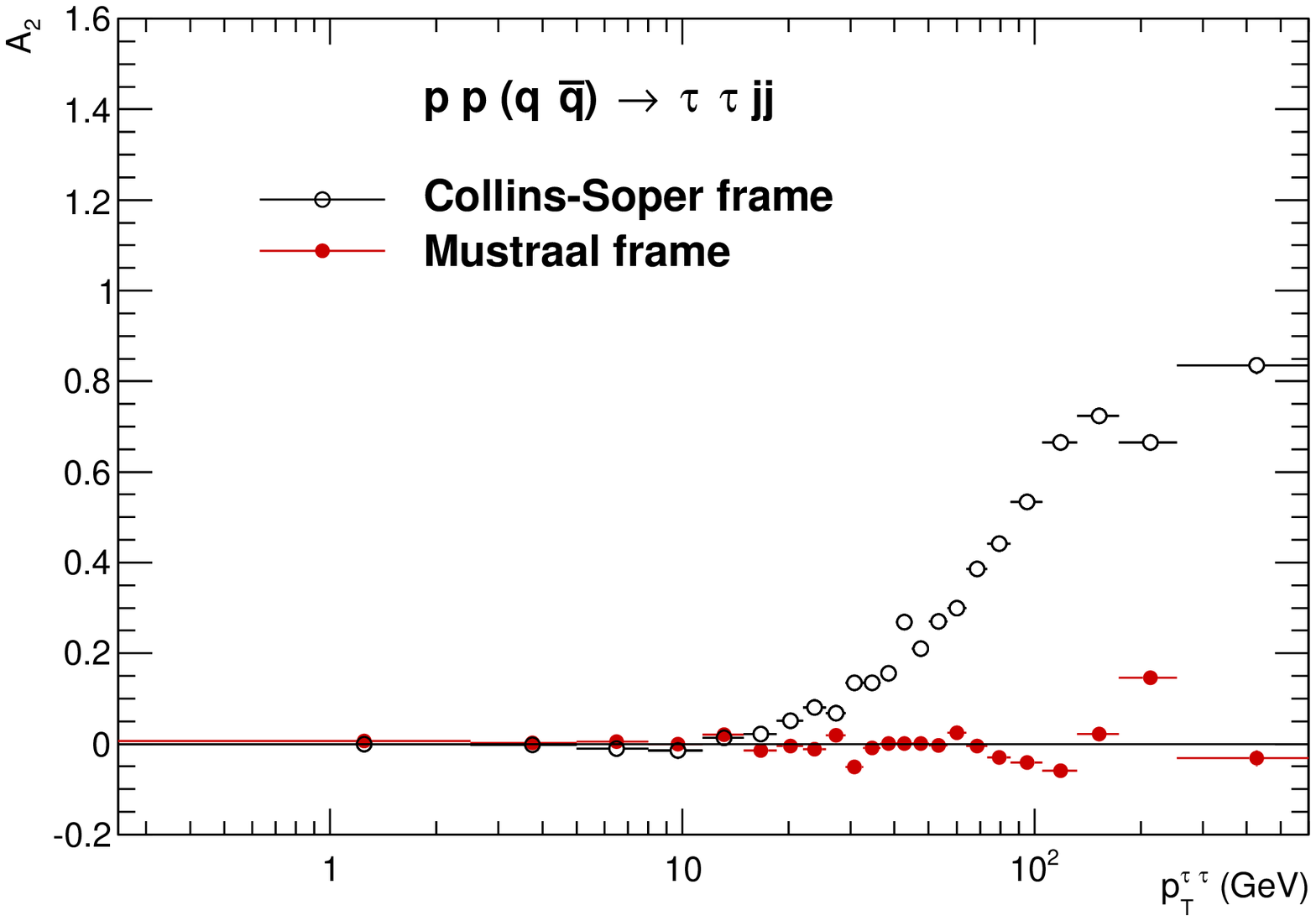}
   \includegraphics[width=7.5cm,angle=0]{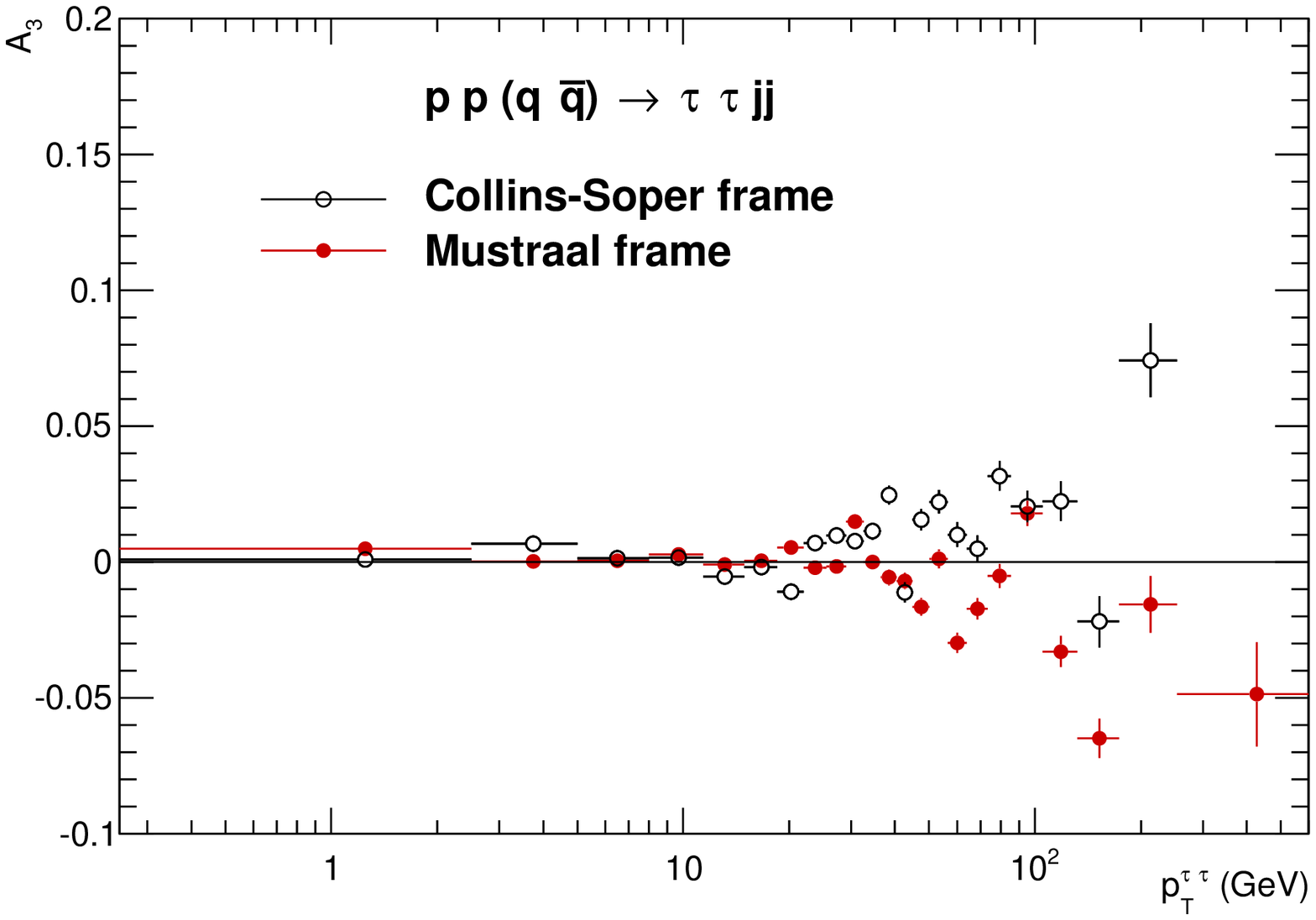}
   \includegraphics[width=7.5cm,angle=0]{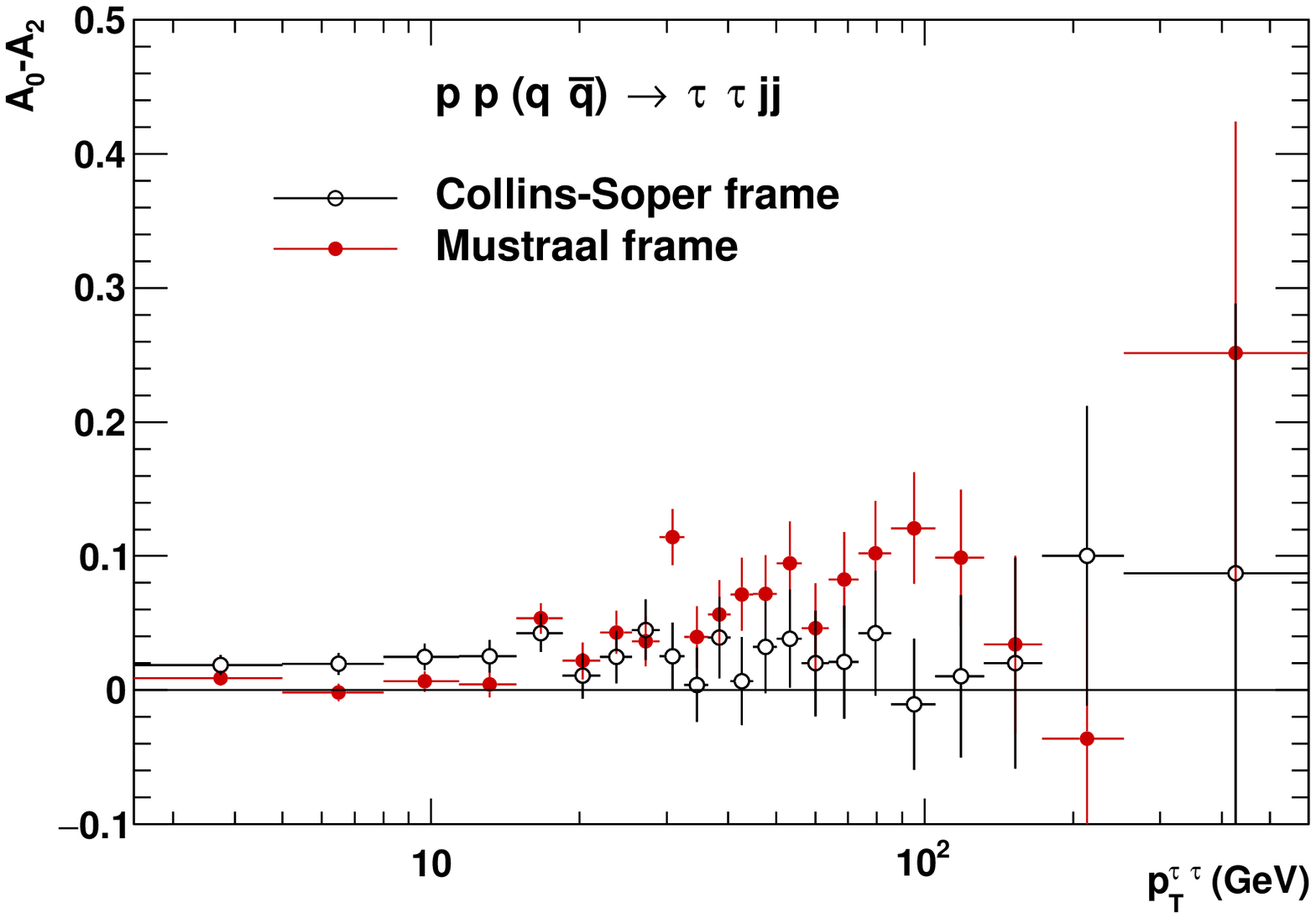}
   \includegraphics[width=7.5cm,angle=0]{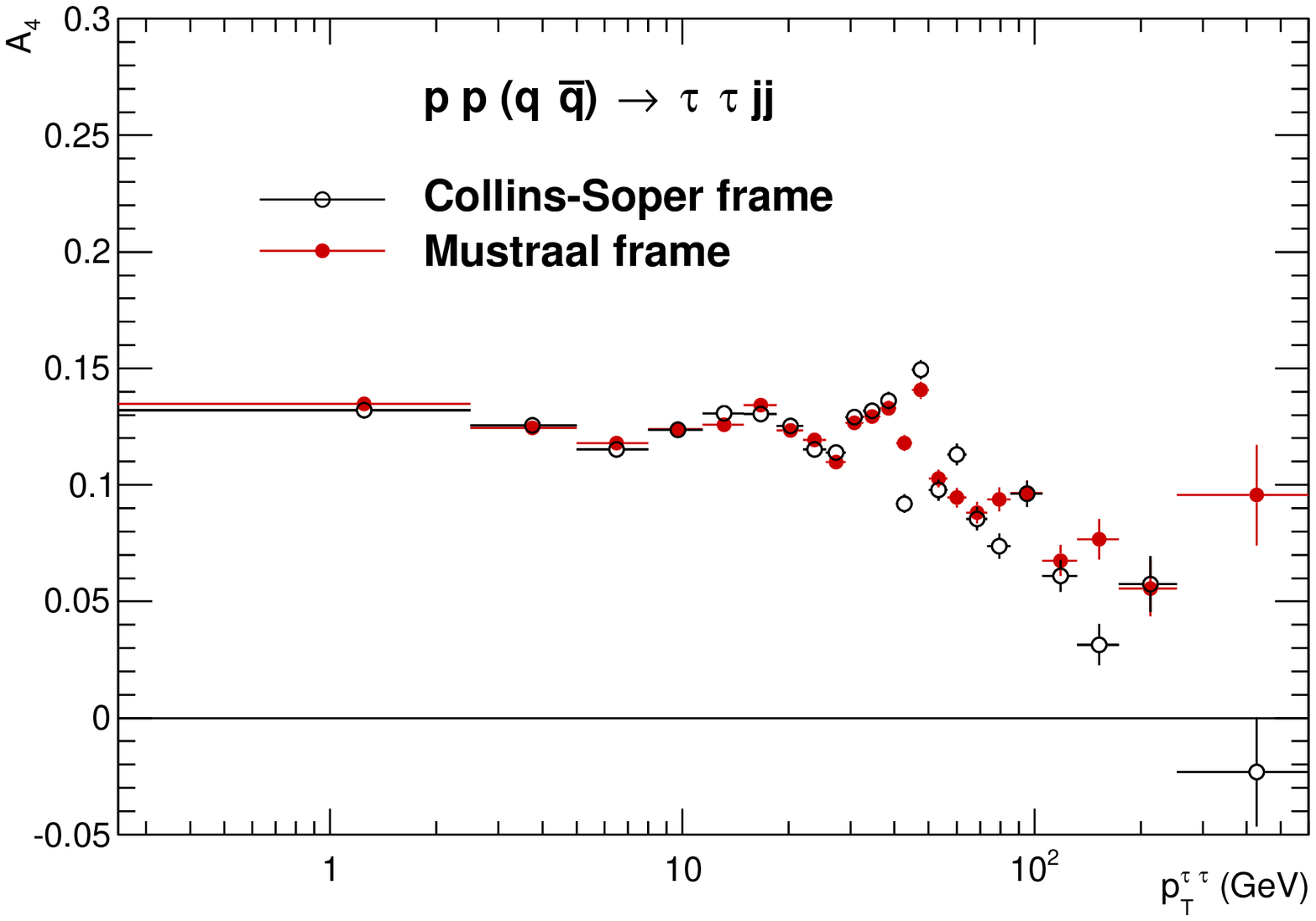}
}
\end{center}
\caption{ 
The $A_i$ coefficients of Eq.~(\ref{Eq:master2}))  calculated in Collins-Soper (black) and in {\tt Mustraal} (red) frames 
for $p p (q \bar q) \to \tau \tau j j$ process generated with {\tt MadGraph}.
Details of initialization are given in Section \ref{sec:numerical}.
\label{fig:Ai2jetsIzo0} }
\end{figure}

\begin{figure}
  \begin{center}                               
{
   \includegraphics[width=7.5cm,angle=0]{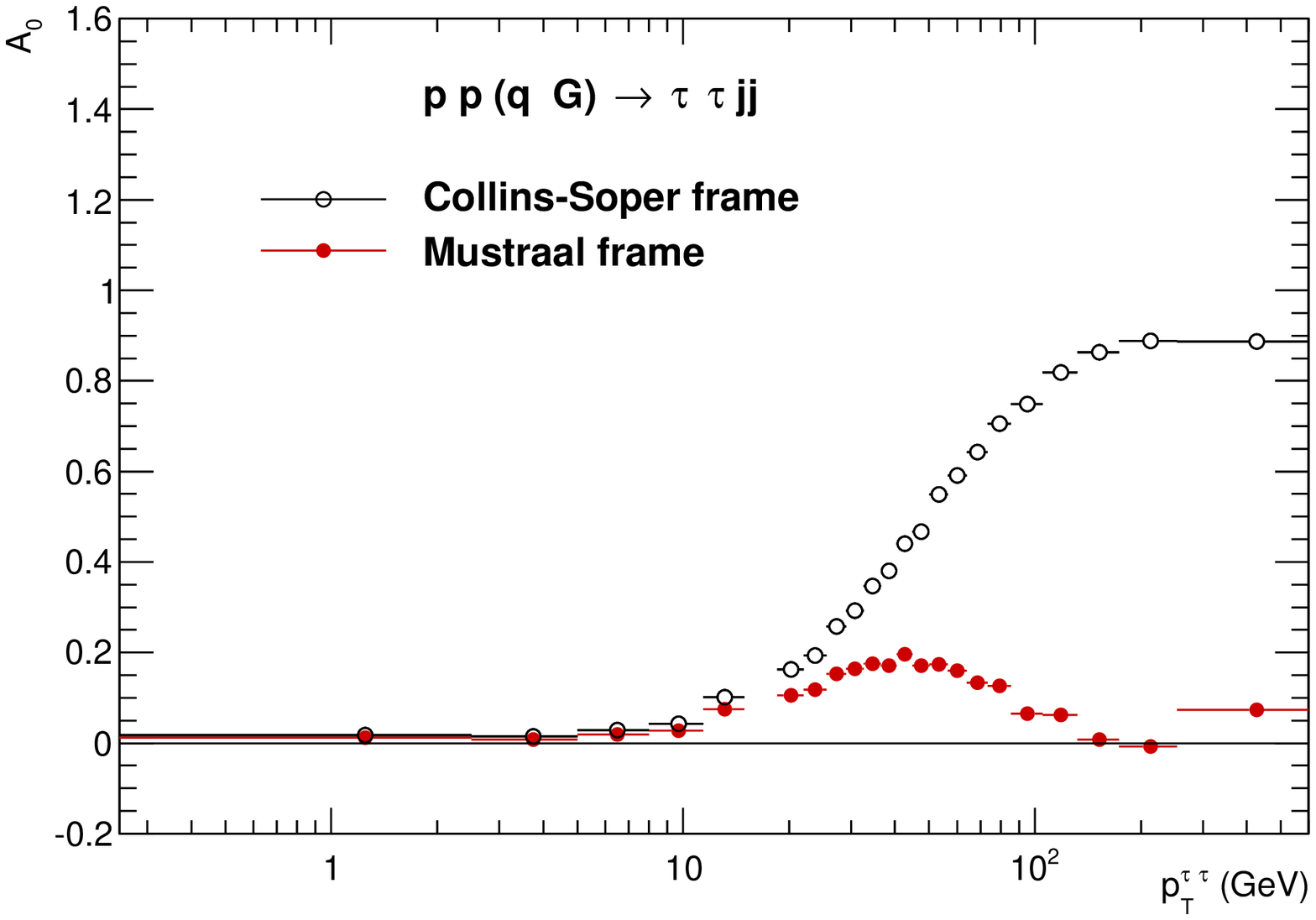}
   \includegraphics[width=7.5cm,angle=0]{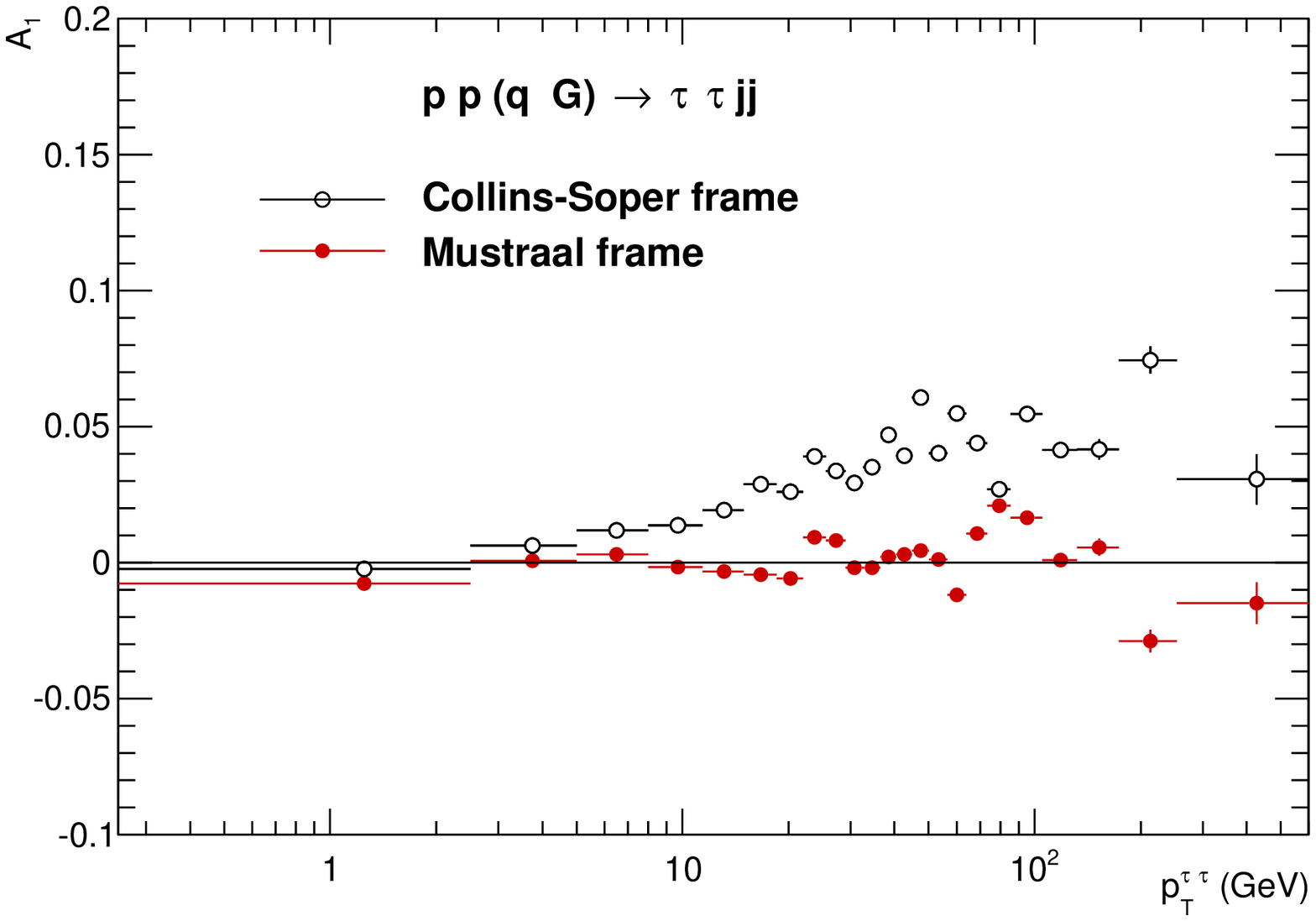}
   \includegraphics[width=7.5cm,angle=0]{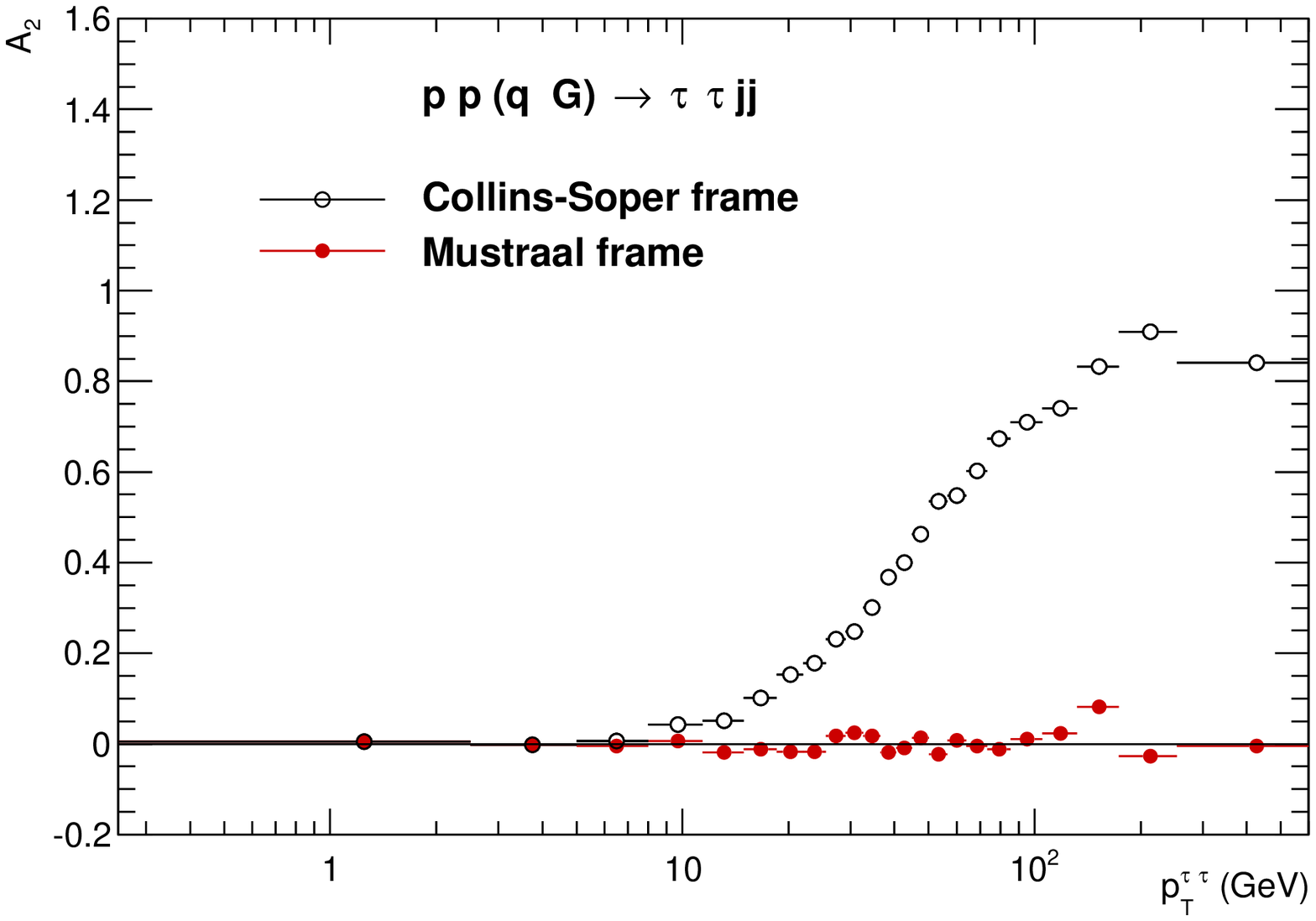}
   \includegraphics[width=7.5cm,angle=0]{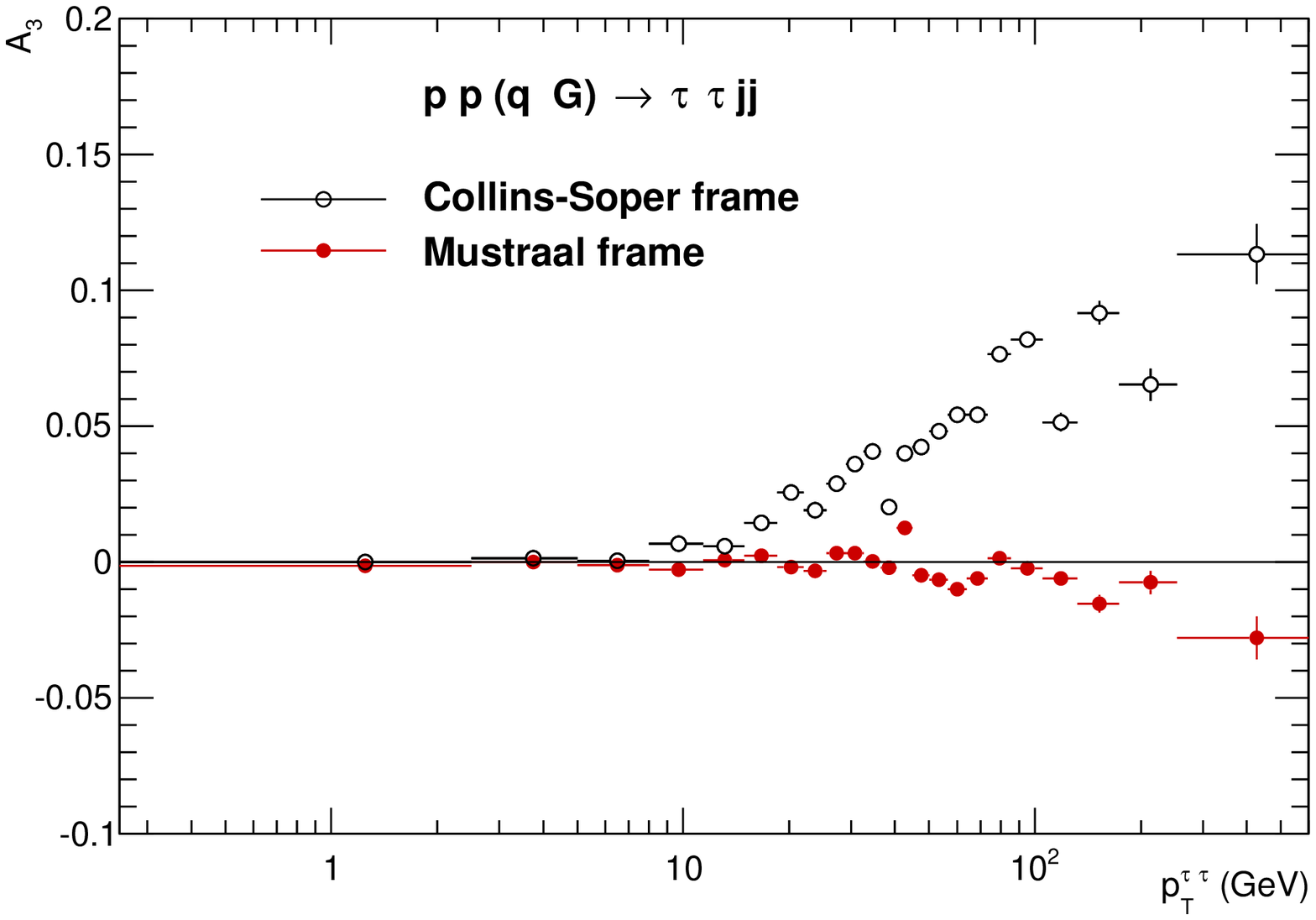}
   \includegraphics[width=7.5cm,angle=0]{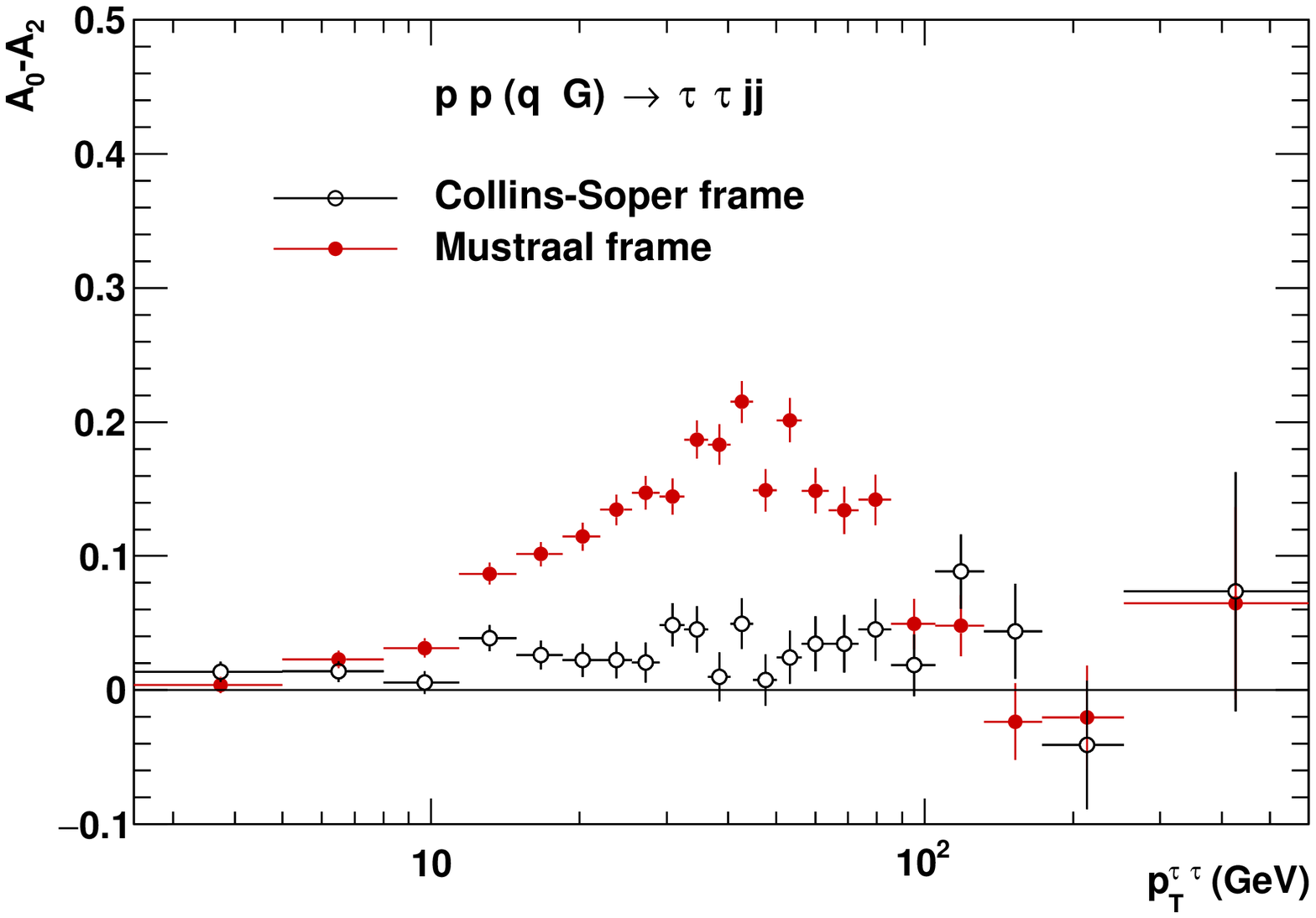}
   \includegraphics[width=7.5cm,angle=0]{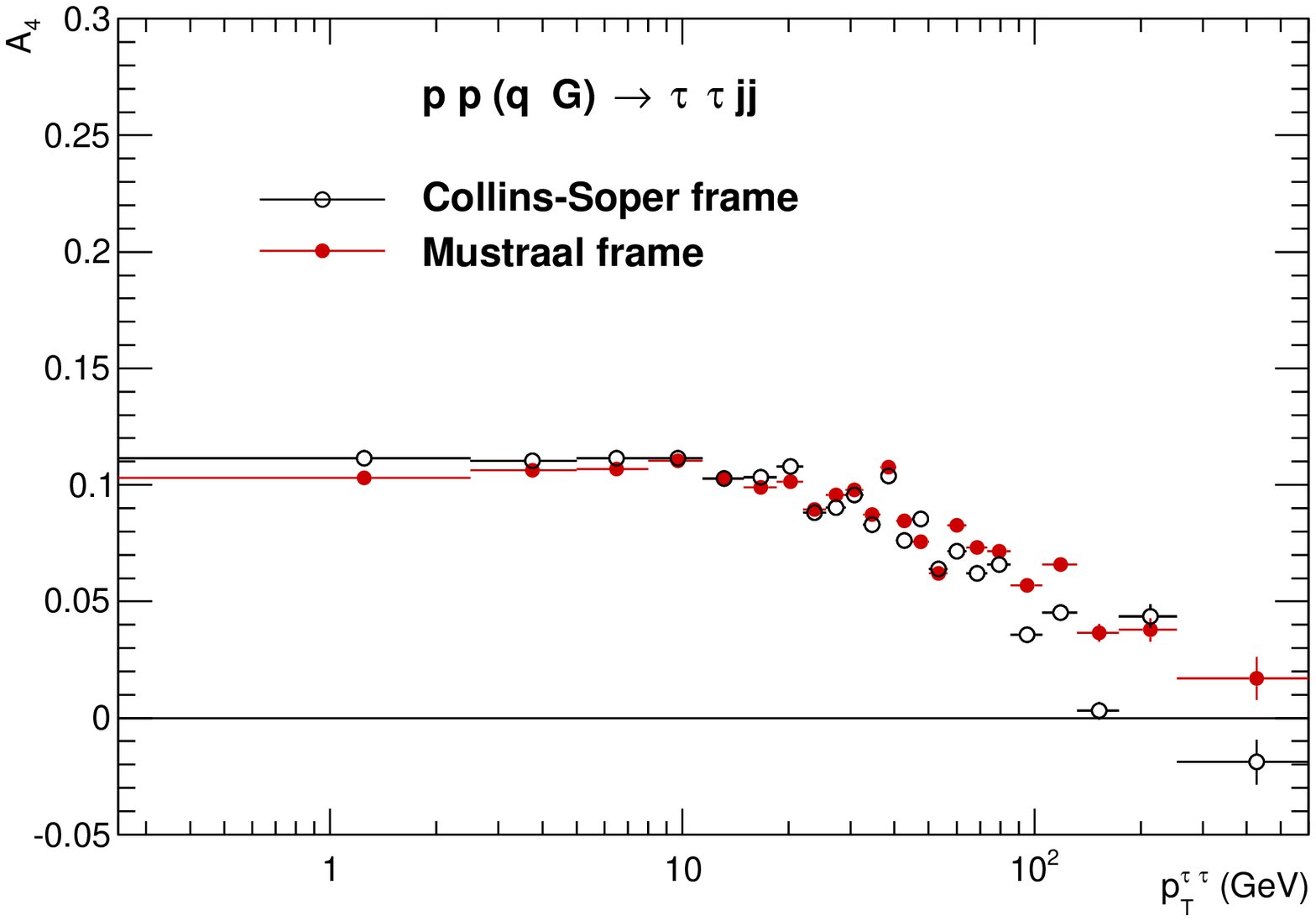}
}
\end{center}
\caption{ 
The $A_i$ coefficients of Eq.~(\ref{Eq:master2})) calculated in Collins-Soper (black) and in {\tt Mustraal} (red) frames 
for $p p (q G) \to  \tau \tau jj $ process generated with {\tt MadGraph}.
Details of initialization are given in Section \ref{sec:numerical}.
\label{fig:Ai2jets1g} }
\end{figure}

\begin{figure}
  \begin{center}                               
{
   \includegraphics[width=7.5cm,angle=0]{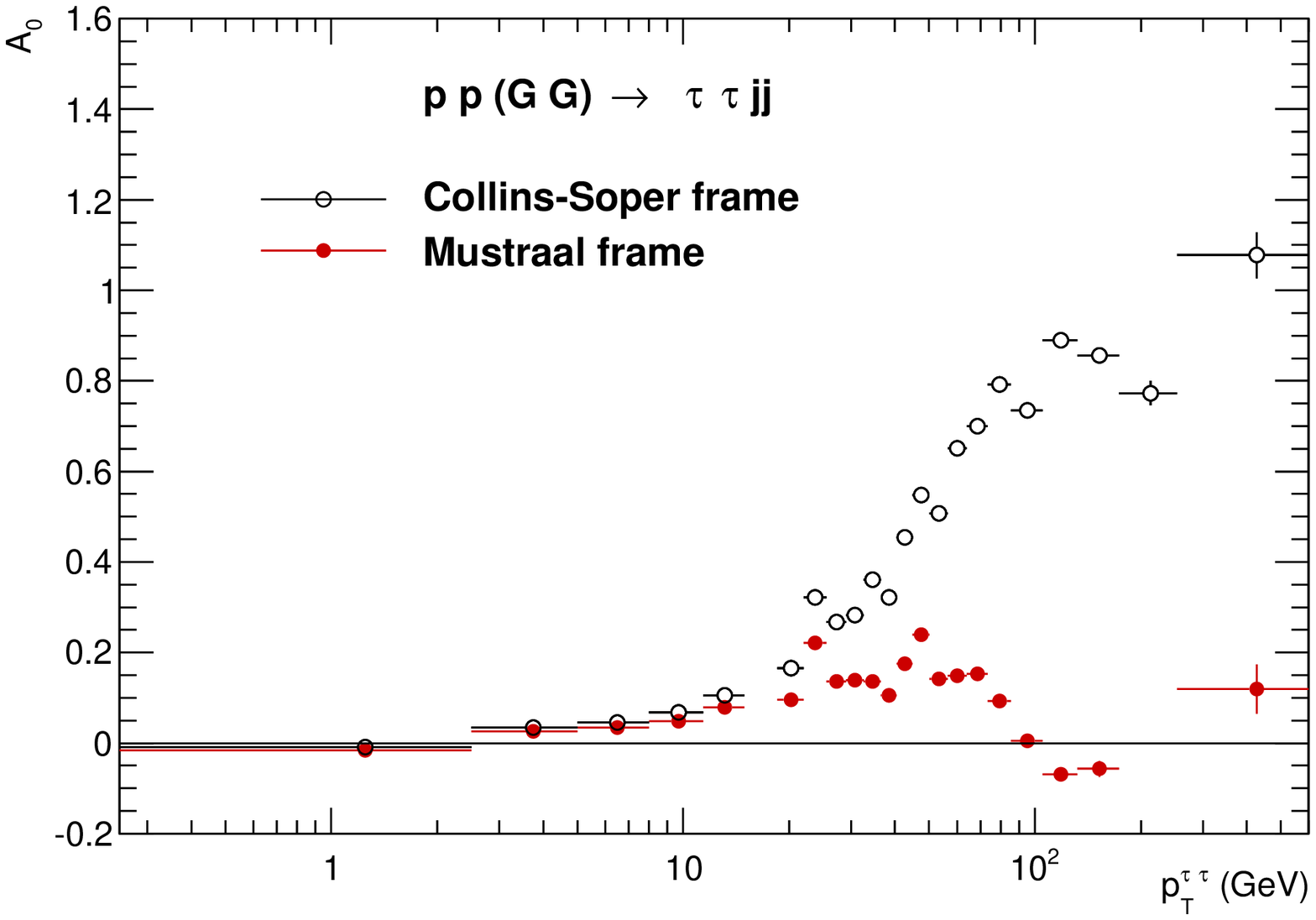}
   \includegraphics[width=7.5cm,angle=0]{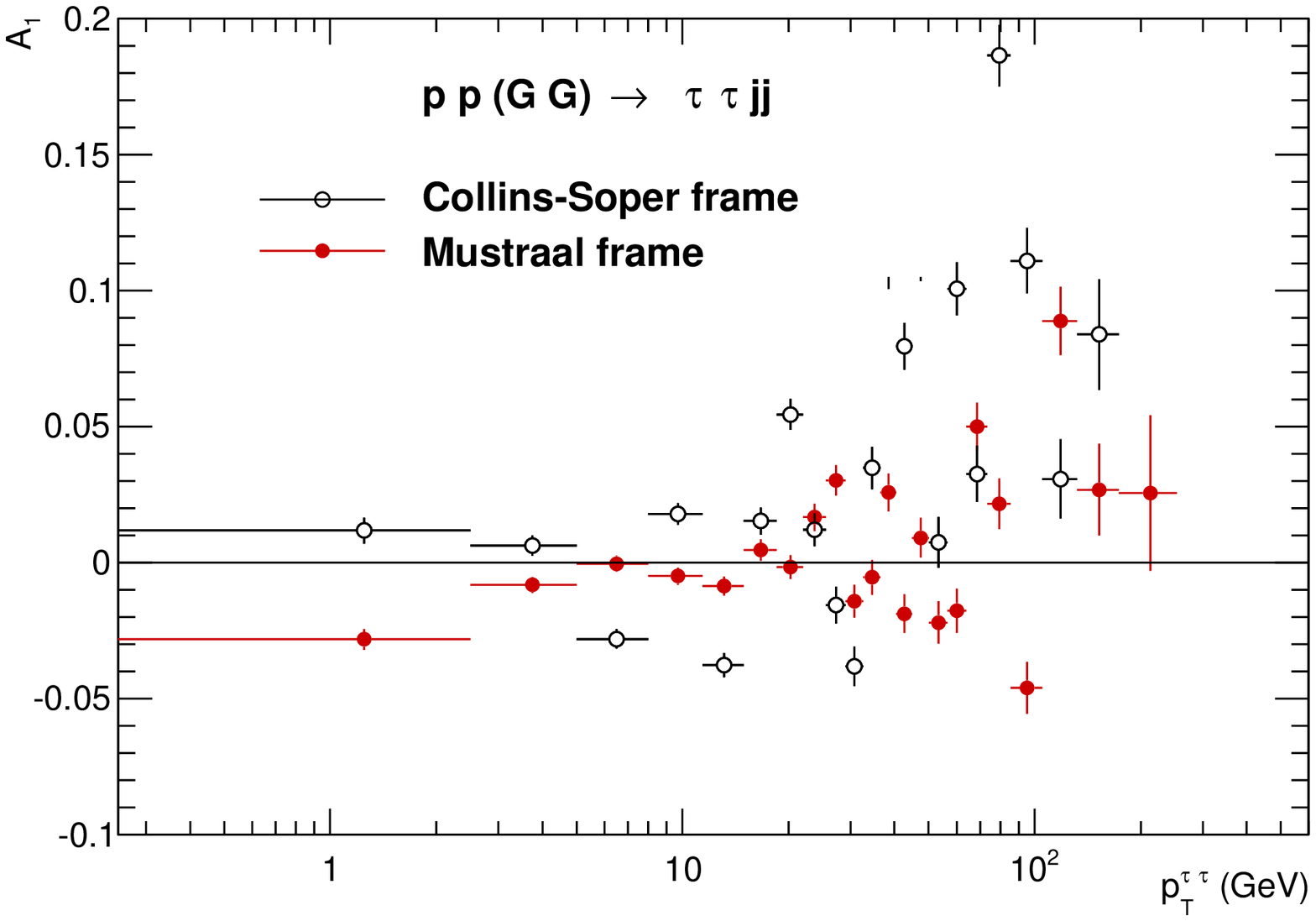}
   \includegraphics[width=7.5cm,angle=0]{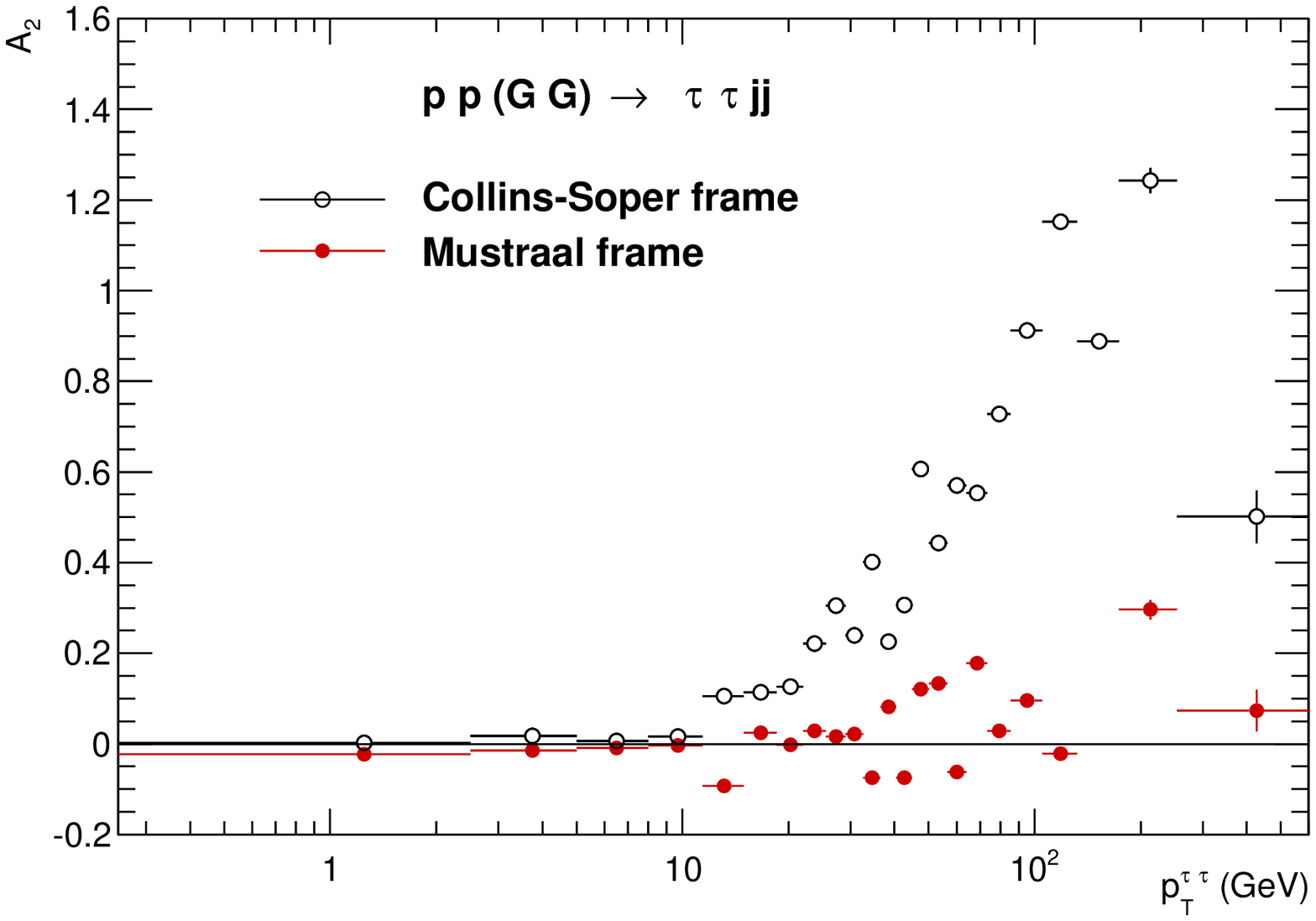}
   \includegraphics[width=7.5cm,angle=0]{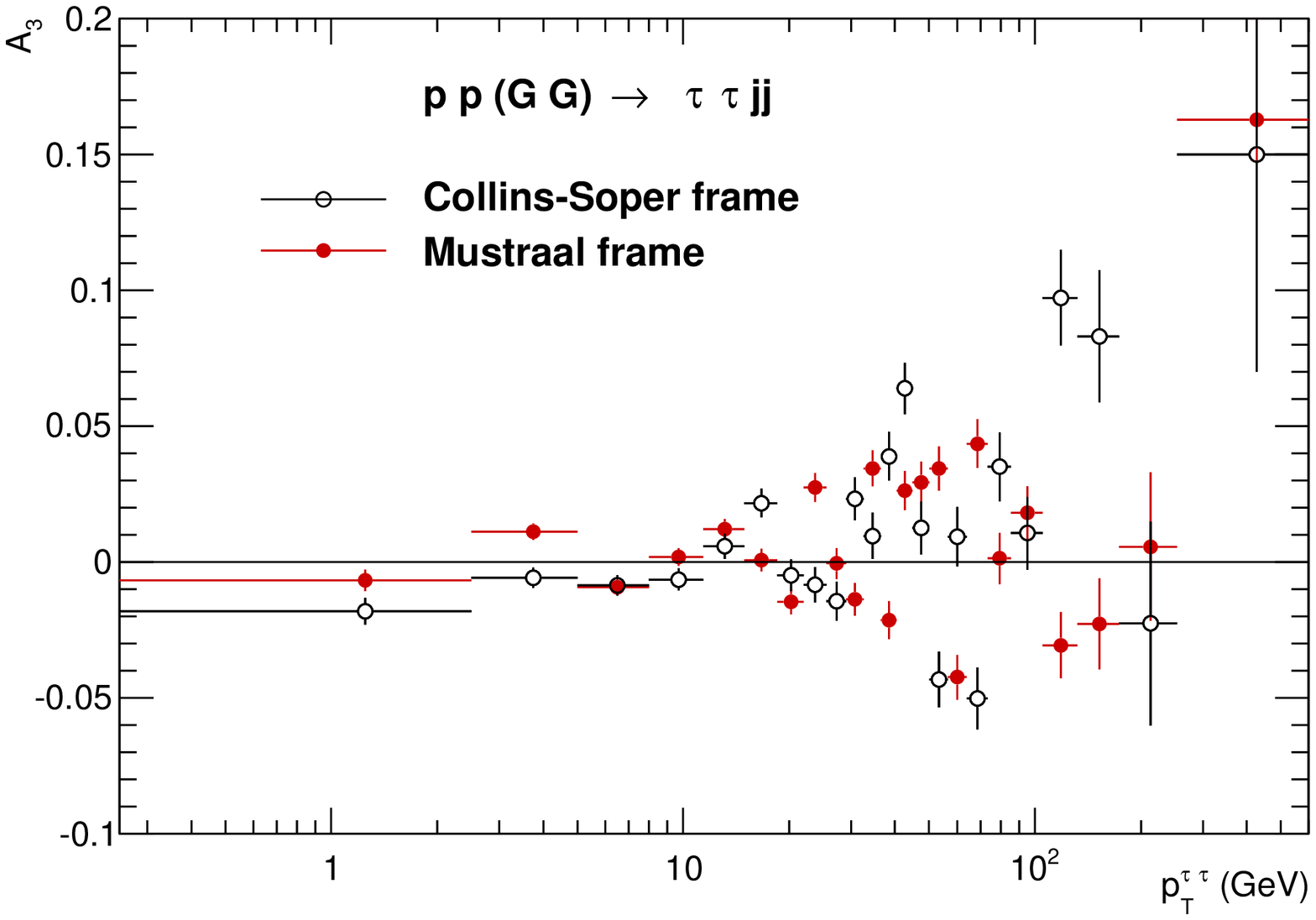}
   \includegraphics[width=7.5cm,angle=0]{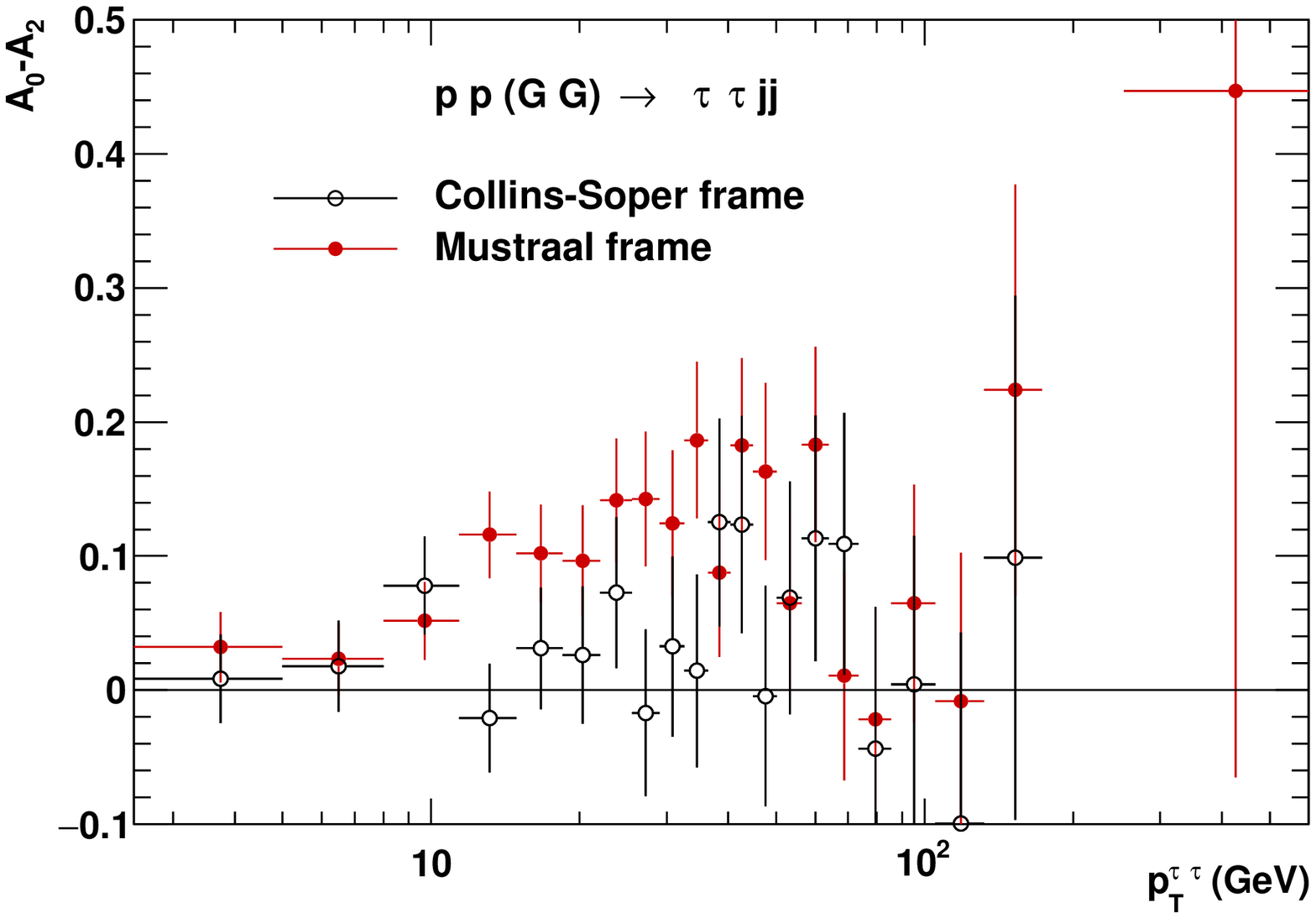}
   \includegraphics[width=7.5cm,angle=0]{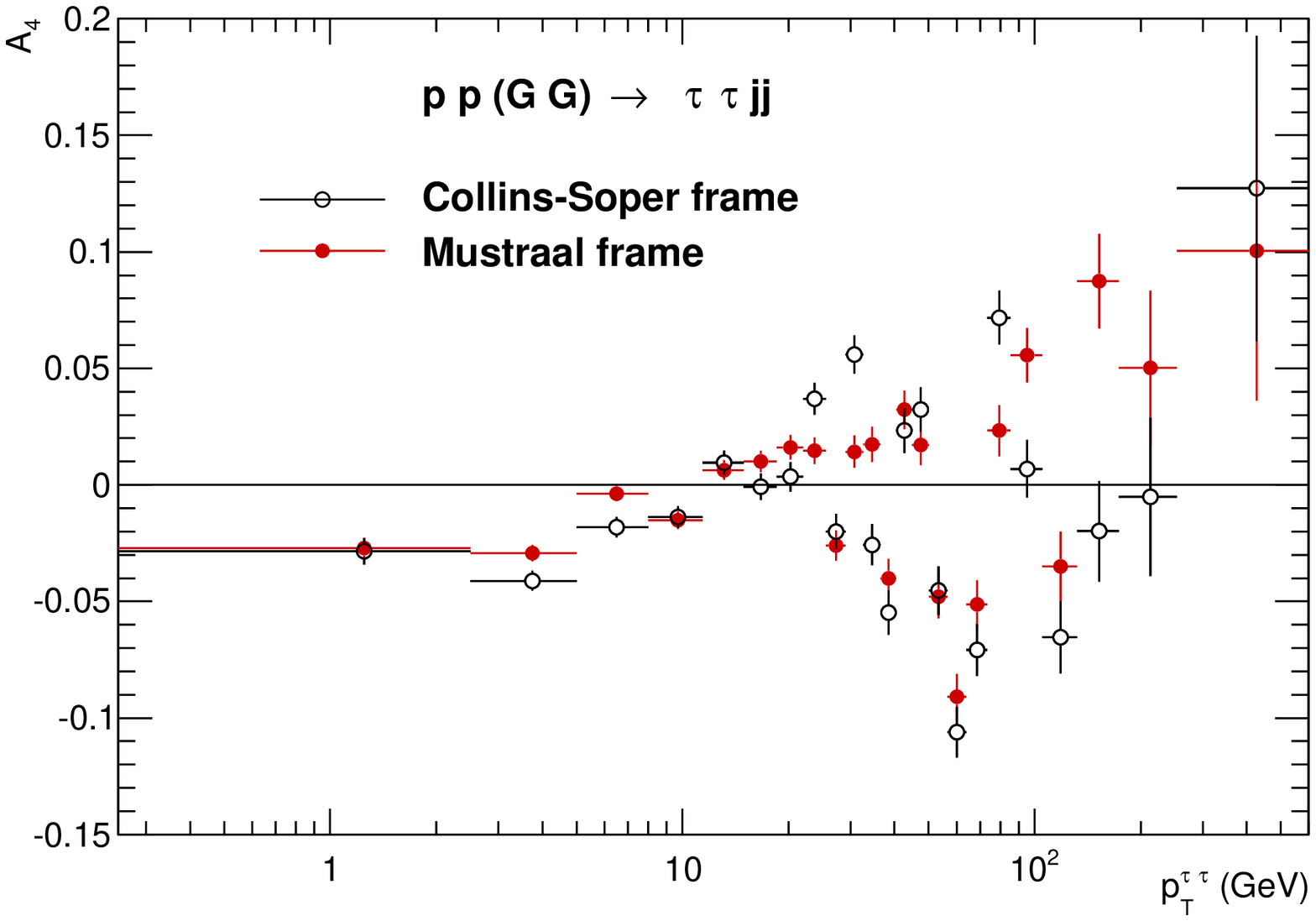}
}
\end{center}
\caption{ 
The $A_i$ coefficients of Eq.~(\ref{Eq:master2})) calculated in Collins-Soper (black) and in {\tt Mustraal} (red) frames 
for $p p (G G)  \to \tau \tau j j $ process generated with {\tt MadGraph}.
Details of initialization are given in Section \ref{sec:numerical}.
\label{fig:Ai2jets2g} }
\end{figure}

As the next step, we turn our attention to processes with two jets in final state, first
of unspecified type, Fig.~\ref{fig:Ai2jets},  and later for processes of quark anti-quark annihilation only,
Fig.~\ref{fig:Ai2jetsIzo0}. Then the final state jets    originate dominantly from a pair of gluons or 
from an intermediate gluon decaying to a quark-antiquark pair.
These  cases are similar from the point of view of spin structure to the single gluon emission.
Finally, the last two figures are for the processes where one, Fig.~\ref{fig:Ai2jets1g}, 
or two gluons, Fig.~\ref{fig:Ai2jets2g}, are present in the initial state.

The separation between processes is based on the flavour of incoming partons (quark, anti-quark or gluon), as stored for each event in the generated sample.
 This allows splitting the sample of two jets events  into groups 
of different spin nature of the intermediate states.
 Such separation cannot be universal, or applied to data of course, where
we work with multiple production mechanisms. It is convenient for 
illustration purposes; it could be explored with  more formal explanations following
reasoning as in Refs.~\cite{Was:2004ig,vanHameren:2008dy}, see Section~\ref{Sec:coefsMUSTvar}. 

\subsection{Results with NLO simulation}
\label{NNLO}
So far, we have discussed results for samples of fixed order tree level matrix elements of single or double parton (jet) emission.
These results represent a particular sector of Drell-Yan processes. In general, configurations with a variable number of jets and
effects of loop corrections and parton shower of initial state should be used to complete our studies. We have performed this task partially only, 
with the help of 40M weighted  $Z+j$ events generated with  {\tt Powheg+MiNLO} Monte Carlo, 
again for $pp$ collisions at 13 TeV, invariant mass of lepton pair restricted
to range $ 80-100$ GeV, and the effective EW scheme~\footnote{Note that the axial coupling and as a consequence $A_4$ coefficient 
will be different  than in case of {\tt MadGraph} initialisation
even in the zero transverse momenta limit.}
 with $\sin^2\theta_W = 0.23147$  but otherwise of default initialization.
The {\tt PowhegBox v2} generator \cite{Nason:2004rx,Alioli:2010xd}, augmented with {\tt MiNLO} method for choices of 
scales~\cite{Hamilton:2012np} and inclusion of Sudakov form factors~\cite{Hamilton:2012rf}, by construction achieves
NLO accuracy for distributions involving finite non-zero transverse momenta of the lepton system.

\begin{figure}
  \begin{center}                               
{
   \includegraphics[width=7.5cm,angle=0]{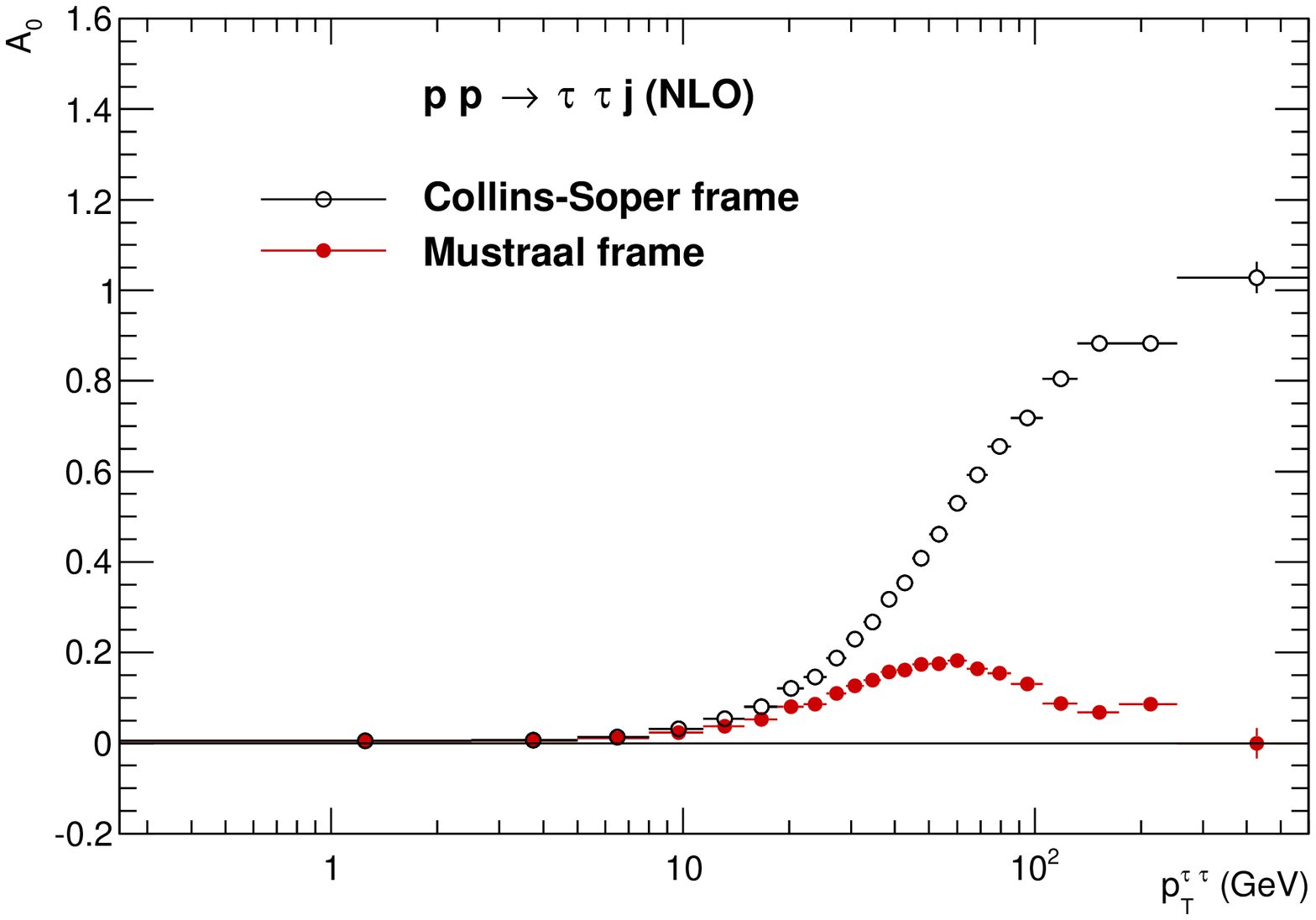}
   \includegraphics[width=7.5cm,angle=0]{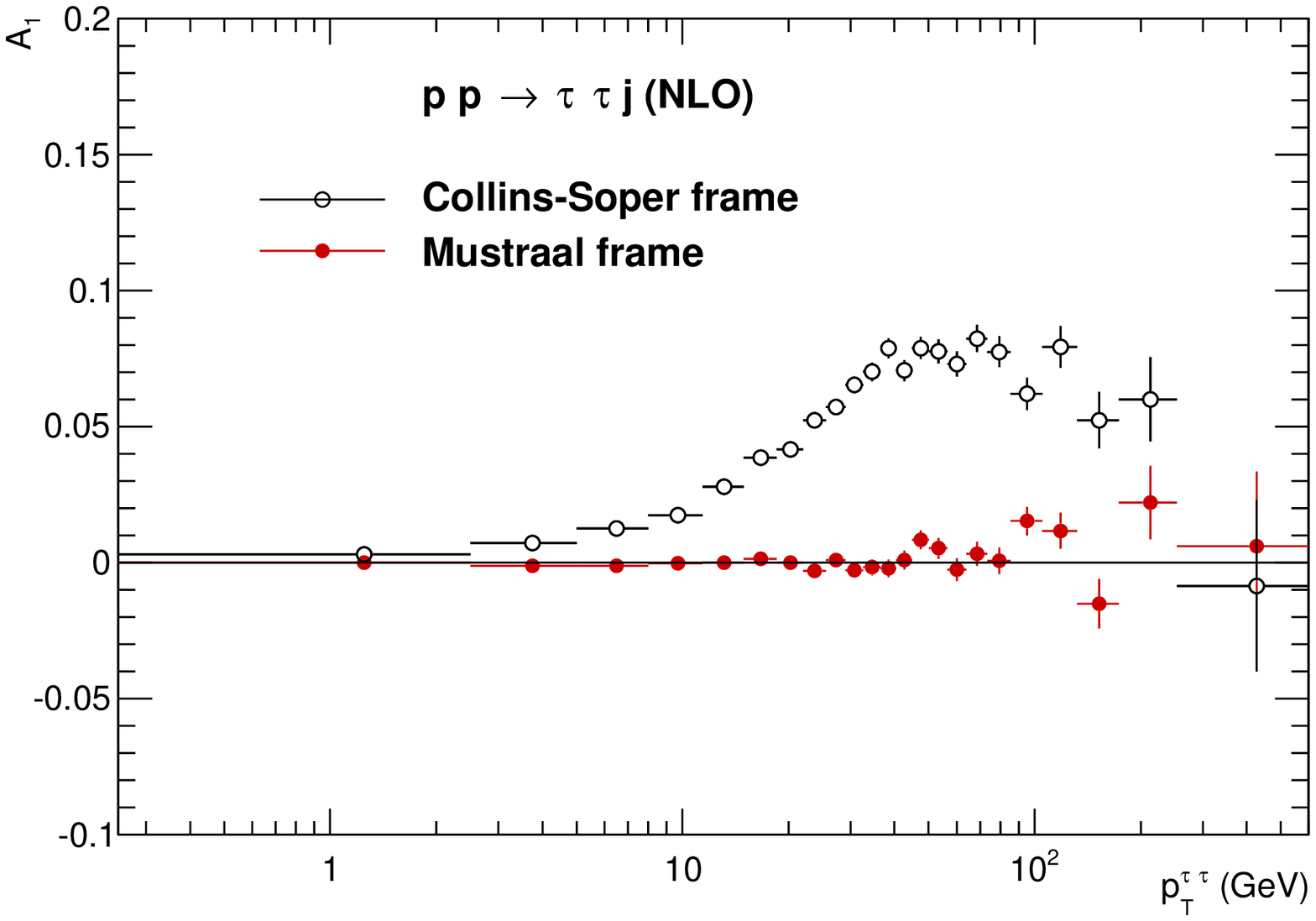}
   \includegraphics[width=7.5cm,angle=0]{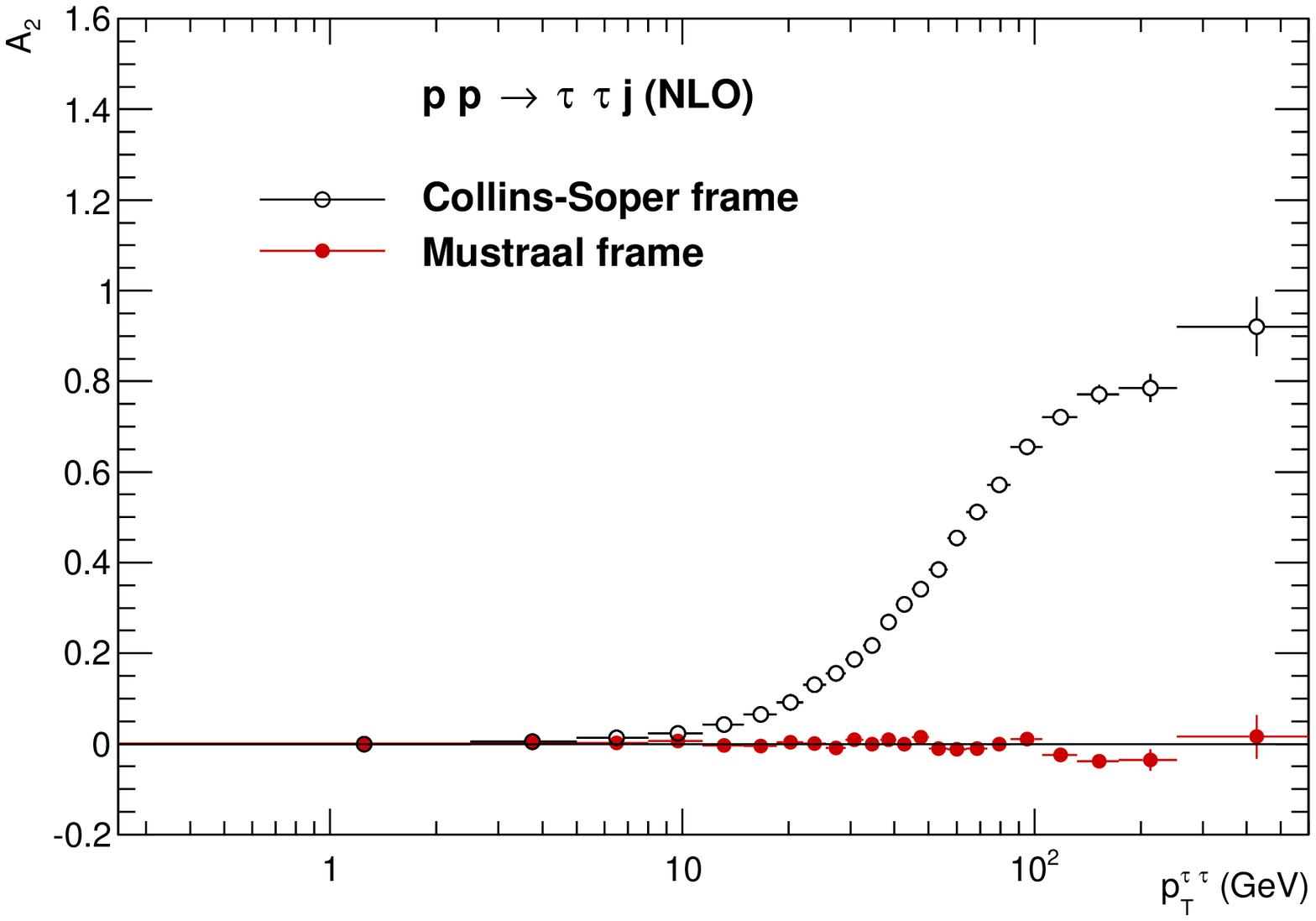}
   \includegraphics[width=7.5cm,angle=0]{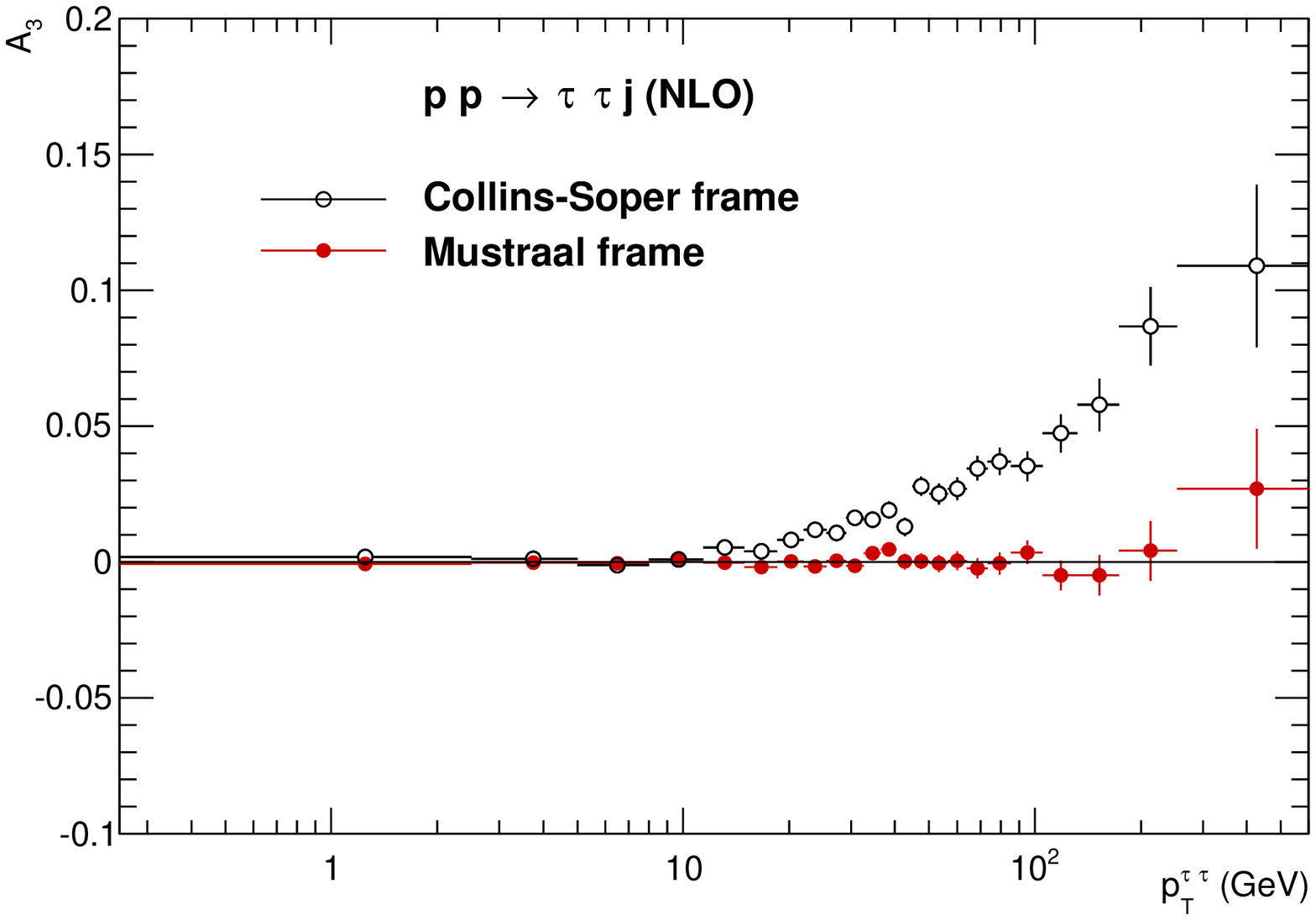}
   \includegraphics[width=7.5cm,angle=0]{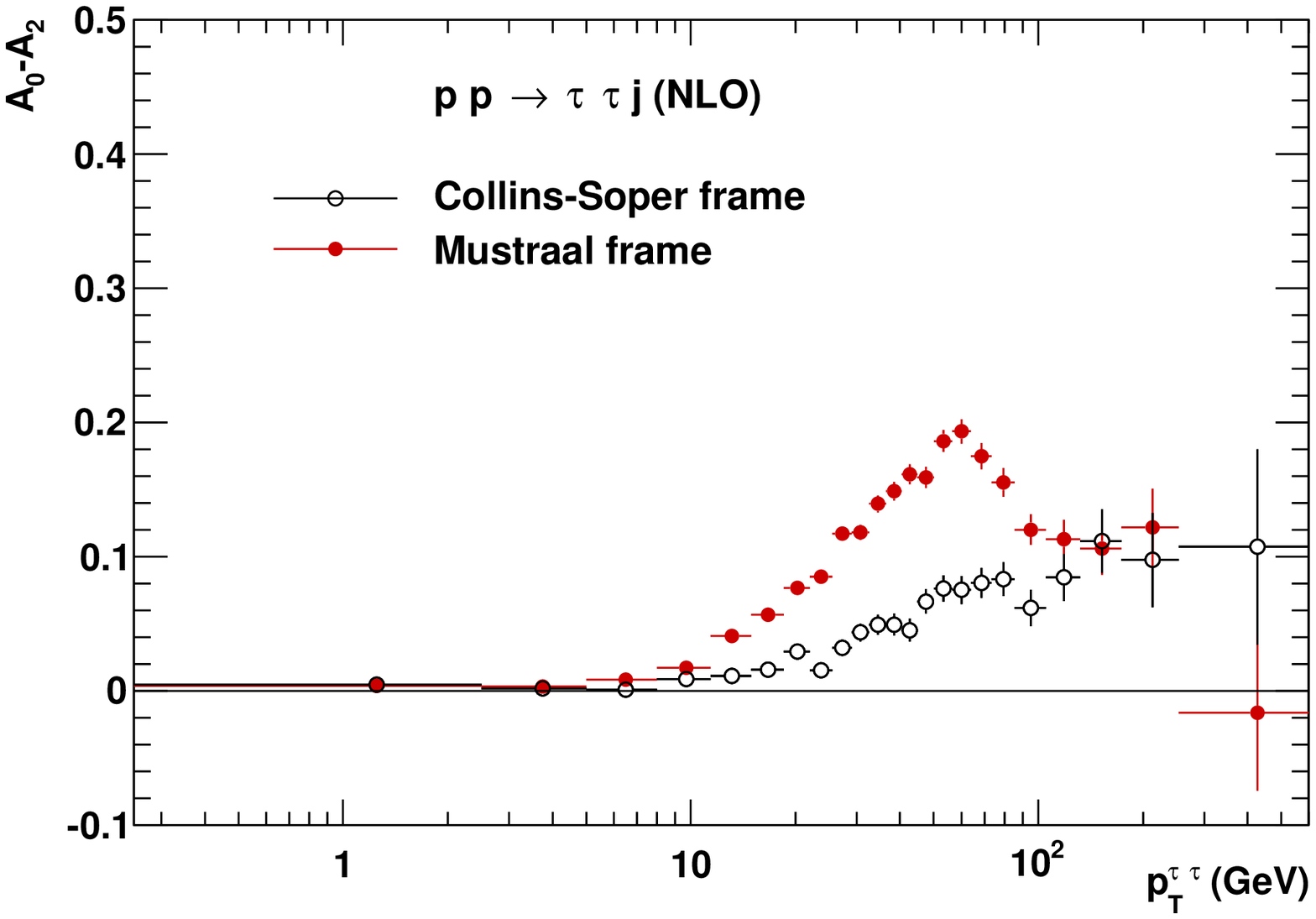}
   \includegraphics[width=7.5cm,angle=0]{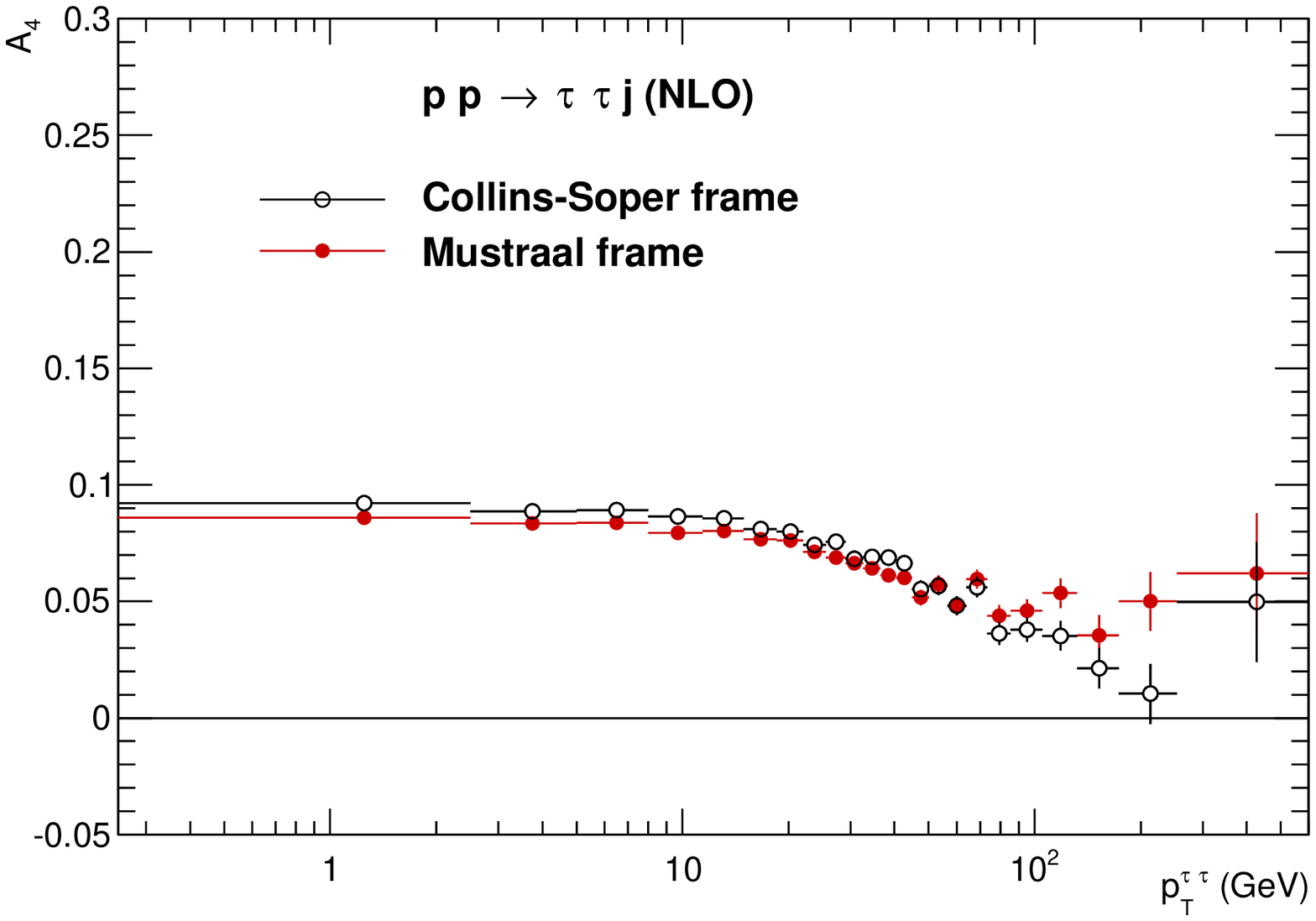}
}
\end{center}
\caption{ 
The $A_i$ coefficients of Eq.~(\ref{Eq:master2}))  calculated in Collins-Soper (black) and in {\tt Mustraal} (red) frames 
for $p p  \to \tau \tau j $ (NLO) process generated with {\tt Powheg+MiNLO}.
Details of initialization are given in Section \ref{sec:numerical}.
\label{Fig:AiPwhegMiNLO} }
\end{figure}

In Fig.~\ref{Fig:AiPwhegMiNLO} results are completed for such a sample, 
 following the same approach as in our previous plots. Comparisons of results
using {\tt Mustraal} and Collins-Soper frames feature again the usual pattern.
This confirms the robustness of our conclusions.
Further studies should include not only parton shower effects in the 
initial and final state but detector reconstruction as well. 
This is however beyond the scope of the present paper. 

\section{ Main features of results}\label{sec:interpretation}
In all collected plots, we have compared results 
for the cases when Collins-Soper and {\tt Mustraal} frames were used
for the cross-section decomposition into angular coefficients and spherical polynomials.
We can conclude that the first choice was better suitable to measure the effects 
of QCD dynamics, whereas in the second choice electroweak effects were far more dominating the picture, even in case 
of high $p_T$ jets. This may indicate which choice is better for the particular use. It is important that in both cases
we rely on precisely measured leptons. 

To confirm the origin of differences between {\tt Mustraal} frame and Collins-Soper frame observed for $A_0$, $A_1$, $A_2$ and $A_3$ coefficients,
we have analysed sample of $pp \to \tau \tau j$ events generated with {\tt MadGraph} configured in the EW sector with
$\sin^2\theta_W = 0.25$. In this way the $A_3$  and $A_4$ are set close zero 
over full $p_T^{\tau \tau}$ range because the dominant contributions from the $Z$-boson peak
was minimised, as they are proportional to $a_v^{\ell} \sim (1/4 - \sin^2\theta_W ) = 0$ (in case of  {\tt Mustraal} $A_3$ was close to zero
even for $\sin^2\theta_W = 0.22222$). We have observed that the differences 
in  $A_0$, $A_1$ and $A_2$ coefficients between Collins-Soper and {\tt Mustraal} frames were quite similar as in the case of 
$\sin^2\theta_W = 1 - m_W^2/m_Z^2 = 0.22222$, which has been used for numerical
results presented for {\tt MadGraph}. This observation suggests
that differences in  $A_0$, $A_1$ and $A_2$ of two frames are due to QCD effects,
and do not depend substantially on the EW sector.
Therefore  differences in  $A_i$
can be used to provide a precision reference point for 
discussing the dynamics of QCD in the Drell-Yan processes, similarly as the
$\phi^*$ observable \cite{Aad:2012wfa,Aad:2015auj}.  

As stated already in \cite{Mirkes:1992hu}, the theoretical uncertainties due to the choice of the 
factorisation and renormalization scales are very small for the cross section ratios $A_i$.
Most of the uncertainties of the structure functions and of the choice of the 
factorisation scheme cancels in the ratios too. 

\subsection{Angular coefficients in Collins-Soper frame}

The coefficients $A_0$ and $A_2$ are increasing functions of $p_T^V$ (where $V$ denotes $W$ or $Z$ boson) and the deviations 
from lowest order expectation ($A_0 = A_2 = 0$) are quite large even at modest value of $p_T^V$,
i.e. for $p_T^V =  20 - 50$ GeV.
 These coefficients are exactly equal at LO, thanks to  Lam-Tung relation~\cite{Lam:1978pu}.
This is no longer true at NLO but predicted corrections are of less than 10\% 
\cite{arXiv9406381}. They were recently measured ~\cite{ATLASDIS2016}, 
and it seems that in fact  larger corrections are needed.
The deviation of $A_1$ and $A_3$ from zero is much smaller even at large $p_T^V$. 

At  the $Z$-boson peak, the $A_4$ coefficient is proportional to the product of vector and axial couplings of the intermediate vector
boson to leptons and quarks, $\simeq(v_{\ell}/ a_{\ell} v_q /a_q)$, and is therefore sensitive to the Weinberg angle~$\theta_W$.

\subsection{Variations of  angular coefficients if {\tt Mustraal} frame is used} \label{Sec:coefsMUSTvar}
In general, for this choice of rest-frame, only $A_4$ is substantially different from zero and is fairly consistent 
with the same coefficient of Collins-Soper frame.
The electroweak sector of the interactions dominates in the decomposition to spherical polynomials. Sensitivity to  $\sin^2 \theta_W$ remains,
but in contrast sensitivity to QCD effects is minimized.

One may wonder, why the property of the matrix element of single gluon emission 
extends to other cases, when final states feature quark or there are 
two outgoing partons, gluonic or quarks. In principle, to understand this, one would 
need to study the properties of all spin amplitudes, e.g. following approach of 
~\cite{Was:2004ig,vanHameren:2008dy}. For the purpose of the present paper,
such an ambitious task is not necessary.
We restrict ourselves   to inspection of  numerical results only.
We are encouraged 
that more formal arguments may be provided if needed, 
up to the precision limits as
explained in~\cite{Kleiss:1990jv}. 

The QCD impact on angular coefficients in the Collins-Soper choice
of frames is dominated by the jet(s) geometry. The observation 
can be made that the effects is due to the choice of frames orientation 
for which angles $\theta$ and  $\phi$ 
are defined. 

On the first look, one could observe that at LO QCD the Lam-Tung relation breaks 
in {\tt Mustraal} case for $q G$ scattering subprocess and holds in the case of Collins-Soper frame. 
For now we don't have explanation of this effect. 
However, at NLO QCD as we can observe from 
Fig.~\ref{Fig:AiPwhegMiNLO}, the breaking of the rule itself 
is comparable for the two choices of the frames, especially at larger 
values of $p_T^{\tau\tau}$. One should not forget that the {\tt Mustraal} 
choice is not, and should not be expected to set $A_0$ and $A_2$ to zero, 
for the parton level processes of incoming gluons, the two coefficients 
are nonetheless sizably smaller in this case.   

\section{Summary}\label{sec:summary} 

The interest in decomposition of results for measurement of final states in Drell-Yan processes at the LHC
into coefficients of second order spherical polynomials for angular
distributions of leptons in the lepton-pair rest-frame, was
recently confirmed by experimental publications \cite{Aaltonen:2011nr,Khachatryan:2015paa,ATLASDIS2016}.
Inspired by those measurements, we have investigated  possible variants of choice for quantization frames orientation.
We have exploited geometrical properties of matrix elements for  vector 
quanta emissions known since early 80's.

We have found that the choice based on those properties 
helps to separate effects due to the Electroweak couplings and due 
to strong interactions: emissions of jets. Most of the coefficients can 
be eliminated or their size can be substantially reduced.
The  approach  of stochastic attribution to each event orientation of reference 
frame, based on the properties of matrix element for single 
gluon emission was further simplified, to eliminate dependence 
on coupling constants and to become  of a purely geometrical nature.

We have checked that such an approach retains sufficiently its nature  and 
that it can be generalized to the case of events 
featuring more than one jet and/or  of quark jets too. 
For that purpose, we have confronted  numerical results of angular coefficients for all parton level processes
with one or two outgoing jets (not necessarily gluonic one), for our
{\tt Mustraal} orientation of frame, with the one of Collins-Soper. 
We have found that with our choice, most of the coefficients 
in the expansion vanish or are reduced substantially, except for the one existing already at partonic Born level.

This may help to separate the interpretation of experimental results  into quantities sensitive to electroweak and strong interaction effects.   
Because directions of leptons are measured with very high precision, measurement of angular coefficients can be more effective 
 than direct measurement of hadronic jets, even if in principle, 
no more physics input is obtained in such a way. The gain of precision may be 
substantial, as in case of $\phi^*_\eta$ observable \cite{Aad:2012wfa,Aad:2015auj}, but it requires more experimentally oriented studies. 
The  approach  of stochastic attribution to each event orientation of reference 
frame may be useful for the optimalization of  {\tt TauSpinner} methods~\cite{Banerjee:2012ez,Kalinowski:2016qcd}, when $(2\to 2)$ Born level amplitudes
are used.

\vskip 1 cm
\centerline{\bf \Large Acknowledgments}
\vskip 0.5 cm
E.R-W would like to thank  A. Armbruster, O. Fedin, D. Froidevaux 
and M. Vincter for numerous inspiring discussions on the angular
decomposition and reference frames for the Drell-Yan process
description.

We would like to thank W. Kotlarski for providing us with samples generated with {\tt MadGraph} which were 
used for numerical results presented here. We  acknowledge  PLGrid Infrastructure of the Academic 
Computer Centre CYFRONET AGH in Krakow, Poland, where majority of numerical calculations were performed.

E.R-W was partially supported by the funds of Polish National Science
Center under decision UMO-2014/15/ST2/00049. 
Z.W was was partially supported by the funds of Polish National Science
Center under decision DEC-2012/04/M/ST2/00240.

\vskip 0.5 cm

\begingroup\begin{thebibliography}{10}

\bibitem{Aad:2008zzm}
{ATLAS} Collaboration, {\em JINST} {\bf 3} (2008)
S08003.

\bibitem{Chatrchyan:2008aa}
{CMS} Collaboration, {\em JINST} {\bf 3} (2008)
S08004.

\bibitem{Aad:2012tfa}
{ATLAS} Collaboration, G.~Aad {\em et al.}, {\em Phys. Lett.} {\bf B716} (2012)
  1--29,
\href{http://www.arXiv.org/abs/1207.7214}{{\tt 1207.7214}}.

\bibitem{Chatrchyan:2012xdj}
{CMS} Collaboration, S.~Chatrchyan {\em et al.}, {\em Phys. Lett.} {\bf B716}
  (2012) 30--61,
\href{http://www.arXiv.org/abs/1207.7235}{{\tt 1207.7235}}.

\bibitem{ATLAS-CONF-2015-044}
{CMS and ATLAS} Collaboration, ``{Measurements of the Higgs boson production
  and decay rates and constraints on its couplings from a combined ATLAS and
  CMS analysis of the LHC $pp$ collision data at $\sqrt{s}$ = 7 and 8~TeV}'',
  preprint ATLAS-CONF-2015-044, CMS-PAS-HIG-15-002 CERN, Geneva (September,
  2015).

\bibitem{Aad:2015baa}
{ATLAS} Collaboration, G.~Aad {\em et al.}, {\em JHEP} {\bf 10} (2015) 134,
\href{http://www.arXiv.org/abs/1508.06608}{{\tt 1508.06608}}.

\bibitem{Aad:2015iea}
{ATLAS} Collaboration, G.~Aad {\em et al.}, {\em JHEP} {\bf 10} (2015) 054,
\href{http://www.arXiv.org/abs/1507.05525}{{\tt 1507.05525}}.

\bibitem{Khachatryan:2016kdk}
{CMS} Collaboration, V.~Khachatryan {\em et al.}, {\em Phys. Lett. B} (2016)
\href{http://www.arXiv.org/abs/1602.06581}{{\tt 1602.06581}}.

\bibitem{Aad:2015auj}
{ATLAS} Collaboration, G.~Aad {\em et al.},
\href{http://www.arXiv.org/abs/1512.02192}{{\tt 1512.02192}}.

\bibitem{Aad:2015uau}
{ATLAS} Collaboration, G.~Aad {\em et al.}, {\em JHEP} {\bf 09} (2015) 049,
\href{http://www.arXiv.org/abs/1503.03709}{{\tt 1503.03709}}.

\bibitem{CMS:2014jea}
{CMS} Collaboration, V.~Khachatryan {\em et al.}, {\em Eur. Phys. J.} {\bf C75}
  (2015), no.~4 147,
\href{http://www.arXiv.org/abs/1412.1115}{{\tt 1412.1115}}.

\bibitem{Khachatryan:2016yte}
{CMS} Collaboration, V.~Khachatryan {\em et al.}, {\em Submitted to: Eur. Phys.
  J. C} (2016)
\href{http://www.arXiv.org/abs/1601.04768}{{\tt 1601.04768}}.

\bibitem{Catani:2015vma}
S.~Catani, D.~de~Florian, G.~Ferrera, and M.~Grazzini, {\em JHEP} {\bf 12}
  (2015) 047,
\href{http://www.arXiv.org/abs/1507.06937}{{\tt 1507.06937}}.

\bibitem{Grazzini:2015wpa}
M.~Grazzini, S.~Kallweit, D.~Rathlev, and M.~Wiesemann, {\em JHEP} {\bf 08}
  (2015) 154,
\href{http://www.arXiv.org/abs/1507.02565}{{\tt 1507.02565}}.

\bibitem{Dittmaier:2015rxo}
S.~Dittmaier, A.~Huss, and C.~Schwinn, {\em Nucl. Phys.} {\bf B904} (2016)
  216--252,
\href{http://www.arXiv.org/abs/1511.08016}{{\tt 1511.08016}}.

\bibitem{Collins:1977iv}
J.~C. Collins and D.~E. Soper, {\em Phys. Rev.} {\bf D16} (1977)
2219.

\bibitem{Collins:1989gx}
J.~C. Collins, D.~E. Soper, and G.~F. Sterman, {\em Adv. Ser. Direct. High
  Energy Phys.} {\bf 5} (1989) 1--91,
\href{http://www.arXiv.org/abs/hep-ph/0409313}{{\tt hep-ph/0409313}}.

\bibitem{Berends:1983mi}
F.~A. Berends, R.~Kleiss, and S.~Jadach, {\em Comput. Phys. Commun.} {\bf 29}
  (1983)
185--200.

\bibitem{Kleiss:1990jv}
R.~Kleiss, {\em Nucl. Phys.} {\bf B347} (1990)
67--85.

\bibitem{Barberio:1990ms}
E.~Barberio, B.~van Eijk, and Z.~W\c{a}s, {\em Comput. Phys. Commun.} {\bf 66}
  (1991)
115.

\bibitem{Mirkes:1992hu}
E.~Mirkes, {\em Nucl. Phys.} {\bf B387} (1992)
3--85.

\bibitem{RichterWas:1994ep}
E.~Richter-Was, {\em Z. Phys.} {\bf C64} (1994)
227--240.

\bibitem{RichterWas:1993ta}
E.~Richter-Was, {\em Z. Phys.} {\bf C61} (1994)
323--340.

\bibitem{Alwall:2011uj}
J.~Alwall, M.~Herquet, F.~Maltoni, O.~Mattelaer, and T.~Stelzer, {\em JHEP}
  {\bf 06} (2011) 128,
\href{http://www.arXiv.org/abs/1106.0522}{{\tt 1106.0522}}.

\bibitem{Nason:2004rx}
P.~Nason, {\em JHEP} {\bf 11} (2004) 040,
\href{http://www.arXiv.org/abs/hep-ph/0409146}{{\tt hep-ph/0409146}}.

\bibitem{Alioli:2010xd}
S.~Alioli, P.~Nason, C.~Oleari, and E.~Re, {\em JHEP} {\bf 06} (2010) 043,
\href{http://www.arXiv.org/abs/1002.2581}{{\tt 1002.2581}}.

\bibitem{DrellYan70}
S.~D. Drell and T.~M. Yan, {\em Phys. Rev. Lett.} {\bf 25} (1970) 316.

\bibitem{Collins82}
J.~C. Collins, D.~E. Soper, and G.~Sterman, {\em Phys. Lett. B} {\bf 109}
  (1982) 388.

\bibitem{Collins84}
J.~C. Collins, D.~E. Soper, and G.~Sterman, {\em Phys. Lett. B} {\bf 134}
  (1984) 263.

\bibitem{Collins85}
J.~C. Collins, D.~E. Soper, and G.~Sterman, {\em Phys. Lett. B} {\bf 261}
  (1985) 105.

\bibitem{Collins88}
J.~C. Collins, D.~E. Soper, and G.~Sterman, {\em Phys. Lett. B} {\bf 308}
  (1988) 833.

\bibitem{Altarelli78}
G.~Altarelli, R.~K. Ellis, and G.~Martinelli, {\em Nucl. Phys. B} {\bf 143}
  (1978) 521.

\bibitem{Altarelli79}
G.~Altarelli, R.~K. Ellis, and G.~Martinelli, {\em Nucl. Phys. B} {\bf 157}
  (1979) 461.

\bibitem{Matsuura89}
T.~Matsuura and et~al., {\em Nucl. Phys. B} {\bf 319} (1989) 570.

\bibitem{Matsuura91}
T.~Matsuura, W.~L. van Neerven, and T.~Matsuura, {\em Nucl. Phys. B} {\bf 359}
  (1991) 343.

\bibitem{Melnikov06a}
K.~Melnikov and F.~Petriello, {\em Phys. Rev. Lett.} {\bf 96} (2006) 231803.

\bibitem{Melnikov06b}
K.~Melnikov and F.~Petriello, {\em Phys. Rev. D} {\bf 74} (2006) 114017.

\bibitem{Catani09}
S.~Catani and et~al., {\em Phys. Rev. Lett.} {\bf 103} (2009) 082001.

\bibitem{Kalinowski:2016qcd}
J.~Kalinowski, W.~Kotlarski, E.~Richter-Was, and Z.~Was,
\href{http://www.arXiv.org/abs/1604.00964}{{\tt 1604.00964}}.

\bibitem{Barze':2013yca}
L.~Barze, G.~Montagna, P.~Nason, O.~Nicrosini, F.~Piccinini, and A.~Vicini,
  {\em Eur. Phys. J.} {\bf C73} (2013), no.~6 2474,
\href{http://www.arXiv.org/abs/1302.4606}{{\tt 1302.4606}}.

\bibitem{Dittmaier:2014qza}
S.~Dittmaier, A.~Huss, and C.~Schwinn, {\em Nucl. Phys.} {\bf B885} (2014)
  318--372,
\href{http://www.arXiv.org/abs/1403.3216}{{\tt 1403.3216}}.

\bibitem{Kulesza:1999gm}
A.~Kulesza and W.~J. Stirling, {\em Nucl. Phys.} {\bf B555} (1999) 279--305,
\href{http://www.arXiv.org/abs/9902234}{{\tt 9902234}}.

\bibitem{Aaltonen:2011nr}
{CDF} Collaboration, T.~Aaltonen {\em et al.}, {\em Phys. Rev. Lett.} {\bf 106}
  (2011) 241801,
\href{http://www.arXiv.org/abs/1103.5699}{{\tt 1103.5699}}.

\bibitem{Khachatryan:2015paa}
{CMS} Collaboration, V.~Khachatryan {\em et al.}, {\em Phys. Lett.} {\bf B750}
  (2015) 154--175,
\href{http://www.arXiv.org/abs/1504.03512}{{\tt 1504.03512}}.

\bibitem{ATLASDIS2016}
{ATLAS} Collaboration, G.~Aad {\em et al.}, Presented at DIS2016.

\bibitem{arXiv9406381}
E.~Mirkes and J.~Ohnemus, {\em Phys. Rev. D} {\bf 50} (1994) 5692,
  \href{http://www.arXiv.org/abs/9406381}{{\tt 9406381}}.

\bibitem{Karlberg:2014qua}
A.~Karlberg, E.~Re, and G.~Zanderighi, {\em JHEP} {\bf 09} (2014) 134,
\href{http://www.arXiv.org/abs/1407.2940}{{\tt 1407.2940}}.

\bibitem{Gavin:2010az}
R.~Gavin, Y.~Li, F.~Petriello, and S.~Quackenbush, {\em Comput.Phys.Commun.}
  {\bf 182} (2011) 2388--2403,
\href{http://www.arXiv.org/abs/1011.3540}{{\tt 1011.3540}}.

\bibitem{CarloniCalame:2007cd}
C.~M. Carloni~Calame, G.~Montagna, O.~Nicrosini, and A.~Vicini, {\em JHEP} {\bf
  10} (2007) 109,
\href{http://www.arXiv.org/abs/0710.1722}{{\tt 0710.1722}}.

\bibitem{koralz4:1994}
S.~Jadach, B.~F.~L. Ward, and Z.~W\c{a}s, {\em Comput. Phys. Commun.} {\bf 79}
  (1994) 503.

\bibitem{Davidson:2010ew}
N.~Davidson, T.~Przedzinski, and Z.~Was,
\href{http://www.arXiv.org/abs/1011.0937}{{\tt 1011.0937}}.

\bibitem{Peng:2015spa}
J.-C. Peng, W.-C. Chang, R.~E. McClellan, and O.~Teryaev,
\href{http://www.arXiv.org/abs/1511.08932}{{\tt 1511.08932}}.

\bibitem{Faccioli:2011pn}
P.~Faccioli, C.~Lourenco, J.~Seixas, and H.~K. Wohri, {\em Phys. Rev.} {\bf
  D83} (2011) 056008,
\href{http://www.arXiv.org/abs/1102.3946}{{\tt 1102.3946}}.

\bibitem{RichterWas:1987gk}
E.~Richter-Was and Z.~Was, ``{Hadron spin in the QCD improved parton model.}'',
  preprint CPT-87/P-2044 (1987).

\bibitem{RichterWas:1988gi}
E.~Richter-Was and Z.~Was, ``{Orbital angular momentum in the QCD
  evolution.}'', preprint CPT-87/P-2080 (1988).

\bibitem{Pumplin:2002vw}
J.~Pumplin, D.~R. Stump, J.~Huston, H.~L. Lai, P.~M. Nadolsky, and W.~K. Tung,
  {\em JHEP} {\bf 07} (2002) 012,
\href{http://www.arXiv.org/abs/hep-ph/0201195}{{\tt hep-ph/0201195}}.

\bibitem{Lam:1978pu}
C.~Lam and W.-K. Tung, {\em Phys.Rev.} {\bf D18} (1978)
2447.

\bibitem{Was:2004ig}
Z.~Was, {\em Eur. Phys. J.} {\bf C44} (2005) 489--503,
\href{http://www.arXiv.org/abs/hep-ph/0406045}{{\tt hep-ph/0406045}}.

\bibitem{vanHameren:2008dy}
A.~van Hameren and Z.~Was, {\em Eur.Phys.J.} {\bf C61} (2009) 33--49,
\href{http://www.arXiv.org/abs/0802.2182}{{\tt 0802.2182}}.

\bibitem{Hamilton:2012np}
K.~Hamilton, P.~Nason, and G.~Zanderighi, {\em JHEP} {\bf 10} (2012) 155,
\href{http://www.arXiv.org/abs/1206.3572}{{\tt 1206.3572}}.

\bibitem{Hamilton:2012rf}
K.~Hamilton, P.~Nason, C.~Oleari, and G.~Zanderighi, {\em JHEP} {\bf 05} (2013)
  082,
\href{http://www.arXiv.org/abs/1212.4504}{{\tt 1212.4504}}.

\bibitem{Aad:2012wfa}
{ATLAS} Collaboration, G.~Aad {\em et al.}, {\em Phys. Lett.} {\bf B720} (2013)
  32--51,
\href{http://www.arXiv.org/abs/1211.6899}{{\tt 1211.6899}}.

\bibitem{Banerjee:2012ez}
S.~Banerjee, J.~Kalinowski, W.~Kotlarski, T.~Przedzinski, and Z.~Was, {\em
  Eur.Phys.J.} {\bf C73} (2013) 2313,
\href{http://www.arXiv.org/abs/1212.2873}{{\tt 1212.2873}}.

\end{thebibliography}\endgroup
\providecommand{\href}[2]{#2}
\clearpage
\appendix
\end{document}